\documentclass[12pt]{article}
\usepackage{epsfig}
\usepackage{rotating}

\title{Light Meson Spectroscopy}
\author{Stephen Godfrey\footnote{{\sf godfrey@physics.carleton.ca}}\\
{\em Ottawa-Carleton Institute for Physics}\\
{\em Department of Physics, Carleton University, Ottawa CANADA, K1S 5B6}\\[0.5in]
Jim Napolitano\footnote{{\sf napolj@rpi.edu}}\\
{\em Department of Physics, Applied Physics, and Astronomy}\\
{\em Rensselaer Polytechnic Institute, Troy, NY USA, 12180-3590}}

\date{Submitted to Review of Modern Physics\\November 12, 1998}

\begin{document}
\maketitle

% ======================================================================
% \include{rev-v6-a}
% ======================================================================
\newpage
\begin{abstract}

We survey the current status of light meson spectroscopy.
%Established meson states are compared to the predictions of the 
%constituent quark model.
%Discrepancies which may point to hadron physics
%beyond the quark model are discussed.
We begin with a general
introduction to meson spectroscopy and and its importance in understanding
the physical states of Quantum Chromo Dynamics (QCD).
Phemenological models
of hadron spectroscopy are described with particular emphasis on
the constituent quark
model and the qualitative features it predicts for the meson spectrum.
We next discuss
expectations for hadrons lying outside the quark model,
such as hadron states with excited gluonic degrees of freedom.
These states include so-called
{\em hybrids} and {\em glueballs}, as well as {\em multiquark} states.
The established meson states are compared to the
quark model predictions and we find that 
most meson states are well described by the quark
model.  However, a number of states in the light-quark sector do not fit
in well, suggesting the existence of hadronic states
with additional degrees of freedom.  We end
with a brief description of future directions in meson spectroscopy.

%\pacs{PACS numbers: 01.30.Rr, 12.39.-x, 14.40.-n}

\end{abstract}

\newpage
\tableofcontents
%\newpage

% ======================================================================
%\include{rev-v6-1}
% ======================================================================
\newpage
\section{INTRODUCTION}

Meson physics and the strong interactions have been intimately connected
since pions were first introduced by Yukawa to explain the inter-nucleon
force (Yukawa, 1935).  
Since that time, our knowledge of mesons and in parallel, our
understanding of the strong interactions, has undergone  several major
revisions.  Our present understanding of the strong interactions
is that it is 
described by the non-Abelian gauge field 
theory Quantum Chromodynamics (QCD) (Fritzch, 1971; Gross, 1973; 
Weinberg, 1973) which 
describes the interactions of quarks and gluons.  Once again, it appears
that mesons are the ideal laboratory for the study of 
strong interactions in the strongly coupled non-perturbative regime.  Even
though in QCD
we have a theory of the strong interactions,  we know very little about
the physical states of the theory.  Until we can both predict the 
properties of the physical states of the theory and confirm these 
predictions by experiment we can hardly claim to understand QCD.  The 
understanding of QCD has implications beyond hadron physics.  
For example, it is possible that at high energies the weak interactions become 
strong, so that strongly interacting field theories may be 
relevant to the mechanism of electroweak symmetry breaking.
In QCD we have an example of such a theory where we 
can test our understanding against experiment.  
The study of meson spectroscopy is the laboratory that
will hopefully elucidate this theory.

To a large extent our knowledge 
of hadron physics is based on phenomenological models and in particular,
the constituent quark model 
(G. Zweig, 1964; M. Gell-Mann, 1964)\footnote{Introductions to the 
quark model are given in (Isgur, 1980) and (Rosner, 1984)}.
Meson and baryon spectroscopy is 
described surprisingly well as composite objects made of constituent
objects --- valence quarks.  We will refer to these hadrons,
described by only valence quark configurations, as ``conventional''.
Most QCD motivated models, however,
predict other types of hadrons with explicit glue degrees of freedom.  These
are the {\em glueballs}, 
which have no constituent quarks in them at all and  are 
entirely described in terms of gluonic fields, and {\em hybrids}
which have both constituent
quarks and excited gluon degrees of freedom\footnote{Some recent 
reviews on this subject are given by 
Close 1988, Godfrey 1989, and Isgur 1989a.}.
It is the prospect of these new forms of
hadronic matter that has led to continued excitement among 
hadron spectroscopists. 

To be able to interpret the nature of new resonances it is important
that we have a template against which to compare observed states with
theoretical predictions.  The constituent quark model offers the most 
complete description of hadron properties and is probably the 
most successful phenomenological 
model of hadron structure.
But to use it as a template to find new 
physics, it is very
important that we test the quark model 
against known states to understand its strengths and weaknesses.
At one extreme, if 
we find that there is too much discrepancy with experiment, we may decide
that it  is not such a good model after all, and we should start over again.
On the other hand, if it gives general agreement with experiment, discrepancies
may indicate the need for new physics; 
either because approximations to the model are
not appropriate, or new types of hadrons which cannot be explained by
the quark model.  To understand our reliance on this very simple, and
perhaps naive, model it is useful to look at the historical evolution of
our understanding of hadron physics. 

Mesons were first introduced by Yukawa (Yukawa 1935) with pions acting 
as the exchange 
bosons responsible for the strong interactions between nucleons.  
With the advent of higher energy accelerators, a whole zoo of mesons 
and baryons
appeared, leading to a confused state of understanding.  
Eventually, by arranging the various 
mesons and baryons into multiplets according to their quantum numbers,
patterns started to emerge. It was recognized that hadrons of
a given $J^{PC}$ arranged themselves into representations of the group
SU(3) although none of the observed states seemed to correspond to
the fundamental triplet representation.  
In an important conceptual leap Zweig (Zweig, 1964) and Gell-Man
(Gell-Man, 1964) postulated that
mesons and baryons were in fact composite objects with mesons made of
a quark-antiquark pair and baryons made of three quarks.  Zweig referred to 
these constituent spin $1\over 2$ fermions as aces and Gell-Man referred to 
them as quarks.  
By taking this simple picture seriously, the qualitative properties
of hadrons were explained quite well.  Serious problems remained however.  
In the
``naive'' quark model the spin $3\over 2$ baryons, the constituent 
quarks'
spin wavefunctions were symmetric as were their flavour wavefunctions.  
Being fermions, the
baryon wavefunction should be antisymmetric in the quark quantum numbers.
This would imply that either quarks obeyed some sort of bizzare para-statistics
or that the ground state spatial wavefunction was antisymmetric.  Yet no
reasonable models could be constructed to give this result.  To avoid this
result, Greenberg 
postulated that quarks had another quantum number which
he named colour, with respect to which the quark wavefunctions could
be antisymmetrized (Greenberg, 1964).  
The serious shortcoming of this model was that
no quarks were observed.  Most physicists took the view that if they
could not be observed they were nothing more than a convenient bookkeeping
device.

By the beginning of the 1970's it was becoming clear that the weak 
interactions could be explained by gauge theories (Glashow 1961, 
Weinberg 1967, Salam 1968).  If this was the case,
it seemed reasonable that the strong interactions should also be described
using the same formalism.  ``Gauging'' the colour degree of freedom leads to
Quantum Chromodynamics, a non-Abelian gauge theory based on the group SU(3),
as the theory of the strong interactions (Fritzch, 1971; Gross, 1973; 
Weinberg, 1973). 

Nevertheless there was still considerable skepticism about the existence
of quarks since they had never been seen.  This situation changed when in 
November 1974 very narrow hadron resonances were discovered simultaneously
at Brookhaven National Laboratory (Aubert 1974)
and the Stanford Linear Accelerator Center (Augustin 1974).
These states, named the $J/\psi$,
were quickly interpreted 
as being bound states of a new heavy quark-- the charm quark.  Quark
models which incorporated the qualitative features of QCD, asymptotic freedom
and confinement, where able to reproduce the charmonium $(c\bar{c})$
spectrum rather well.
(Appelquist 1975a, 1975b, 1975c, Eichten 1975).
These developments, both experimental and theoretical, 
convinced all but a few
that quarks were real objects and were the building blocks of hadronic matter.
In a seminal paper on the subject,
deRujula, Georgi, and Glashow, (De R\'ujula 1975)
showed that these ideas could successfully be used to describe the 
phenomenology of light quark spectroscopy.

With the acceptance that QCD is the theory of the strong 
interactions comes the need to understand its physical states. 
Understanding the spectrum of hadrons reveals information on the 
non-perturbative aspects of QCD.
Unfortunately, calculating the properties of hadrons from the QCD Lagrangean
has proven to be a very difficult task in this strongly coupled non-linear
theory.  In the long term, the most promising technique is formulating the
theory on  a discrete space-time lattice (Creutz 1983a, 1983b, Kogut 
1979, 1983, Montvay 1994).  
By constructing interpolating
fields with the quantum numbers of physical hadrons and evaluating
their correlations on the lattice one is able to calculate hadron
properties from first principles.  Although a great deal of progress 
has been made, these calculations take
enormous amounts of computer time and progress has been slow.  
Additionally, a disadvantage of this approach is that one may obtain 
numerical results without any corresponding physical insight.

A less rigorous approach which has proven to be quite useful and
reasonably successful,
has been to use phenomenological models of hadron structure
to describe hadron properties.  These models
predict, in addition to the conventional 
$q\bar{q}$ mesons and $qqq$ baryons of the quark model;
multiquark states, glueballs, and hybrids.  
%The latter are 
%new forms of hadronic matter which have explicit glue degrees of freedom.
Probably the most pressing question in hadron spectroscopy
is whether these states do in fact exist and what their properties are.
However, the predictions of the various models can differ appreciably so 
that
experiment is needed to point the model builders in the right direction.

We will often refer to glueballs and hybrids as exotics because they lie 
outside the constituent quark model.  However they are not exotics in 
the sense that if they exist they are simply additional hadronic states
expected from QCD.  Nevertheless,
from the historical development of the field we 
see that the quark model provides 
a good framework on which to base further study.  If we find discrepancies 
everywhere it obviously fails and we should abandon it as a tool.  On the
other hand, since it does work reasonably well it gives us a criteria
on which to decide if we have discovered the new forms of hadronic
matter we are interested in; namely glueballs and hybrids.  

The present situation in light meson spectroscopy is that the constituent 
quark model works surprisingly well in describing most observed states.  
At the same time there are still many problems and puzzles that need to 
be understood and that might signal physics beyond the quark model.  
Although most QCD based models expect glueballs and hybrids and there
is mounting evidence that some have been found, thus
far no observed state has unambiguously been identified as one.
The best candidates are states with ``exotic quantum numbers'', that is
states which cannot be formed in the quark model.
Part of the problem and confusion is that the conventional mesons are
not understood well enough to rule out new states 
as conventional states and part of the problem
is that these exotics may  have properties
which have made them difficult to find up to now.  Despite 
these qualifications, there has been considerable recent progress in 
understanding the properties of exotic mesons that could help 
distinguish between conventional and exotic mesons.  With sufficient 
evidence,  a strong case can be made to label an experimentally 
observed state as an exotic hadron.
Thus, to have
any hope of distinguishing between conventional and exotic mesons it is
crucial that we understand conventional meson spectroscopy very well. 

The purpose of this review is to summarize the present status
of meson spectroscopy and identify puzzles, perhaps
pointing out measurements which could help resolve them.  To this end we will
begin with a discussion of the  theoretical 
ingredients relevant to this article.
In the course of this review we will refer to numerous experiments so 
in section III we briefly survey relevant experiments along with 
the attributes that contribute useful information to the study of 
mesons.
Since the eventual goal is to identify discrepancies between the observed
meson spectrum and conventional quark model predictions, in section IV
we will compare the predictions of 
one specific quark model with experiment.  This will allow us to identify 
discrepancies between the quark model and experiment which may signal
physics beyond conventional hadron spectroscopy.  In section
V we will go over these puzzles in detail
%, sector by sector,
to help decide whether the
discrepancy  is most likely a problem with the model, with a
confused state in experiment, or whether it most likely signals some
interesting new physics.  In section VI we will briefly outline some 
future facilities for the study of meson spectroscopy
that are under construction or that are being considered.
Finally, in section VII we will attempt to summarize our most interesting
findings.
Our hope is that the reader will see that meson spectroscopy is a
vibrant field.

Because of the breadth of this review we can only touch the surface of 
many interesting topics.  
There are a number of recent reviews of meson spectroscopy and related 
topics 
with emphasis somewhat different from that of the present one.  We 
strongly encourage the interested reader to refer to these reviews for 
further details.  
The reader is referred to the reviews by 
F. Close (1988), S. Cooper (1988), B. Diekmann (1988), 
T. Burnett and S. Sharpe (Burnett 1990), 
N. T\"{o}rnqvist (1990), C. Amsler and F. Myhrer (1991), K. Konigsman (1991), 
R. Landau (1996), C. Amsler (1998), and Barnes (1998).
In addition, the Review of 
Particle Physics (Particle Data Group (PDG), Caso 1998)
contains a wealth of information on the properties of mesons in its
tables of properties and mini-reviews on topics of special interest and 
should be consulted for further information.

% ======================================================================
%\include{rev-v6-2}
% ======================================================================
\newpage
\section{THEORETICAL OVERVIEW}
\label{sec:THEORY}

\subsection{Quantum Chromodynamics}

Quantum Chromodynamics (QCD),  the theory of the strong interactions,
(Fritzch 1971, Gross 1973, Weinberg 1973)
may be thought of as a generalization of
quantum electrodynamics (QED), our most successful physical theory.  QCD is 
described by the Lagrangean;
\begin{equation}
{\cal L}_{QCD} =\bar{q}_i (i\partial_\mu\gamma^\mu \delta_{ij}
 +g {\lambda_{ij} ^a\over 2}A_\mu^a\gamma^\mu -m\delta_{ij})
\gamma^\mu q_j -{1 \over 4} F_{\mu\nu}^a F^{a \mu\nu} 
\end{equation}
where
\begin{equation}
F^{\mu\nu}_a =\partial^\mu A^\nu_a -\partial^\nu A^\mu_a +gf_{abc} 
A_b^\mu A_c^\nu , 
\end{equation}
$ A^\mu_a $ are the gluon fields which transform according to the adjoint
representation of SU(3) with a=1,...,8, 
$ q_i$ are the quark fields with colour indices 
i={1,2,3}, $ g$ is the bare coupling, 
$m$ is the quark mass, and ${\lambda^i \over 2}$ are the generators of SU(3).
One immediately observes that quarks couple to gluons in much the same way
as the electron couples to photons with $e\gamma^\mu$ of QED
replaced by $g\gamma^\mu {\lambda\over 2}$ of QCD.  The significant difference
between QED and QCD is that in QCD the quarks come in coloured triplets
and the gluons in a colour octet where colour is
labelled by the Latin subscripts.
The non-Abelian group structure of SU(3) leads to nonlinear terms in the
field strength $F^{\mu\nu}$, which give rise to trilinear and quadratic
vertices in the theory so that gluons couple to themselves in addition
to interacting with quarks. This makes the theory nonlinear, very difficult 
to solve and leads to the confinement of colour.  
A consequence of this behavior appears to be the existence 
of new forms of hadronic matter
with excited gluonic degrees of freedom known as glueballs and hybrids 
(Close 1988).

Because of the difficulties in solving QCD exactly to obtain the properties of
the physical states of the theory, we have resorted to various approximation
methods.  The most promising of these is to redefine the problem on a
discrete spacetime
lattice, in analogy to the approach one might take in the numerical
solution of a difficult differential equation (Wilson 1974, Creutz 
1983a, 1983b, Kogut 1979, 1983, C. Michael 1995, 1997, Montvay 1994).  
For QCD, one
formulates the problem in terms of the path integral in Euclidean 
space-time  and evaluates expectation values of the appropriate operators
using a Monte-Carlo integration over the field configurations. 
Although progress
is being made on the problem, it requires enormous computer capacity 
so that progress is slow in making precise, detailed predictions of 
the properties of 
the physical states of the theory.  As a consequence, our 
understanding of hadrons continues to rely on 
insights obtained from experiment and QCD motivated models in addition 
to lattice QCD results.

\begin{figure}
\centerline{\epsfig{file=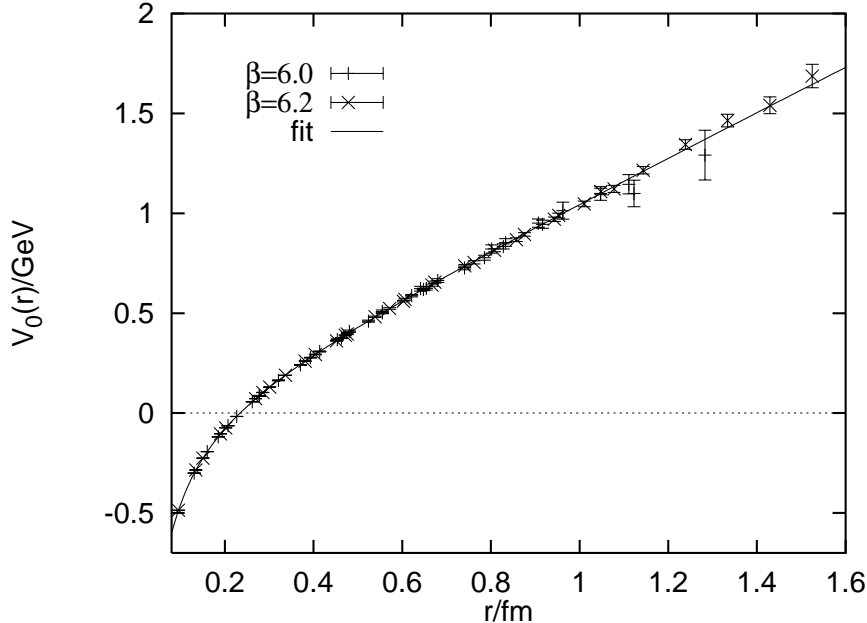,width=12.0cm,clip=}}
\caption[]{
The static $Q\bar{Q}$ potential from Bali, Schilling, and Wachter (Bali 1997).}
\label{fig:potential}
\end{figure}

In later sections we will use the predictions of QCD inspired  
models as the basis for interpreting the nature of the observed 
mesons.\footnote{For a recent review see Barnes 1996.}
It is therefore useful to sketch the QCD motivation for these models. 
We start with the quark-antiquark ($Q\bar{Q}$) 
potential in the limit of infinitely massive quarks
which can be used in the Schr\"odinger equation to
obtain the spectroscopy of heavy quarkonium.  This is analogous to the 
adiabatic potentials for diatomic molecules in molecular physics with 
the heavy quarks corresponding to slow moving nuclei and the gluonic 
fields corresponding to fast moving electrons.
The $Q\bar{Q}$ potential is found by calculating the
energy of a fixed, infinitely heavy, quark-antiquark pair given by the 
expectation value of what is known as the Wilson loop operator 
(Wilson 1974).  
%For finite $m_Q$, corrections of order $m_Q$ are expected.
The resulting potential is 
referred to as the static potential since the massive quarks do not move.
Limiting cases of the static potential are given by:
\begin{equation}
V(r) = br \quad \hbox{for}  \quad r>> {1\over\Lambda}
\end{equation}
where the constant $b$, the ``string tension'', is numerically found to 
be $b \simeq 0.18$~GeV$^2$ $\simeq$ 0.9~GeV/fm
and
\begin{equation}
V(r) \sim {-4\over 3} {\alpha_s \over r} \quad \hbox{for} 
\quad r<< {1\over \Lambda}
\end{equation}
where $\alpha_s $ is the strong coupling.
The result of one such lattice calculation is shown in
Fig.~\ref{fig:potential}, taken from (Bali, 1997).  
The lattice potential, $V(r)$, can then be used to determine the spectrum 
of $b\bar{b}$ mesons by solving the Schr\"odinger equation since the 
$b$-quark motion is approximately non-relativistic.

\begin{figure}
\centerline{\epsfig{file=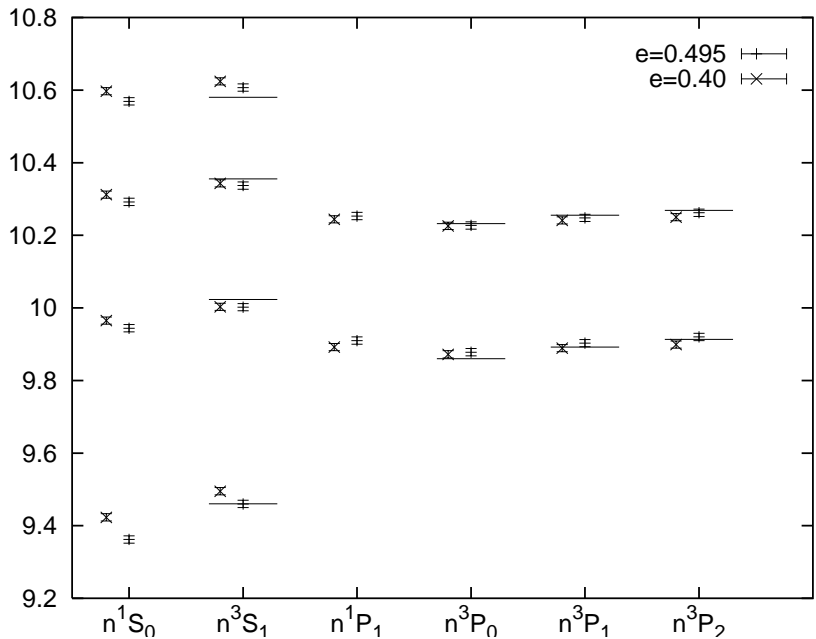,width=12.0cm,clip=}}
\caption[]{The bottomonium spectrum (in GeV) calculated using potentials from
lattice QCD.  The horizontal lines are experimental results. From Bali,
Schilling, and Wachter (Bali 96).}
\label{fig:lat-b}
\end{figure}

There are also spin dependent forces between the quarks analogous
to the spin dependent forces in the hydrogen atom which give rise to the
fine and hyperfine structure in atomic spectroscopy.  To obtain the spin
dependent potentials in QCD the Wilson loop is expanded 
in the inverse quark mass which gives the terms next order in an 
expansion in $v^2/c^2$ (Eichten 1979, 1981, Gromes
1984a, 1984b):
\begin{eqnarray}
V_{spin}(r)  & = & \left( {1\over {2m_1^2}} \vec{L}\cdot \vec{S}_1
+{1\over {2m_2^2}} \vec{L}\cdot\vec{S}_2 \right) {1\over r} {d \over{dr}}
\left( V(r) + 2V_1 (r) \right) \nonumber \\
& & \quad + {1\over {m_1 m_2}} \vec{L}\cdot (\vec{S}_1 +\vec{S}_2 ) {1\over r}
{{d V_2 (r) }\over {dr}}  \\
& & \quad + {1\over{m_1 m_2}} (\hat{r}\cdot\vec{S}_1 \; \hat{r}\cdot\vec{S}_2
-{1\over 3} \vec{S}_1\cdot \vec{S}_2 ) V_3 (r) 
 + {1\over{3m_1 m_2}} \vec{S}_1 \cdot \vec{S}_2 \; V_4 (r) 
\nonumber
\end{eqnarray}
where $\vec{S}_1$, $\vec{S}_2$, and $\vec{L}$ are the quark and 
antiquark spins and 
relative orbital angular momentum and $V(r)$ is the 
interquark potential defined by the Wilson loop operator.
The spin dependent potentials, $V_1 - V_4$, can be related
to  correlation functions of the colour electric and colour magnetic
fields.  For example $V_3$ and $V_4$ are given by (Eichten
1979, 1981,  Gromes 1984a, 1984b):
\begin{equation}
( {{r_i r_j}\over {r^2}} - {1\over 3} \delta_{ij} ) V_3(r)
+ {1\over 3} \delta_{ij} V_4 (r) ={\lim_{T\to \infty}} {{g^2}\over T}
\int^T_0 {{ dtdt' \langle B_i (0,t) B_j (r,t' ) \rangle } \over
{\langle 1 \rangle}} 
\end{equation}
which can be evaluated using nonperturbative techniques, in particular
lattice QCD.  The lattice results can be compared to the phenomenological
expectations that the magnetic correlations are short range as
expected from one gluon exchange:
\begin{equation}
 V_3(r) = {{4\alpha_s}\over {r^3}} \qquad \hbox{and} \qquad
V_4(r)  ={{8\pi}\over 3} \alpha_s \delta^3 (r) 
\end{equation}
The lattice results do indeed agree 
with these spin dependent potentials (Bali 1997,
Campostrini 1986, 1987).  Bali Schilling and Wachter 
(Bali 1997) have studied the $b\bar{b}$ and $c\bar{c}$ spectra 
using the potentials 
calculated using lattice QCD.  They solved the Schr\"odinger equation 
using the spin-independent potential and treated the spin-dependent 
and other relativistic corrections (not discussed here) as 
perturbations. The resulting beautyonium spectrum is shown in 
Fig~\ref{fig:lat-b} and is found to be in reasonable agreement with 
the experimental $\Upsilon$ spectrum.  The main 
deviations between experiment and prediction are 
due to the quenched approximation and the neglect of higher order 
relativistic corrections.
Direct lattice calculations of spin-dependent splittings also agree
with the measured splittings (Davies 1998).

Although historically, the spin dependent potentials were obtained 
phenomenologically by comparing the observed quarkonium spectrum 
($c\bar{c}$ and $b\bar{b}$) 
to the predictions of potential models, it turns out that the resulting 
potentials are in reasonable agreement with those obtained from lattice 
QCD. 
The phenomenological spin-dependent potential typically
assumes a Lorentz vector one gluon exchange for the short distance 
piece which results in terms analogous to the 
Breit-Fermi Hamiltonian in atomic physics,
and a Lorentz scalar linear confining piece.  
The resulting spin dependent Hamiltonian is then of the form:
\begin{equation}
H_{spin} = H^{hyp}_{ij} + H^{s.o.(cm)}_{ij} + H^{s.o.(tp)}_{ij} 
\end{equation}
where
\begin{equation}
H^{hyp}_{ij}={{4\alpha_s (r)}\over{3 m_i m_j}}
\left\{ { { {8\pi} \over{3} }
\vec{S}_i \cdot \vec{S}_j \, \delta^3 (\vec{r}_{ij})
+{1\over{r^3_{ij}} }
\left[ { { {3\vec{S}_i\cdot\vec{r}_{ij} \vec{S}_j \cdot\vec{r}_{ij} }
\over{r_{ij}^2}}-\vec{S}_{i} \cdot\vec{S}_{j} }\right] }\right\}
\label{eq:hyp}
\end{equation}
is the colour hyperfine interaction,
\begin{equation}
H^{s.o.(cm)}_{ij}={{4\alpha_s(r)}\over{3r^3_{ij}}}
\left( { {1\over{m_i} } + {1\over{m_j} } }\right)
\left( { {{\vec{S}_i}\over{m_i} } +{{\vec{S}_j}\over{m_j}} }\right) 
\cdot\vec{L}
\label{eq:socm}
\end{equation}
is the spin-orbit 
colour magnetic piece arising from the one-gluon exchange and
\begin{equation}
H^{s.o.(tp)}_{ij} = {-1\over{2r_{ij}} } \; 
{ {\partial V(r) } \over {\partial r_{ij}} }\;
\left( { {{\vec{S}_i}\over{m_i^2}} + {\vec{S}_j\over
{m^2_j}} }\right)\cdot\vec{L}
\label{eq:sotp}
\end{equation}
is the spin-orbit  Thomas precession term where $V(r)$ is the interquark 
potential given by the Wilson loop.  Note that the contribution 
arising from one gluon exchange is of opposite sign to the 
contribution from the confining potential.
$\alpha_s(r)$ is the running coupling constant of QCD.  

\subsection{Colour Singlets in QCD}

Because of confinement only colour singlet objects can exist as 
physical hadrons.  Coloured quarks form the fundamental triplet $(3)$
representation of the SU(3) colour gauge group and antiquarks the 
conjugate $\bar{3}$ representation.  Therefore, a quark-antiquark pair can 
combine to form a colour singlet as can three quarks
while a quark-quark pair cannot.  Other states are also possible, for 
example $q\bar{q}q\bar{q}$, and it is dynamical question whether such 
multiquark systems are realized in nature as single multiquark states, 
as two distinct $q\bar{q}$ states, or as a loosely associated system 
of colour singlet mesons analogous to a diatomic molecule.  Colour 
singlets can also be constructed with gluons ($g$).  Glueballs are hadrons 
with no valence quark content and hybrids are made up of valence 
quarks and antiquarks and an explicit gluon degree of freedom.  Of 
course, life is not so simple, and there is apriori no reason that the 
physical mesons cannot be linear combinations of $q\bar{q}$, 
$q\bar{q}q\bar{q}$, $gg$, and $q\bar{q}g$.  

\subsection{The Constituent Quark Model}

In the constituent quark model, conventional mesons
are bound states of a spin $1\over 2$ quark and
a spin $1\over 2$ antiquark bound by a QCD motivated phenomenological 
potential such as the one described above. 
The quark and antiquark spins  combine into a spin singlet or triplet with 
total spin $S = 0$ or 1 respectively.
$S$ is coupled to the orbital angular momentum $L$ resulting in
total angular momentum $J=L$ for the singlet state and $J=L-1,\; L, \; L+1$
for the triplet states.  In spectroscopic notation the resulting state 
is denoted by $n^{2S+1}L_J$ 
with S for $L=0$, P for $L=1$, D for $L=2$, and F, G, H, for $L=3,4,5$ etc.
Parity is given by $ P(q\bar{q},L) = (-1)^{L+1} $ and C-Parity is also 
defined for neutral self-conjugate mesons and is given by
$ C(q\bar{q},L,S)= (-1)^{L+S} $.
Thus, the ground state vector meson with $J^{PC}=1^{--}$ is the $1^3S_1$ 
quark model state.

    The light-quark quarkonia are composed of $u$, $d$ or $s$ quarks.
Since the $u$ and $d$ quarks are quite similar in mass, $\sim 5-10$~MeV,
which is much smaller than the intrinsic mass scale of 
QCD, it is convenient to treat them as members of an ``isosopin'' 
doublet with the resulting SU(2) isospin 
an approximate symmetry of the strong 
interactions.  Combining $u,d$ and $\bar{u},\;\bar{d}$  
into mesons forms isospin singlet and triplet multiplets.
We will use the symbol $n$
(for non-strangeness) to generically stand for $u$ or $d$. 
Thus,  one should read
\begin{equation}
 n\bar{n} = (u\bar{u}\pm d\bar{d})/\sqrt{2}
\end{equation}
with $+$ for the isoscalar mesons  and  $-$ for
the neutral member of the isovector multiplet.

When dealing with the charged members of an isospin triplet, it is
customary to refer to their $C$-parity as the $C$-parity of the neutral
member of the multiplet, $C_n$.  
It is convenient, however, to introduce a new
quantum number, $G\equiv C_n(-1)^I=\pm1$.  The so-called $G$-parity is
defined, and has the same value, for all members of the multiplet. It is
important to note, however, that unlike $C$-parity, $G$-parity is
{\em not} an exact symmetry of the strong interaction because of the
inherent approximate nature of isospin.

Hadrons containing $s$ quarks have similar properties to the $(u, \; 
d)$ systems so that mesons are arranged into $SU(3)$ flavour nonets; three 
isovector states $(u\bar{d}, \; u\bar{u}-d\bar{d}, \; d\bar{d})$, two 
isoscalar states $u\bar{u}+d\bar{d}, \; s\bar{s}$, and four strange 
$I=1/2$ states $u\bar{s},\; s\bar{u}, \; d\bar{s}, \; s\bar{d}$.
With the heavier strange quark mass, the $s\bar{s}$ isoscalar
states are sufficiently heavier than the $(u, \; d)$ $q\bar{q}$ states 
that there is little mixing between $s\bar{s}$ and the light $n\bar{n}$ 
states with the exception of the $\eta-\eta'$ system where
\begin{equation}
| \eta \rangle = \cos(\phi) |n\bar{n}\rangle 
- \sin(\phi) |s\bar{s}\rangle
\end{equation}
\begin{equation}
| \eta' \rangle = \sin(\phi) |n\bar{n} \rangle 
+ \cos(\phi) |s\bar{s}\rangle .
\end{equation}
where the flavour mixing angle $\phi \simeq 45^o$.
These states are often also expressed as linear combinations of flavour 
SU(3) octet and singlet states
\begin{equation}
| \eta \rangle = \cos(\theta) |8\rangle 
- \sin(\theta) |1\rangle
\end{equation}
\begin{equation}
| \eta' \rangle = \sin(\theta) |8\rangle 
+ \cos(\theta) |1\rangle .
\end{equation}
The two angles are trivially related by $\phi=\theta 
+\tan^{-1}(\sqrt{2})$.

Combining the spin and orbital angular momentum wavefunctions with 
the quark flavour wavefunctions
results in the meson states of Table \ref{tb:qn} where we have used the 
Particle Data Group naming conventions (PDG, Caso 1998).  
States not fitting into this picture are considered to be ``exotics''.
Thus, a meson with $J^{PC}=1^{-+}$
would be forbidden in the constituent quark model as would a doubly
charged meson $m^{++}$.
\begin{table}
\caption{The quantum numbers and names of conventional $q\bar{q}$ 
mesons.}
\label{tb:qn}
\begin{center}
\begin{tabular}{|l|l|l|c|c|c|c|} 
\hline
	&	&	$J^{PC}$ & I=1 & I=0  $(n\bar{n})$ 
	& I=0 $s\bar{s}$ & Strange	\\ \hline
 L=0	& S=0	&	$0^{-+}$ & $\pi$ & $\eta$ &$\eta'$ & $K$ \\
	& S=1	&	$1^{--}$ & $\rho$ & $\omega$ & $\phi$ & $K^*$ \\
\hline
 L=1   & S=0  &	$1^{+-}$ & $b_1$ & $h$   & $h'$   & $K_1$ \\
 	& S=1	&	$0^{++}$ & $a_0$ & $f_0$ & $f_0'$ & $K_0$ \\
	& 	&	$1^{++}$ & $a_1$ & $f_1$ & $f_1'$ & $K_1$ \\
	&	&	$2^{++}$ & $a_2$ & $f_2$ & $f_2'$ & $K_2^*$ \\
\hline
 L=2	& S=0	& $2^{-+}$ & $\pi_2$ & $\eta_2$ & $\eta_2'$ & $K_2$ \\
	& S=1	& $1^{--}$ & $\rho$  & $\omega$ & $\phi$ & $K_1^*$\\
	&	& $2^{--}$ & $\rho_2$ & $\omega_2$ & $\phi_2$ & $K_2$  \\
	&	& $3^{--}$ & $\rho_3$ & $\omega_3$ & $\phi_3$ & $K_3^*$\\
\hline
   .	& .	& .	& .		&	.	& .	&	.\\
   .	& .	& .	& .		&	.	& .	&	.\\
   .	& .	& .	& .		&	.	& .	&	.\\
\hline
\end{tabular}
\end{center}
\end{table}

\begin{figure}
\centerline{\epsfig{file=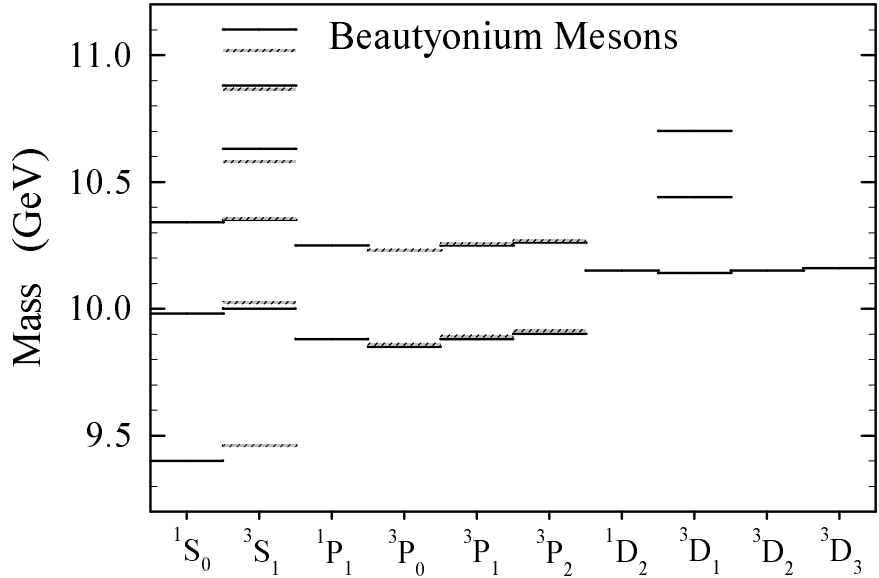,width=12.0cm,clip=}}
\caption[]{The $b\bar{b}$ spectrum from a quark potential model
(Godfrey, 1985a). The solid lines are the quark model 
predictions and the shaded regions are the experimental states.}
\label{fig:bb-mass}
\end{figure}

To obtain the meson spectrum one solves for the eigenvalues of 
the Schr\"{o}dinger
equation with a $q\bar{q}$ potential, including the spin dependent potentials,
and tunes the constituent quark masses to give agreement with experiment.
There is nothing fundamental about the values assigned
to the constituent quark masses, specific values are chosen simply 
to improve the
predictions of the model.  The relative positioning of the multiplets, i.e.
the 1S, 1P, 1D, 2S, 2P, ..., levels
is sensitive to the details of the potential.
Fig.~\ref{fig:bb-mass} shows the $b\bar{b}$ spectrum predicted by 
one representative model which gives reasonably
good agreement with experiment. The phenomenological, QCD motivated, 
linear plus Coulomb potential gives the observed multiplet positioning
and is consistent with the lattice potential described above.
The spin dependent potentials split the multiplets by
giving the spectrum ``fine'' and ``hyperfine'' structure analogous to
their counterparts in QED and reproduce the observed $b\bar{b}$ 
spectrum quite well..

The preceeding discussion may seem like a lengthy digression but it 
demonstrates an important result; that phenomenological models starting with
experimental measurements, and lattice calculations starting from the
underlying theory, find a common ground in the language of potential models.
That phenomenological models of heavy quarkonia work well and agree, at least
qualitatively, with the potentials predicted by quantum chromodynamics
using lattice QCD, is strong support for this approach, at least
for heavy quark systems.  The success for heavy quarkonia begs the question
about extending potential models to light-quark systems where the use
of the static potential is questionable.

\begin{figure}
\centerline{\epsfig{file=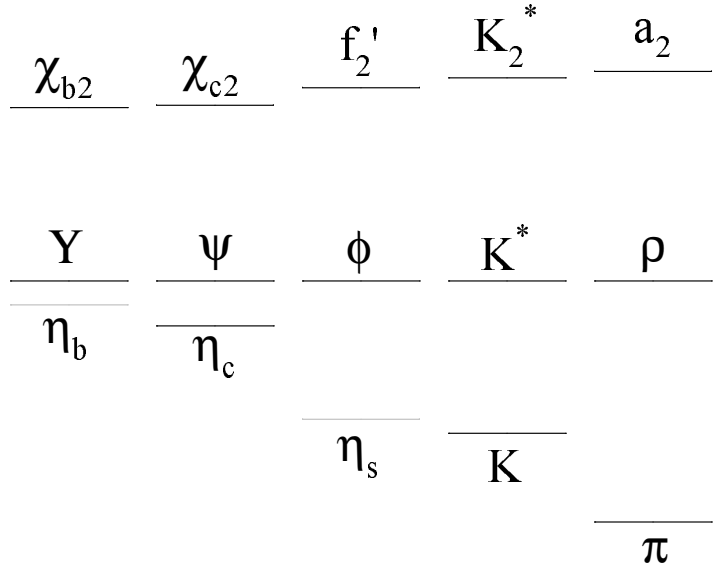,width=10.0cm,clip=}}
\caption[]{
The evolution of the $1^3P_2 - 1^3S_1$ and the $1^3S_1-1^1S_0$
splittings as a function of quark masses (Godfrey 1985a).
The splittings are drawn to scale.  Note that the $\eta_b$ and
the $\eta_s$ (dotted lines) are in fact calculations.}
\label{fig:splittings}
\end{figure}

In Fig.~\ref{fig:splittings}, we show
the evolution of the $1^3P_2 - 1^3S_1$ and the $1^3S_1-1^1S_0$
splittings as a function of quark masses (Godfrey 1985a).
In heavy quark systems the former
splitting is a non-relativistic orbital excitation analogous to the Lyman
$\alpha$ line in hydrogen while the latter is a Breit-Fermi (order p/m) 
relativistic correction analogous to the 21 cm line of hydrogen.  One can
see that there is a smooth evolution going from the heavy $b\bar{b}$ system,
where we believe potential models to be approximately valid, to the 
relativistic light quark systems.  We take this as evidence that 
qualitatively,
the basic structure in heavy and light systems are identical, the
main difference being that in the light quark systems the relativistic 
splittings are comparable to the orbital splittings.  Thus, to describe
the light quark hadrons we should include relativistic effects and the
characteristics expected from QCD (Capstick 1986). 
Ideally this would be done by deriving
the correct relativistic equations from QCD and solving them.  A less
ambitious approach is to model the exact  equations by including various
parameters.  In the following sections,
to interpret the spectrum of mesons with light quark content,
we will use the predictions of one such attempt at a 
``relativized'' quark model (Godfrey, 1985a, 1985c, 1986), 
which we take to be 
representative of the many similar models in the literature
(Stanley 1980, Carlson 1983a, 1983b, Gupta 1986, 
Brayshaw 1987, Crater 1988, Olson 1992, Fulcher 1993, Jean 1994).

In quark models, mesons are approximated by the valence
quark sector of the Fock space, in effect integrating out the degrees of
freedom below some distance scale, $\mu$.  This results in an
effective potential $V_{q\bar{q}}(\vec{p},\vec{r})$, whose dynamics
are governed by a Lorentz vector one-gluon-exchange
interaction at short distance and a linear Lorentz scalar
confining interaction.
In the relativized quark model mesons are 
described by the relativistic rest frame Schr\"odinger-type equation:
\begin{equation}
H | \Psi \rangle = [ H_0 +V_{q\overline{q}}(\vec{p},\vec{r}) +H_A ]
| \Psi \rangle = E | \Psi \rangle
\end{equation}
where $H_0$ is the kinetic energy operator, 
$H_A$ is the annihilation amplitude which we must consider in self conjugate
mesons where $q\overline{q}$ annihilation via gluons can contribute to the 
masses, and $V_{q\overline{q}} (\vec{p},\vec{r})$ is the effective 
quark-antiquark potential
which is found by equating the scattering
amplitude of free quarks with the potential, $V_{\rm eff}$, 
between bound quarks inside a meson (Gromes 1984b).
To first order in $(v/c)^2$ 
$V_{q\overline{q}} (\vec{p},\vec{r})$ reduces to the standard non-relativistic
result:
\begin{equation}
V_{q\overline{q}}(\vec{p},\vec{r})\to V(\vec{r}_{ij})=H^{conf}_{ij} +
H^{hyp}_{ij} +H^{s.o.}_{ij}
\end{equation}
where
\begin{equation}
H^{conf}_{ij} = - \frac{4}{3}\frac{\alpha_{s}(r)}{r} + br +C
\label{eq:conf}
\end{equation}
includes the spin independent linear confinement and Coulomb like interaction,
and $H^{hyp}_{ij}$ and $H^{s.o.}_{ij}$ are given by equations 
\ref{eq:hyp}-\ref{eq:sotp}.

The confinement potential, Eqn.~\ref{eq:conf},
includes the spin independent linear confinement and Coulomb like 
interactions.  The Coulomb piece dominates at short distance 
while the linear piece dominates at large distance. 
Because heavy quarkonium have smaller radii they are 
more sensitive to the short range colour-Coulomb interaction 
while the light quarks, especially the orbitally excited mesons, have 
larger radii and are more sensitive to the long range confining 
interaction.  Thus, measurement of both heavy quarkonium and mesons 
with light quark content probe different regions of the confinement 
potential and complement each other.  
The linear character of the trajectories of the orbital excitations
is a direct consequence of the linear
confining potential so that the experimental measurement of these masses
is a measure of the slope of the potential and 
will give information about the nature of confinement (Godfrey, 1985c).

The spin dependent parts of the potential consist of the colour hyperfine
interaction (Eqn.~\ref{eq:hyp}) and the spin-orbit interaction 
(Eqn.~\ref{eq:socm},\ref{eq:sotp}).
The colour hyperfine interaction 
is responsible for $^3S_1 - ^1S_0$ splitting in  $\rho -\pi$, $K^* -K$,
and $J/\Psi -\eta_c$.  In addition to multiplet splitting, 
the tensor term can cause mixings between states with the same quantum 
numbers related by $\Delta L=2$ such as $^3S_1$ and $^3D_1$.
The spin-orbit terms contribute to the splitting of the $L\ne 0$ 
multiplets.  For states with unequal mass quark and antiquark
where $C$-parity and $G$-parity are no longer good quantum numbers,
the spin-orbit terms can also contribute to $^1L_J-^3L_J$ mixing.
The spin-orbit interaction has two contributions,
$H^{s.o.(cm)}_{ij}$ and $H^{s.o.(tp)}_{ij}$ given by Eqns.~\ref{eq:socm}
and \ref{eq:sotp}.  Since the hyperfine term is
relatively short distance, it becomes less important for larger radii 
so that multiplet splittings become a measure of the 
spin-orbit splittings, with contributions of opposite sign coming from 
the short range Lorentz vector one-gluon-exchange 
and the long range Lorentz scalar linear confinement potential.
The ordering
of states within a multiplet of given orbital angular momentum gives
information on the relative importance of the two pieces (Schnitzer 
1984a, 1984b, Godfrey 1985b, 1985c, Isgur 1998).  
Thus, the multiplet splittings act as a probe of the
confinement potential,  providing information on non-perturbative QCD.

\subsection{Meson Decays}

While the quark potential model makes mass predictions for 
$q\bar{q}$ mesons, the couplings of these states are sensitive to the 
details of the meson wavefunctions and consequently provide an 
important test of our understanding of the internal structure of 
these states.  Knowledge of expected decay modes is also useful 
for meson searches and comparing the observed decay 
properties of mesons to the expectations of different interpretations, 
$q\bar{q}$ vs hybrid for example, is an important means of 
determining what they are.
Thus, the strong, 
electromagnetic, and weak  couplings of mesons can give 
important clues to the nature of an observed state.

\begin{figure}
\centerline{\epsfig{file=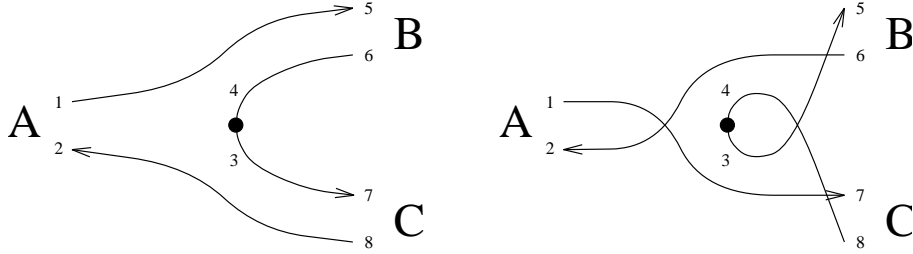,width=12.0cm,clip=}}
\caption[]{Diagrams contributing to the meson decay 
$A\to BC$. In many cases only one of these diagrams will contribute.}
\label{fig:decays}
\end{figure}

As a consequence, a successful 
model of strong decays would be a very useful tool in determining the nature of 
observed resonances.  A large number of models exist in the literature.  
In an important subset of 
models a quark-antiquark pair materializes and combines with 
the quark and antiquark of the original meson to form two new mesons.  
This process is described in Fig. \ref{fig:decays}.
The models differ in the details of how the quark pair creation
process occurs.  In one 
variant the $q\bar{q}$ pair originates in an intermediate gluon and is 
therefore formed in a $^3S_1$ state with $J^{PC}=1^{--}$ while in the 
other major variant, the $^3P_0$ model of LeYaouanc {\it et al.} 
(LeYaouanc 1973, 1974, 1975), 
the quark pair creation process is viewed as an 
inherantly nonperturbative process where the $q\bar{q}$ pair is 
formed with the quantum numbers of the vacuum, $J^{PC}=0^{++}$, and 
therefore is in a $^3P_0$ state.  Geiger and Swanson (Geiger 1994) 
performed a 
detailed study of these models and concluded that the decay width 
predictions of the $^3P_0$ model give better agreement with experiment.
A variation of this model is the flux-tube breaking model of Isgur and 
Kokoski (Kokoski 1987)
which assumes that the $q\bar{q}$ pair is most likely to be 
created in the region between the original quark and antiquark.  In 
practice the predictions of this variation do not differ significantly 
from those of the $^3P_0$ model as the overlap of the original and 
final state mesons is greatest in this central region.  Detailed 
decay predictions are given by Kokoski and Isgur (Kokoski 1987), 
Blundell and Godfrey (Blundell 1996), and Barnes Close Page and 
Swanson (Barnes 1997) which can be used to compare with experiment.
Comparing the partial decay widths of  non-$q\bar{q}$ candidates to 
quark model predictions provides an important tool for understanding
the nature of observed resonances when we 
discuss candidates for non-$q\bar{q}$ ``exotic'' mesons.

Electromagnetic couplings are another source of useful information 
about resonances.  A first example is two photon couplings which are 
measured in the reaction $e^+e^- \to e^+e^- + \hbox{hadrons}$.  In 
the cross section to the final state $\pi^0\pi^0$, for example, the 
$f_2(1270)$ can be seen as a clear bump.  From the cross section, the 
two-photon width times the branching fraction, 
$\Gamma_{\gamma\gamma}(f_2) \cdot B(f_2 \to \pi^0\pi^0)$, can be 
determined.  In principle the absolute two-photon widths can be calculated 
from  quark model wavefunctions (Godfrey 1985a) but for light 
$q\bar{q}$ mesons the results are sensitive to relativistic effects 
(Ackleh 1992).  To some extent this can be evaded when testing for 
$q\bar{q}$ candidates by comparing the relative rates of the possible 
members of a $u$, $d$, $s$ multiplet with the same $J^{PC}$.  The 
decay amplitude for $\gamma\gamma$ couplings involves the charge 
matrix element of two electromagnetic vertices
\begin{equation}
A(q\bar{q}\to \gamma\gamma) \propto \langle q\bar{q} | e_q^2 | 
0\rangle
\end{equation}  
This gives, for example, the relative amplitudes of 
\begin{equation}
\langle f : a : f' | e_q^2 | 0 \rangle
= \frac{(2/3)^2 +(-1/3)^2}{\sqrt{2}} : \frac{(2/3)^2 -(-1/3)^2}{\sqrt{2}}
: (-1/3)^2 
\end{equation}
which results in the relative $\gamma\gamma$ decay 
rates for I=0 : I=1 : $s\bar{s}$ mesons of the same state and 
neglecting phase space differences of
\begin{equation}
\Gamma_{\gamma\gamma} (f:a:f') = 25:9:2.
\end{equation}
The $L=2$ $\pi_2(1670)$ and probably the $\eta_2$ state have been 
observed in two-photon production (Antreasyan 1990, Behrend 1990, 
Karch 1992).  Because glueballs do not have valence quark content, the 
observation or non-observation of a state in two photon production 
provides another clue about the nature of an observed state.

Single photon transitions $(q\bar{q})_i\to \gamma (q\bar{q})_f$ 
is another useful measurement for identifying $q\bar{q}$ states.  
These have the characteristic pattern of rates based on flavour;
\begin{equation}
\Gamma ((q\bar{q})_i \to \gamma (q\bar{q}_f) = 9:4:1
\end{equation} 
for $\Delta I=1 : s\bar{s} : \Delta I=0$.
Although measurements of radiative transitions would be useful for the 
classification of higher $q\bar{q}$ and non-$q\bar{q}$ states only two
such transitions have been measured, $a_2\to \pi \gamma$ and $a_1 \to 
\pi \gamma$.  Since measurements of radiative transitions could 
determine the nature of controversial states such as the $f_1(1420)$, 
which will be discussed in Section V, they should be carried out if 
possible.  These measurements could be made in electroproduction, 
the inverse reaction.

\subsection{Mesons With Gluonic Excitations}

In addition to conventional hadrons it is expected that
other forms of hadronic matter exist with excited gluonic degrees of
freedom; glueballs which are made primarily of gluons,
and  hybrids which have both valence quarks and gluonic degrees of 
freedom (Jaffe 1976, Barnes 1984, 1985a, Chanowitz 1983a, 1983b, 
Close 1988, Godfrey 1989, Isgur 1989a).

\subsubsection{Glueballs}
\label{secII:glueballs}

Many different QCD based models and calculations make predictions for such
states; Bag Models (Barnes 1977, 1983c, DeTar 1983, Chanowitz 1983a,
Hasenfratz 1980).
Constituent Glue Models (Horn 1978), Flux Tube Models (Isgur 1983, 
1985a, 1985b), QCD Sum Rules (Latorre 1984), and Lattice Gauge Theory.  
Recent Lattice QCD calculations are converging towards 
agreement (Schierholz 1989, Michael 1989,
Sexton 1995b, Bali 1993, Teper 1995, Morningstar 1997)
although there is still some variation between the various 
calculations.  We expect that ultimately, 
the lattice results will be the most relevant 
since they originate from QCD. 

\begin{figure}
\begin{minipage}{3.0in}
\centerline{\epsfig{file=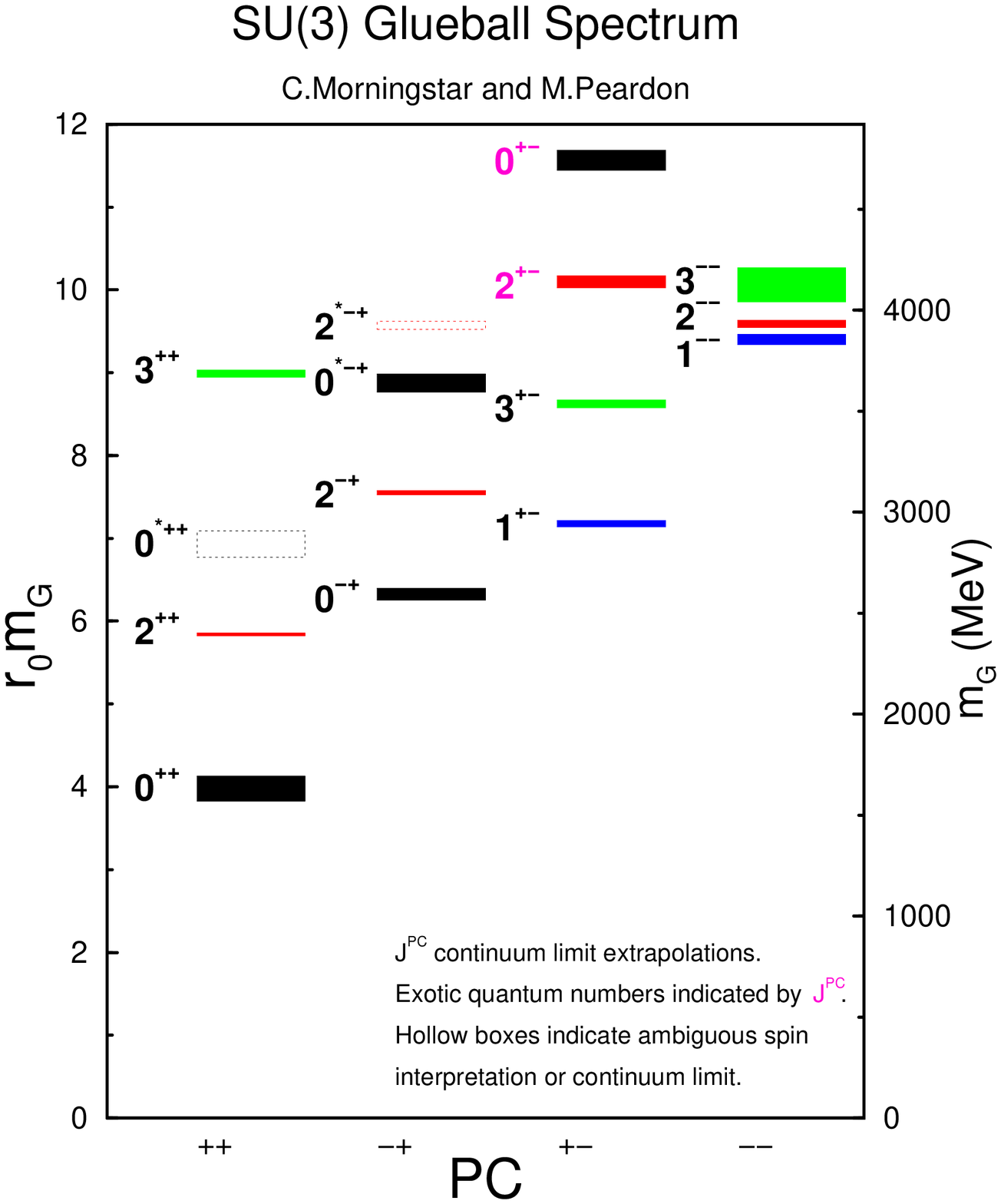,height=3.5in,clip=}}
\end{minipage}
\hfill
\begin{minipage}{3.0in}
\caption[]{The mass of the glueball states.  The scale is set by 
$r_0$ with $1/r_0=410(20)$~MeV (Morningstar 1997, Peardon 1998). }
\label{fig:glueballs}
\end{minipage}
\end{figure}

Lattice QCD predictions for glueball masses 
from one representative calculation are shown in Fig.~\ref{fig:glueballs}.  
One should be cautioned that these results are in the so 
called quenched approximation which neglects internal quark loops.  
The lightest glueball is found to be a $0^{++}$ state with the 
following masses from the different collaborations:
$1550 \pm 50$~MeV (Bali 1993), $1600\pm 160$~MeV (Michael 1998),
$1648\pm 58$ (Vaccarino 1998), and $1630\pm 100$~MeV (Morningstar 1997). 
The difference between these results lies mainly in how the mass scale 
is set. 
The next lightest states are the $2^{++}$ with mass estimates
$2232\pm 220$~MeV (Michael 1998), $2270\pm 100$~MeV (Bali 1993), and 
$2359\pm128$~MeV (Chen 1994) and the 
$0^{-+}$ state with a similar mass.  Mixings with $q\bar{q}$ and
$q\bar{q}q\bar{q}$ states could modify these predictions.

We concentrate on glueballs with conventional quantum numbers since
the first glueballs with exotic quantum numbers, ``oddballs'', 
($0^{+-}$, $2^{-+}$, $1^{-+}$) do not appear 
until $\sim$3~GeV. Because the lowest glueballs have conventional quantum 
numbers with masses situated in a dense background of 
conventional $q\bar{q}$ states it is difficult to distinguish them 
from conventional mesons.  It is therefore a painstaking process to 
identify a glueball by comparing a glueball candidate's properties 
to the expected properties of glueballs and conventional mesons.

Sexton {\it et al.} (Sexton 1996)
have estimated the width of the $0^{++}$ glueball 
to all possible pseudoscalar pairs to be $108\pm 29$~MeV.  This number 
combined with reasonable guesses for the effect of finite lattice 
spacing, finite lattice volume, and the remaining width to multibody 
states yields a total width small enough for the lightest scalar 
glueball to be easily observed. The significant property of glueball 
decays is that  one expects them to have flavour-symmetric couplings 
to final state hadrons.  This gives the characteristic flavour-singlet 
branching fraction to pseudoscalar pairs (factoring out phase space)
\begin{equation}
\Gamma (G \to \pi\pi \; : \; K\bar{K} \; : \; \eta \eta \; : \;
\eta \eta' \; : \; \eta'\eta' )/(\hbox{phase space}) = 3 : 4 : 1 : 0 : 1 .
\end{equation}
Of course, one should also expect some modifications from phase space, 
the glueball wavefunction, and the decay mechanism (Sexton 1996).

Measurements of electromagnetic couplings to glueball candidates 
would be extremely useful for the clarification of the nature of these 
states.  The radiative transition rates of a relatively pure glueball 
would be anomalous relative to the expectations for a conventional 
$q\bar{q}$ state and similarly, 
a glueball should have suppressed couplings to $\gamma\gamma$.  The 
former could be measured in electroproduction experiments, at say, an 
energy upgraded CEBAF while the latter would be 
possible at B-factories or a Tau-Charm Factory.  

There are three production mechanisms that are considered optimal for 
finding glueballs.  The first is the radiative decay 
$J/\Psi\to \gamma G$ where the glueball is formed from intermediate 
gluons (Cakir 1994, Close 1997c).  
The second is in central production $pp \to p_f (G) p_s$ away 
from the quark beams and target where glueballs are produced from 
pomerons which are believed to be multigluon objects.
The third is in proton-antiproton 
annihilation where the destruction of quarks can lead to the creation 
of glueballs.   Because gluons do not carry electric charge, glueball 
production should be suppressed in $\gamma\gamma$ collisions.  
By comparing two photon widths to $J/\psi$ production of a state 
Chanowitz created a measure of glue content he calls ``stickiness'',
(Chanowitz 1984):
\begin{equation}
S= \frac{\Gamma(J/\psi \to \gamma X)}{PS(J/\psi \to \gamma X)}
\times \frac{PS(\gamma\gamma \to X)}{\Gamma(\gamma\gamma\to  X)}
\label{eq:sticky}
\end{equation}
where $PS$ denotes phase space.  A large value of $S$ reflects 
enhanced glue content.  The idea of stickiness has been further 
developed by Close Farrar and Li (Close 1997c).

The simple picture presented above is likely to be muddied by 
complicated mixing effects between the pure LGT 
glueball and $q\bar{q}$ states with the same $J^{PC}$ quantum numbers
(Amsler and Close 1995, 1996).  Lee and Weingarten (Lee 1998a, 1998b) have 
calculated the mixing energy between the lightest $q\bar{q}$ scalar 
state and the lightest scalar glueball in the continuum limit of the 
valence approximation on the lattice.  With this motivation they 
perform a phenomenological fit which finds the $f_0(1710)$ 
to be $\sim 74$\% glueball and the $f_0(1500)$ 
to be $\sim 98$\% quarkonium, mainly $s\bar{s}$. 
Although these results are not rigorous,
they do remind us that physical states are most 
likely mixtures of underlying components with the same quantum 
numbers.  As we will see in subsequent sections, mixings can 
significantly alter the properties of the underlying states which
makes the 
interpretation of observed states difficult, and often controversial.

\subsubsection{Hybrids}

Given the discussion of the previous subsection, the conventional wisdom 
is that it would be more fruitful 
to search for low mass hybrid mesons with exotic quantum numbers 
than to search for glueballs.  Hybrids have the additional attraction 
that, unlike glueballs, they span complete flavour nonets and hence 
provide many possibilities for experimental detection.  In addition, 
the lightest hybrid multiplet includes at least one $J^{PC}$ exotic.
The phenomenological properties of hybrids have been reviewed
elsewhere (Barnes 1984, 1985a, Chanowitz 1987, Close 1988, Godfrey 
1989, Barnes 1995, Close 1995a, Barnes 1996, Page 1997c).
In this section we briefly summarize hybrid properties,  such
as quantum numbers, masses, and decays which may help in their 
discovery. 

In searching for hybrids there are two ways of distinguishing them
from conventional states.  One approach 
is to look for an excess of observed states over the number predicted
by the quark model.  The drawback to this method is that it depends on a 
good understanding of hadron spectroscopy in a mass region that is still
rather murky; the experimental situation is sufficiently unsettled that the
phenomenological  models have yet to be tested  to the extent that a given
state can be reliably ruled out as a conventional meson.  
The situation is further muddied by expected mixing between 
conventional $q\bar{q}$ states and hybrids with the same $J^{PC}$ 
quantum numbers.  The other
approach is to search for quantum numbers which cannot be accomodated 
in the quark model. The discovery of exotic quantum
numbers would be irrefutable evidence of something new.  

To enumerate the hybrid $J^{PC}$ quantum numbers 
in a model independent manner  obeying gauge invariance
one forms gauge invariant operators (Barnes 1985b, Jaffe 1986)
from a colour octet $q\bar q$ operator and a gluon field strength.
The resulting lowest lying $q\bar{q}g$ states with exotic quantum numbers 
not present in the constituent quark model have 
$J^{PC}=2^{+-}$, $1^{-+}$, $0^{+-}$, and $0^{--}$.
We label these states with the
same symbol as the conventional meson with all the same quantum numbers 
except for the $C$ parity and add a {\it hat} to the symbol.
For example, an isospin 1 $J^{PC}=0^{--}$ meson would be a $\hat{\pi}$
and an isospin 0 $J^{PC}=0^{--}$ meson would be a $\hat{\eta}$ or 
$\hat{\eta}'$,
an isospin 1 $J^{PC}=1^{-+}$ meson would be a $\hat{\rho}$, an isospin 1 
$J^{PC}=2^{+-}$ an $\hat{a}_2$, etc.
The discovery of mesons with these exotic quantum numbers would
unambiguously signal hadron spectroscopy beyond the quark model.

\begin{figure}
\begin{minipage}{3.0in}
\centerline{\epsfig{file=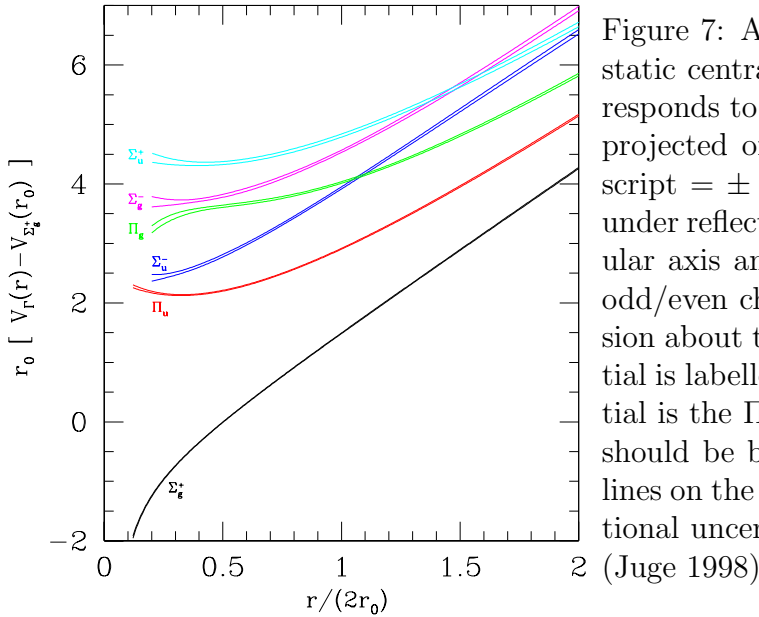,height=3.5in,clip=}}
\end{minipage}
\begin{minipage}{3.4in}
\caption[]{A set of hybrid adiabatic surfaces for static central 
potentials.  $\Lambda = \Sigma ,\; \Pi , \; \Delta , \ldots$ 
corresponds to the magnitude of $J_{glue}=0, \; 1, \; 2, \ldots$ 
projected onto the molecular axis.  The superscript $=\pm$ corresponds 
to the even or oddness under reflections in a plane containing the 
molecular axis and the subscript $u/g$ corresponds to odd/even charge 
conjugation plus spatial inversion about the midpoint.  The familiar 
$q\bar{q}$ potential is labelled as $\Sigma^+_g$ and the first-excited 
potential is the $\Pi_u$ so the lowest lying hybrid mesons should 
be based on this potential. 
The double lines on the excited surfaces indicate the calculational 
uncertainty in determining the potential.  
(Juge 1998)}
\label{fig:gluesurfaces}
\end{minipage}
\end{figure}

To gain some physical insights into hybrids, it 
is useful to turn to lattice results in the heavy quark limit 
before turning to predictions of specific models and calculations.  
A useful approach is to use the leading Born-Oppenheimer approximation 
to map out the adiabatic surfaces corresponding to the non ground 
state gluon configurations (Griffiths 1983, Perantonis 1990, 
Morningstar 1997).
This is analogous to the calculation of the nucleus-nucleus potential 
in diatomic molecules where the slow heavy quarks and fast gluon 
fields in hybrids correspond to the nuclei and electrons in diatomic 
molecules.  One treats the quark and antiquark as spatially fixed 
colour sources and determines the energy levels of the glue as a 
function of the $Q\bar{Q}$ separation.  Each of these energy levels 
defines an adiabatic potential $V_{Q\bar{Q}}(r)$.  
The ground state potential has cylindrical symmetry about 
the interquark axis while less symmetric configurations correspond to 
excitations of the gluonic flux joining the quark-antiquark pair.
For example, the lowest lying gluonic excitation corresponds to a 
component of angular momentum of one unit along the quark-antiquark 
axis.  The adiabatic potentials are determined using lattice QCD.   
One such set of adiabatic surfaces is shown in 
Fig.~\ref{fig:gluesurfaces}.
The quark motion is 
then restored by solving the Schr\"odinger equation in each of these 
potentials.  Conventional mesons are based on 
the lowest lying potential and hybrid states emerge from the excited 
potentials. 
Combining the resulting flux-tube spatial 
wave functions which have $L^{PC}=1^{+-}$ and $1^{--}$ with the quark 
and antiquark spins yields a set of eight degenerate hybrid states with 
$J^{PC}=1^{--}$, $0^{-+}$, $1^{-+}$, $2^{-+}$, 
and $1^{++}$, $0^{+-}$, $1^{+-}$, $2^{+-}$ respectively.  
These contain the $J^{PC}$ exotics with $J^{PC}=1^{-+}$, 
$0^{+-}$, and $2^{+-}$.  
The degeneracy of the eight $J^{PC}$ states is expected to be broken by 
the different excitation energies in the $L^{PC}=1^{+-}$ (magnetic) 
and $1^{-+}$ (pseudo-electric) gluonic excitations, spin-orbit terms, 
as well as mixing between hybrid states and $q\bar{q}$ mesons with 
non-exotic spins.  

While this picture is appropriate for heavy quarkonia it is not at all 
clear that it can be applied to light quark hybrids.  Nevertheless, 
given that the constituent quark model works so well for light quarks, 
it is not unreasonable to also extend this flux tube description to light 
quarks.  The flux tube model,
developed by Isgur and Paton, is based on 
the strong coupling expansion of lattice QCD (Isgur 1983, 1985a, 1985b).  
It predicts 8 nearly 
degenerate nonets around 2 GeV, $J^{PC}=$ $2^{\pm\mp},$ $1^{\pm\mp}$,
$0^{\pm\mp}$, and $1^{\pm\pm}$. 
In the flux tube model,
the glue degree of freedom manifests itself as excited phonon modes
of the flux tube connecting the $q\bar{q}$ pair, so the first
excited state is doubly degenerate. This picture of gluonic excitations 
appears to be supported by lattice calculations 
(Perantonis 1990, Michael 1994).

Other models exist, for example the bag model
(Chanowitz 1983a, 1983b, Barnes 1983a, 1983b, 1983c)
which in contrast to the flux tube model,  expects only four 
degenerate states, the $2^{-+}$, $1^{-+}$, $1^{--}$, and $0^{-+}$.
This difference is
symptomatic of the differences between the two models.  
In the bag model, the gluon degrees of
freedom are either transverse electric (TE) or transverse magnetic (TM) modes 
of the bag, with the TM mode considerably higher in mass than the TE mode. 

A recent Hamiltonian Monte Carlo study of the flux tube 
model (Barnes 1995) finds the lightest $n\bar{n}$ hybrid masses 
to be 1.8-1.9~GeV. 
This result is consistent with the lattice QCD 
(quenched approximation) results 
of the UKQCD Collaboration (Perantonis 1990, Lacock 1996, 1997)
who find $M_{\hat{\rho}} \simeq 1.88$~GeV and 
$M_{\hat{\phi}}\simeq 2.09$~GeV and Bernard {\it et al.} (Bernard 
1996, 1997) who find $M_{\hat{\rho}} \simeq 1.97$~GeV and 
$M_{\hat{\phi}}\simeq 2.17$~GeV.  Lacock {\it et al.} (Lacock 1996)
also find $M_{\hat{a}_0}\sim 2.09$~GeV and $M_{\hat{a}_2}\sim 2.09$~GeV 
for the next lightest hybrids.

Hybrid decays appear to follow the almost universal
selection rule that gluonic excitations cannot transfer angular
momentum to the final states as relative angular momentum. Rather, it must
instead appear as internal angular momentum of the $q\bar{q}$ pairs 
(Page 1997a, 1997b, Kalashnikova 1994).
The selection rule suppresses decay channels likely to be large,
and may make hybrids stable enough to appear as conventional resonances. 
Unfortunately, this selection rule is not absolute; in the flux tube 
and constituent glue models it can be broken by wave 
function and relativistic effects while the bag model adds 
a qualifier that
it is also possible that the excited quark loses its angular momentum to 
orbital angular momentum.  In this case 
the $1^{-+}$ could decay to two S-wave mesons such as $\pi\eta$ or 
$\pi\rho$ in a relative P-wave which would dominate due to the large 
available phase space.
In any case, if we take this selection rule seriously, it explains why
hybrids with exotic $J^{PC}$ 
have yet to be seen;  they do not couple strongly to simple final 
states.
Thus, in the list of possible  $\hat{\rho}$ decays,
\begin{equation}
 \hat{\rho} \to [\underline{\pi\eta},\underline{\pi\eta'},\pi\rho,
K^*K,\eta\rho,\ldots ]_P, \quad
  [\pi b_1, \pi f_1, \eta a_1, KK_1 \ldots ]_S, 
\end{equation}
most models expect the $b_1\pi$ and $f_1\pi$ modes to dominate.
(The underlined modes to two distinct pseudoscalars provide a unique 
signature of the $1^{-+}$ state.)  For states with conventional 
quantum numbers we expect mixing between hybrids and conventional 
$q\bar{q}$ states, even in the quenched approximation, which could 
significantly modify the properties of these states.

The decay predictions of the flux tube model (Isgur 1985b, Close 1995a)
are given in Table~\ref{tab:hybriddecays}.
\begin{table}
\caption{The dominant hybrid decay widths for $A\to [BC]_L$ for 
partial wave $L$ calculated 
using the flux tube model. From Close and Page (Close 1995a). 
Hybrid masses before spin 
splitting for $n\bar{n}$
are 2.0~GeV except for $0^{+-}$ (2.3 GeV), $1^{+-}$ (2.15 GeV),
and $2^{+-}$ (1.85 GeV)  
and for $s\bar{s}$ are 2.15~GeV except for $0^{+-}$ (2.25 GeV) 
following Merlin and Paton (Merlin 1987).}
\label{tab:hybriddecays}
\begin{center}
\begin{tabular}{|l|l|r|l|r|l|r|}
\hline
	& \multicolumn{2}{c|}{$I=1$} 
	& \multicolumn{2}{c|}{$I=0 \; n\bar{n}$} 
	& \multicolumn{2}{c|}{$I=0 \; s\bar{s}$} \\ \hline
$A$ & $[BC]_L$ & $\Gamma$ & $[BC]_L$ & $\Gamma$ &
	$[BC]_L$ & $\Gamma$ \\ \hline
$2^{-+}$ & $[f_2(1270)\pi]_S$ & 40 
		& $[a_2(1320)\pi]_S$ & 125 
		& $[K_2^*(1430)K]_S$ & 100 \\
	& $[f_2(1270)\pi]_D$ & 20 
		& $[a_2(1320)\pi]_D$ & 60 
		& $[K_1(1270)K]_D$ & 20 \\
	& $[b_1(1235)\pi]_D$ & 40 
		& $[f_2(1270)\eta]_S$ & $\sim 50$ 
		& & \\
	& $[a_2(1320)\eta]_S$ & $\sim 40$ 
		& $[K_2^*(1430) K]_S$ & $\sim 30$ 
		& &  \\
	& $[K_2^*(1430)K]_S$ & $\sim 30$ 
		&  &  
		&  &  \\
	& $[\rho\pi]_P$ & 8 & $[K^*K]_P$ & 2 & $[K^*K]_P$ & 6 \\
	& $[K^*K]_P$ & 2 & & & & \\ 
\hline
$2^{+-}$ & $[a_2(1320)\pi]_P$ & 200 
		& $[b_1(1235)\pi]_P$ & 250 
		& $[K_2^*(1430)K]_P$ & 90 \\
	& $[a_1(1260)\pi]_P$ & 70 
		& $[h_1(1170)\eta]_P$ & 30 
		& $[K_1(1270)K]_P$ & 30 \\
	& $[h_1(1170)\pi]_P$ & 90 
		& &
		& $[K_1(1400)K]_P$ & 70 \\
	& $[b_1(1235)\eta]_P$ & $\sim 15$ 
		& & & &  \\
	& & & $[\rho\pi]_D$ & 1 & $[K^*K]_D$ & 1 \\
\hline
$0^{+-}$ & $[a_1(1260)\pi]_P$ & 700 
		& $[b_1(1235)\pi]_P$ & 300 
		& $[K_1(1270)K]_P$ & 400 \\
	& $[h_1(1170)\pi]_P$ & 125
		& $[h_1(1170)\eta]_P$ & 90 
		& $[K_1(1400)K]_P$ & 175 \\
	& $[b_1(1235)\eta]_P$ & 80 
		& $[K_1(1270)K]_P$ & 600
		&  &  \\
	& $[K_1(1270)K]_P$ & 600 
		& $[K_1(1400)K]_P$ & 150 & &  \\
	& $[K_1(1400)K]_P$ & 150 &  & & & \\
\hline
$1^{+-}$ & $[a_2(1320)\pi]_P$ & 175 
		& $[b_1(1235)\pi]_P$ & 500 
		& $[K_2^*(1430)K]_P$ & 70 \\
	& $[a_1(1260)\pi]_P$ & 90
		& $[h_1(1170)\eta]_P$ & 175
		& $[K_1(1270)K]_P$ & 250 \\ 
	& $[h_1(1170)\pi]_P$ & 175
		& $[K_2^*(1430)K]_P$ & 60
		& $[K_0^*(1430)K]_P$ & 125 \\
	& $[b_1(1235)\eta]_P$ & 150 
		& $[K_1(1270)K]_P$ & 250  & & \\
	& $[K_2^*(1430)K]_P$ &  60 
		& $[K_0^*(1430)K]_P$ & 70 & & \\
	& $[K_1(1270)K]_P$ & 250 & & & & \\
	& $[K_0^*(1430)K]_P$ & 70 & & & &  \\
	& $[\omega \pi]_S$ & 15 & $[\rho\pi]_S$ & 40 & $[K^*K]_S$ & 20 \\
	& $[\rho\eta]_S$ & 20 & $[\omega\eta]_S$ & 20 & $[\phi\eta]_S$& 40 \\
	&$[\rho\eta']_S$& 30 & $[\omega\eta']_S$& 30 & $[\phi\eta']_S$& 40 \\
	& $[K^*K]_S$ & 30 & $[K^*K]_S$ & 30 & & \\
\hline
\end{tabular}
\end{center}
\end{table}
\begin{table}
\centerline{Table~\ref{tab:hybriddecays}, continued.}
\begin{center}
\begin{tabular}{|l|l|r|l|r|l|r|}
\hline
	& \multicolumn{2}{c|}{$I=1$} 
	& \multicolumn{2}{c|}{$I=0 \; n\bar{n}$} 
	& \multicolumn{2}{c|}{$I=0 \; s\bar{s}$} \\ \hline
$A$ & $[BC]_L$ & $\Gamma$ & $[BC]_L$ & $\Gamma$ &
	$[BC]_L$ & $\Gamma$ \\
\hline
$1^{++}$ & $[f_2(1270)\pi]_P$ & 175 
		& $[a_2(1320)\pi]_P$ & 500 
		& $[K_2^*(1430)K]_P$ & 125 \\
	& $[f_1(1285)\pi]_P$ & 150
		& $[a_1(1260)\pi]_P$ & 450
		& $[K_1(1270)K]_P$ & 70 \\ 
	& $[f_0(1300)\pi]_P$ & $\sim 20$
		& $[f_2(1270)\eta]_P$ & 70
		& $[K_1(1400)K]_P$ & 100 \\ 
	& $[a_2(1320)\eta]_P$ & 50
		& $[f_1(1285)\eta]_P$ & 60 & & \\
	& $[a_1(1260)\eta]_P$ & 90
		& $[K_2^*(1430)K]_P$ & $\sim 20$ & & \\
	& $[K_2^*(1430)K]_P$ & $\sim 20$ 
		& $[K_1(1270)K]_P$ & 40 &  & \\
	& $[K_1(1270)K]_P$ & 40
		& $[K_1(1400)K]_P$ & $\sim 20$ & & \\
	& $[K_1(1400)K]_P$ & $\sim 20$ & & & & \\
	& $[\rho\pi]_S$ & 20 & $[K^*K]_S$ & 15 & $[K^*K]_S$ & 10 \\
	& $[K^*K]_S$ & 15 & & & & \\
\hline
$1^{-+}$ & $[f_1(1285)\pi]_S$ & 40 
		& $[a_1(1260)\pi]_S$ & 100 
		& $[K_1(1270)K]_S$ & 40 \\
	& $[f_1(1285)\pi]_D$ & 20 
		& $[a_1(1260)\pi]_D$ & 70 
		& $[K_1(1270)K]_D$ & 60 \\
	& $[b_1(1235)\pi]_S$ & 150 
		& $[f_1(1285)\eta]_S$ & 50 
		& $[K_1(1400)K]_S$ & 25 \\
	& $[b_1(1235)\pi]_D$ & 20 
		& $[K_1(1270)K]_S$ & 20 & & \\
	& $[a_1(1260)\eta]_S$ & 50 
		& $[K_1(1400)K]_S$ & $\sim 125$ & & \\ 
	& $[K_1(1270)K]_S$ & 20  & & & & \\
	& $[K_1(1400)K]_S$ & $\sim 125$  & & & & \\
	& $[\rho\pi]_P$ & 8 & $[K^*K]_P$ & 2 & $[K^*K]_P$ & 6 \\
	& $[K^*K]_S$ & 2 & & & & \\
\hline
$0^{-+}$ & $[f_2(1270)\pi]_D$ & 20 
		& $[a_2(1320)\pi]_D$ & 60 
		& $[K_2^*(1430)K]_D$ & 20 \\
	& $[f_0(1300)\pi]_S$ & $\sim 150$ 
		& $[f_0(1300)\eta]_S$ & $\sim 200$ 
		& $[K_0^*(1430)K]_S$ & 400 \\
	& $[K_0^*(1430)K]_S$ & 200 
		& $[K_0^*(1430)K]_S$ & 200 & & \\
	& $[\rho\pi]_P$ & 30 & $[K^*K]_P$ & 8 & $[K^*K]_P$ & 30 \\
	& $[K^*K]_P$ & 8 & & & & \\
\hline
$1^{--}$ & $[a_2(1320)\pi]_D$ & 50 
		& $[K_1(1270)K]_S$ & 40 
		& $[K_2^*(1430)K]_D$ & 20 \\
	& $[a_1(1260)\pi]_S$ & 150
		& $[K_1(1400)K]_S$ & 60 
		& $[K_1(1270)K]_S$ & 60 \\
	& $[a_1(1260)\pi]_D$ & 20 
		& & 
		& $[K_1(1400)K]_S$ & 125 \\
	& $[K_1(1270)K]_S$ & 40  & & & & \\
	& $[K_1(1400)K]_S$ & $\sim 60$  & & & & \\
	& $[\omega \pi]_P$ & 8 & $[\rho\pi]_P$ & 20 & $[K^*K]_P$ & 15 \\
	& $[\rho\eta]_P$ & 7 & $[\omega\eta]_P$ & 7 
		& $[\phi\eta]_P$ & 8 \\
	& $[\rho\eta']_P$ & 3 & $[\omega\eta']_P$ & 3
		& $[\phi\eta']_P$ & 2 \\
	& $[K^*K]_P$ & 4 & $[K^*K]_P$ & 4 & & \\
\hline
\end{tabular}
\end{center}
\end{table}
These predictions suggest that many hybrids 
are too broad to be distinguished as a resonance while a few hybrids 
should be narrow enough to be easily observable.  
In particular, of the hybrids with exotic quantum numbers
the flux tube model
predicts that the $\hat{a}_0$, $\hat{f}_0$, 
and $\hat{f}'_0$ are probably too broad to
appear as resonances. 
The $\hat{\omega}_1$ decays mainly to 
$[a_1 \pi]_S$  with $\Gamma\approx 100$ MeV,
which would make it difficult to 
reconstruct the original hybrid given the broad width of the final state
mesons.  Similar problems could also make the 
$\hat{\phi}_1$ difficult to find.  According to the flux tube model,
the best bets for finding hybrids are:
\begin{equation}
\begin{array}{cll}
\hat{\rho}_1 & \to [b_1 \pi ]_S & (\Gamma \approx 150 \hbox{MeV}) \\
		    & \to [f_1 \pi ]_S & (\Gamma \approx 50  \hbox{MeV}) \\
\hat{a}_2 & \to [a_2 \pi]_P & (\Gamma \approx 200 \hbox{MeV}) \\
\hat{f}_2 & \to [b_1 \pi ]_P & (\Gamma \approx 250 \hbox{MeV}) \\
\hat{f}'_2 &\to [K^*(1430)_2\bar{K}]_P &(\Gamma\approx 90\hbox{MeV})\\
		    &\to [\bar{K}K_1 ]_P     &(\Gamma\approx 100\hbox{MeV}) 
\end{array}
\end{equation}
Finally, some ``forbidden'' decays such as $\hat{\rho}(1900) \to \rho 
\pi$ have small but finite partial widths due to differences in the 
final state spatial wavefunctions.  Thus, it may be possible to 
observe hybrids in these simpler decay modes in addition to the 
favoured  but more difficult to reconstruct final states such as 
$b_1\pi$ and $K_1 K$. 

So far we have concentrated on the $J^{PC}$ exotic members of the 
lowest flux-tube hybrid multiplet and one might wonder whether the 
nonexotic hybrids might be narrow enough to be observables.  According 
to the results of Close and Page (Close 1995a)
reproduced in Table~\ref{tab:hybriddecays} 
many of the nonexotic hybrids are also so broad as to be 
effectively unobservable.  There are several notable exceptions.  
The first is a $1^{--}$ $\omega$ hybrid with a total width of only $\sim 
100$~MeV which decays to $K_1(1270)K$ and $K_1(1400)K$. The $\phi$ is 
also relatively narrow with  $\Gamma_{tot}\sim 225$~MeV. Two more
interesting hybrids are the $\pi_2$ with $\Gamma_{tot}\sim 170$~MeV. 
and its $s\bar{s}$ partner, the $\eta_2'$, with
$\Gamma_{tot}\sim 120$~MeV decaying dominantly to $K_2^*K$.
In addition, there are several other hybrids that 
have total widths around 300~MeV and so should also be observable.  

To determine whether an observed state with non-exotic quantum 
numbers is a conventional $q\bar{q}$ state or a hybrid one would make 
use of the detailed predictions we have described above for the two 
possibilities (Barnes 1997).
For example, a selection rule of the $^3P_0$ decay model 
forbids the decay of a spin-singlet $q\bar{q}$ state to two 
spin-singlet mesons in the final state (Page  1997a).  
This selection rule forbids the decay  
$\pi_2(^1D_2) \to b_1  \pi $ while in contrast the decay is 
allowed for the hybrid $\pi_2$ and is in fact rather large.  
A second illustration is a $0^{-+}$ state with $M\simeq 
1800$~MeV.  The largest decay modes for the $\pi(3^1S_0)$ and a hybrid 
with the same mass and quantum numbers, $\pi_H$, are shown in 
Table~\ref{tab:pi}.
\begin{table}
\caption{Decay of quark model and hybrid $\pi(1800)$.}
\label{tab:pi}
\begin{center}
\begin{tabular}{|c|cccccc|}
\hline
State & \multicolumn{6}{c|}{Partial Widths to Final States} \\ \hline
	& $\pi\rho$ & $\omega\rho$ & $\rho(1465)\pi$ & $f_0(1300)\pi$ 
	& $f_2\pi$ & $K^*K$ \\ \hline
$\pi_{3s}(1800)$ & 30 & 74 & 56 & 6 & 29 & 36 \\
$\pi_{H}(1800)$ & 30 & --- & 30 & 170 & 6 & 5 \\
\hline
\end{tabular}
\end{center}
\end{table} 
Both states decay to most of the same final states albeit with much 
different partial widths.  A discriminator between the two 
possibilities is the $\rho\omega$ channel which is dominant for 
$\pi(3^1S_0)$ whereas it is predicted to be absent for the $\pi_H$.  
There are many such examples.  
The essential point is that although the two states may have the same 
$J^{PC}$ quantum numbers they have different internal structure which 
will manifest itself in their decays.  
Unfortunately,  nothing is simple and we once again point out that 
strong mixing is expected between hybrids with conventional quantum 
numbers and $q\bar{q}$ states with the same $J^{PC}$ so that the decay 
patterns of physical states may not closely resemble those of either 
pure hybrids or pure $q\bar{q}$ states. 

The final ingredient in hybrid searches is the production mechanism.  
Just as in glueball searches,
the best place to look is in the gluon rich $J/\psi$ decays.
A second reaction which has attracted interest is in $p\bar{p}$
annihilation.
% Notwithstanding these preferred channels, the only
%exotic candidate has been seen in the reaction $\pi^- p$.
%** NO. CBAR sees 1400 state too - JN
Finally, photoproduction is potentially an 
important mechanism for producing hybrids so that hybrids could be 
produced copiously at an upgraded CEBAF at TJNAF via an off-shell 
$\rho$, $\omega$, or $\phi$ via vector meson dominance interacting 
with an off-shell exchanged $\pi$ (Close 1995b).
The moral is that what is really needed is careful high statistics experiments
in all possible reactions.

\subsection{Multiquark Hadrons}

The notion of colour
naturally explained nature's preference for $q\bar{q}$ and $qqq$ 
colourless sytems.  However, it also appears to predict multiquark states such
as $q^2\bar{q}^2$ and $q^3 q\bar{q}$ which could have exotic quantum numbers,
thus indicating non $q\bar{q}$ and $qqq$ states (Jaffe 1977a, 1977b, 
1978, Lipkin 1978).  
Upon considering $qq\bar{q}\bar{q}$ systems we find that
the colour couplings are not unique as they are in mesons and baryons.
For example, we can combine two colour triplet $q$'s into a colour 6 
or $\bar{3}$.  Likewise we can combine two antitriplet $\bar{q}$'s 
into a 3 or a $\bar{6}$.  Therefore, there are two possible ways of 
combining $qq\bar{q}\bar{q}$ into a colour singlet: $3\bar{3}$ or 
$6\bar{6}$.  In addition, since we could have combined a $q$ and $\bar{q}$ 
into a colour 1 or 8 we could also have combined the 
$qq\bar{q}\bar{q}$ into colour 11 and 88.  Since two free mesons (in a 
colour 11) are 
clearly a possible combination of $qq\bar{q}\bar{q}$ the $3\bar{3}$ 
$6\bar{6}$ couplings can mix to give the 11 - 88 colour 
configurations which further complicates the details of the calculation.
Thus, whether or not  multiquark states exist is a dynamical question. 
It is possible that multiquark states exist as bound states,
but it is also possible that 
$qq\bar{q}\bar{q}$ configurations lead to hadron-hadron potentials 
(Barnes 1987, Weinstein 1990, Swanson 1992). Both
must be taken into account when attempting to unravel the hadron spectrum.

As in the case of hybrid mesons, states not fitting into the 
$q\bar{q}$ framework are the most unambiguous signature for multiquark 
states.  In particular, flavour exotics are our best bet for finding genuine
multiquark states.
There is a large literature on the physics of multiquark states which 
attempts to predict their masses, explain their properties and 
interpret observed hadron structures as multiquark states.  One should 
take much of what exists in the literature with a grain of salt as 
few of these predictions are based on full dynamical calculations.
An exception is a quark model 
study of the $J^{PC}=0^{++}$ sector of the $qq\bar{q}\bar{q}$ system
(Weinstein 1982, 1983).  It  found
that weakly bound $K\bar{K}$ ``molecules'' exist in the isospin-zero and
-one sectors in analogy to the deuteron.  It was suggested that these
two bound states be identified 
with the $f_0 (975)$ and $a_0 (980)$. 
The meson-meson potentials which come from this picture,
when used with a coupled channel Schr\"odinger equation, reproduce the
observed phase shifts for the $a_0$ and $f_0$ in $\pi\pi$ scattering.
The $K\bar{K}$ molecules are the exception however, as the model predicts 
that in general the $qq\bar{q}\bar{q}$ ground states are two unbound mesons.

So far only  pseudoscalar mesons in the final state have been 
considered in detail 
so the next logical step is to extend the analysis to vector-vector 
(Dooley 1992)
and pseudoscalar-vector channels (Caldwell 1987, Longacre 1990).  
One such possibility is a threshold 
enhancement in $K^*K$ just above threhold.

There are a number of distinctive signatures for the 
multiquark interpretation of a resonance.  For molecules one expects 
strong couplings to constituent channels.  For example, the 
anomalously large coupling of the $f_0(980)$ to $K\bar{K}$ despite 
having almost no phase space is a hint that it is not a conventional 
$q\bar{q}$ state.
Electromagnetic couplings are another clue to nonstandard origins of a 
state.  
Barnes found that the two-photon widths for 
a $q\bar{q}$ state are expected to be much larger than 
that of a $K\bar{K}$ molecule (Barnes 1985b).  Radiative transitions 
can also be used to distinguish between the two possibilities.  
For the case of a $f_1(K^*K)$ object one would expect the dominant 
radiative mode to arise from the radiative transition of the $K^*$ 
constituent in $K^*\to K\gamma$ 
 while an $f_1(s\bar{s})$ state would be 
dominated by the transition $f_1(s\bar{s}) \to \gamma \phi$.  The two 
cases would be distinguished by different ratios of the $\gamma K^0 
\bar{K}^0$ and $\gamma K^+ K^-$ final states.  Likewise, Close Isgur 
and Kumano (Close 1993) suggest a related test for the $f_0(980)$ and 
$a_0(980)$ involving the radiative decays  $\phi \to \gamma f_0$ and 
$\gamma a_0$.

Whether or not multiquark states exist 
it is still extremely important to understand hadron-hadron potentials
arising from multiquark configurations 
(Swanson 1992) so that the observed experimental
structure can be unravelled and understood.
There is, in fact, evidence for meson-meson potentials.
In the reaction $\gamma\gamma\to \pi^0 \pi^0$,
the meson-meson potentials are needed 
to reproduce the  $\gamma\gamma \to \pi^0\pi^0$ 
cross section data (Blundell 1998). Enhancements in the production of low
invariant mass $\pi\pi$ pairs have been observed in other processes 
as well; $\eta' \to \eta \pi\pi$, $\psi' \to J/\psi \pi\pi$, 
$\Upsilon(nS) \to \Upsilon (mS) \pi\pi$, and $\psi\to \omega\pi\pi$.
Similar enhancements have also been seen in some $K\pi$ channels
in $\bar{p}p\to K\bar{K}\pi$.  The lesson 
is that final state interactions arising from hadron-hadron
potentials will play an important role in understanding the 1 to 2 GeV mass
region.  

% ======================================================================
%\include{rev-v6-3}
% ======================================================================
\newpage
\section{EXPERIMENTS}

The meson spectrum consists of an increasingly large
number of broad, overlapping
states, crowded into the mass region above around one~GeV/$c^2$.
Experiments disentagle these states using a combination of three 
approaches.
First, states have different sensitivity to the various available production
mechanisms.  This can be particularly useful when trying to enhance one
state relative to another, comparing the results from two or more sources.
Second, experiments using the same production mechanism (or at least the
same beam and target) may be sensitive to a number of different final states.
This naturally leads to a sensitivity to differing isospin and $G$-parity
channels, and can provide consistency checks for states within one
particular experiment.

\begin{figure}[hbt]
\begin{minipage}{3.0in}
\centerline{\epsfig{file=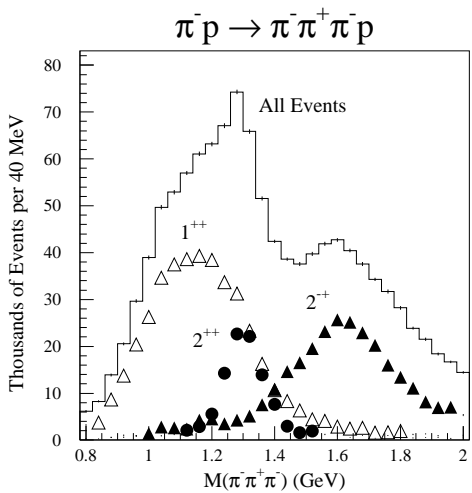,height=3.0in}}
\end{minipage}
\hfill
\begin{minipage}{3.0in}
\caption{Three-pion mass distribution for the reaction
$\pi^-p\to\pi^-\pi^+\pi^-p$ at 18~GeV/$c$, from experiment E852 at BNL.
A partial wave analysis is used to decompose the spectrum into its
dominant $J^{PC}$ components, clearly showing the $a_1(1260)$, the
$a_2(1320)$, and the $\pi_2(1680)$.}
\label{figIII:e852_3pi}
\end{minipage}
\end{figure}

The third technique aims at unraveling the different $J^{PC}$ combinations
within a specific experiment and final state. Such a ``Partial Wave Analysis''
is crucial when overlapping states are produced in the same reaction
(Chung and Trueman, 1975; Aston, 1985;
Sinervo, 1993; Cummings and Weygand, 1997).
In all cases, it is important to fully understand
the ``acceptance'' of the detector over the whole of phase space, so
that the basis orthogonality conditions are correctly exploited.
One clear example is demonstrated (Adams, 1998)
in Fig.~\ref{figIII:e852_3pi},
where the $\pi^-\pi^+\pi^-$ mass spectrum from the reaction
$\pi^-p\to\pi^-\pi^+\pi^-p$ shows a clearly rich structure of strongly
overlapping peaks.  In this case, the partial wave analysis is able to
cleanly separate the contributions from the various $J^{PC}$.
Furthermore, one can examine the relative phase motion of one ``wave''
relative to the other, and demonstrate that the intensity peak is in
fact resonant.

Different experiments measuring different properties are
taken together to unravel the meson spectrum.
In the following sections, we describe the 
various types of experiments, along with their advantages and disadvantages.
We also mention some specific
experiments whose data will be described in later chapters.

\subsection{Hadronic Peripheral Production}

Most of the data on light meson spectroscopy has come from multi-GeV pion
and kaon beams on nucleon or nuclear targets, where the beam particle is
excited and continues to move forward, exchanging momentum and quantum
numbers with a recoiling nucleon.

\begin{figure}
\centerline{\epsfig{file=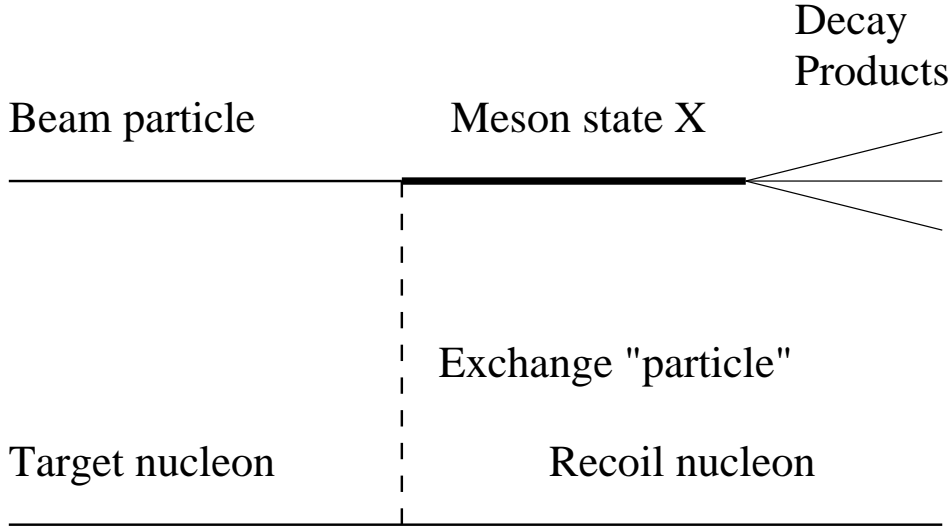,width=5.0in}}
\caption{Schematic diagram of a hadronic peripheral production process.
Momentum is exchanged through an off-mass-shell particle, which may or
may not be charged.}
\label{figIII:periph}
\end{figure}

Meson-nucleon scattering reactions at high energy are strongly forward
peaked, in the direction of the incoming meson.  Typically, the forward
going products are mesons, with a ground- or excited-state baryon
recoiling at large angle.  This mechanism is shown schematically in
Fig.~\ref{figIII:periph}.
The excited meson state $X$ has quantum
numbers determined by the exchange, and subsequently decays to two or
more stable particles.  Two typical examples in modern experiments
include $K^-p\to K^-K^+\Lambda$ by the LASS collaboration
(Aston, 1988d) and $\pi^-p\to X^-p\to\rho^0\pi^-p\to\pi^-\pi^+\pi^-p$
by the E852 collaboration (Adams, 1998).
Other recent experiments with $\pi^-$ beams include VES at IHEP/Serpukhov
(Beladidze, 1993; Gouz, 1993) and BENKEI at KEK
(Fukui, 1991; Aoyagi, 1993). Of particular note is the GAMS collaboration
(Alde, 1986, 1988a, 1988b, 1992, 1997) which detected all neutral final states.
Peripheral reactions with positively charged beams on proton targets
are also possible.  There is no apriori reason to expect
that any particular type of hadronic state ($q\bar{q}$, multiquark,
glueball, or hybrid) should be preferred over any other in this mechanism.

Peripheral reactions are characterized by the square of the four-momentum
exchanged, called $t\equiv(p_{\rm Beam}-p_X)^2<0$.
(See Fig.~\ref{figIII:periph}.)
The forward-peaking nature is seen in an approximately exponentially falling
cross section with $t$, i.e. $e^{bt}$ with $b\sim3-8$~GeV$^{-2}$.
For example, in charge exchange reactions at
small values of $-t$, one pion exchange (OPE) dominates and is fairly well
understood.  It provides access only to
states with $J^{PC}$ = even$^{++}$ and odd$^{--}$, the so called ``natural
parity'' states.
Other states such as $J^{PC}=0^{-+}$ can be produced by neutral
$J^{PC}=0^{++}$ ``Pomeron'' exchange, or $\rho^+$
exchange but these are not as well undertood. Often the analysis is performed
independently for several ranges of $t$, to try to understand the nature of
the production (exchange) mechanism.

In the spirit of both Regge phenomenology and field theory, the diagram in
Fig.~\ref{figIII:periph} is taken literally when interpreting a partial
wave analysis of peripheral reactions.  That is, the result of the PWA is
used to infer the exchange particle and that it couples to the beam particle
and excited meson state, conserving angular momentum, parity, and charge
conjugation.  As shown by Chung and Trueman (Chung 1975), the
analysis is naturally divided into two sets of non-interfering waves on the
basis of positive or negative ``reflectivity'', which in turn corresponds
to ``natural'' ($P=(-1)^J$) or ``unnatural'' ($P=(-1)^{J+1}$) parity of
the exchange particle with spin $J$.

The generality of this production mechanism and the high statistics
available result in several advantages.  One possibility is to use 
as unrestrictive triggers as possible to give large, uniform acceptance
and choose particular final states in the analysis stage
(Aston, 1990).  On the other hand,
many experiments design the trigger to choose only a particular final state
since the events of interest may occur very infrequently.  
A difficulty with this approach is that the detection efficiency for the
final state is usually not uniform in the kinematic variables and
one must be careful in modeling the experiment when performing the
partial wave analysis.

Detection of final state photons, to identify $\pi^0$ and $\eta$ and the
objects which decay to them, has come of age in recent experiments.
GAMS made use of all-photon final states in particular, but fine-grained
electromagnetic calorimeters have been combined with charged particle
tracking and particle identification in E852 and in VES.  This opens up
a large number of final states that can be studied in a single experiment
simultaneously during the same run. This can be very powerful by
comparing decay branches of various states, as well as searching for
decay modes that were not previously accessible.

\subsection{Peripheral Photoproduction}
\label{secIII:photons}

Peripheral hadronic reactions have been the workhorse of meson spectroscopy,
mainly because of the wide range of kinematics available, along with the
accessibility and high cross section of hadron beams.  Unfortunately,
however, there is little selectivity for specific meson states.  Except
for the ability to do some selection on $t$ and to bring in strange quarks
by using $K$ instead of $\pi$ beams, one is limited to exciting the
spin singlet ground state $q\bar{q}$ combination (i.e. pions and kaons)
because only these are
stable against strong decay and therefore live long enough to produce
beams for experiments.

Peripheral {\em photo}-production reactions provide a qualitative alternative.
The hadronic properties of the photon are essentially given by vector
dominance (Bauer, 1978).
That is, the photon couples to hadrons as if it
were a superposition of {\em vector} meson states.
In this case, Fig.~\ref{figIII:periph} still applies, but the incoming
``beam'' particle is a spin {\em triplet} ground state $q\bar{q}$.
Consequently, the series of preferred excitations is likely to be quite
different.  This mechanism has, in fact, been argued to be the most likely
way to produce hybrid 
mesons with exotic quantum numbers by means of flux tube
excitation (Isgur, Kokoski, and Paton, 1985b; Afanasev and Page, 1998).

Peripheral photoproduction has further advantages.  The vector dominance
model allows non-OZI suppressed excitation of heavy quark states, such as
$s\bar{s}$ and $c\bar{c}$, through production of the associated vector
meson(s), the $\phi$ and $\psi$ states respectively.

Unfortunately, there is a dearth of data from peripheral photoproduction.
This is mainly due to the lack of high quality, high intensity photon beams
and associated experimental apparatus, although this situation will change
in the near future.  (See Sec.~\ref{sec:JLab}.)
A thorough review of the experimental situation through the mid-1970's is
available in (Bauer 1978).  
The most significant
contributions to meson spectroscopy since that time are by the LAMP2
experiment at Daresbury (Barber 1978, 1980) with
photon beam energies up to 5~GeV, and the $\Omega$-Photon collaboration
at CERN (Aston, 1982) using energies between 20 and 70~GeV.
Spectroscopy in exclusive photoproduction has also been carried out by
E401 at Fermilab (Busenitz, 1989), also an
electronic detector with a relatively open trigger;
E687 at Fermilab (Frabetti, 1992), an evolution of E401 which
concentrated on heavy quark physics; and
by the SLAC hybrid bubble chamber and a laser-backscattered
photon beam (Condo, 1993).

\subsection{$\bar{p}p$ and $\bar{N}N$ Reactions}

Annihilation of antiquarks on quarks can be accomplished straightforwardly
using antiproton beams on hydrogen or deuterium targets. States which decay
directly to $\bar{p}p$ and $\bar{p}n$ can be studied by measuring inclusive
and exclusive annihilation cross sections as a function of the beam energy.
This is obviously limited to states with masses greater than 2~GeV/$c^2$
and has been used quite effectively to study the charmonium system by the
Fermilab E769 collaboration (Armstrong, 1997, and references therein)
as well as some relatively massive light-quark states in the JETSET experiment
at CERN (Bertolotto, 1995; Evangelista, 1997; Buzzo, 1997; Evangelista, 1998).
However, most of the
contributions to light meson spectroscopy have come from $\bar{p}p$
annihilations at rest with the Crystal Barrel experiment (Aker, 1992)
at CERN. In fact, this has been reviewed quite recently (Amsler, 1998).
Significant contributions have also come from the OBELIX experiment
(Bertin, 1997a, 1997b, 1998) in particular using the $\bar{n}p$ annihilation
reaction.  There is also data from the older ASTERIX collaboration
(May 1989, 1990a, 1990b; Weidenauer, 1993).

The annihilation process is clearly complicated at a microscopic level.
However, in annihilation at rest in which a state $X$ recoils against a light,
stable meson, one may expect many different components to the wave function
of $X$.  This would include non $q\bar{q}$ degrees of freedom. In fact,
this process has been suggested as a fine way to excite gluonic degrees of
freedom in which case $X$ might have large glueball or hybrid content, so
long as the mass is not much larger than $\sim1700$~MeV/$c^2$.

Antiproton annihilation in liquid hydrogen proceeds almost entirely through
a $\bar{p}p$ relative $S$-state (Amsler, 1998).  This lends enormous power
to the partial wave analysis because the initial state is tightly constrained.
Annihilation into three, stable, pseudoscalar mesons
(for example $\bar{p}p\to\pi^0\pi^0\pi^0$ and $\bar{p}p\to\eta\eta\pi^0$)
has been particularly fruitful.  In these reactions, one studies the two-body
decay of a meson resonance which recoils off of the third meson.
These data have had their greatest impact on the scalar meson sector.

\subsection{Central Production}
\label{secIII:central}

\begin{figure}
\centerline{\epsfig{file=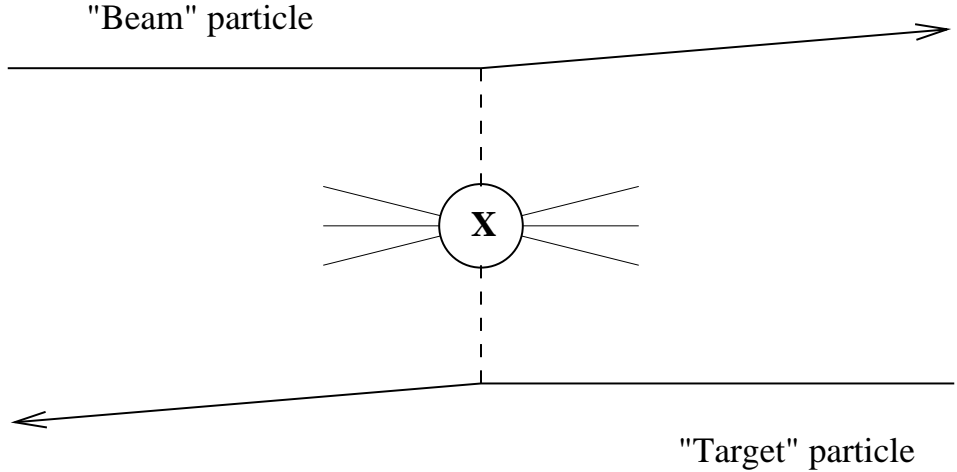,width=5.0in}}
\caption{Schematic diagram of a central production process.
Two off-mass-shell particles collide, while the ``beam'' and ``target''
move off essentially unscathed.  The exchange particles are understood to
be very rich in ``glue'' when the transverse momentum kick to the beam
and target is small.}
\label{figIII:central}
\end{figure}

Peripheral processes are viewed as exciting the ``beam'' particle by means
of an exchange with the ``target'' particle, leaving the target more or less
unchanged.  Central production refers to the case where there is a collision
between exchange particles.  This is shown schematically in
Fig.~\ref{figIII:central}.
Experimentally, using proton beam and target, one observes the reaction
$pp\to p_f (X^0) p_s$ where $p_s$ and $p_f$ represent the slowest and fastest
particles in the laboratory frame.  At high energies, and low transverse
momentum to $p_s$ and $p_f$,  this process is believed to
be dominated by double Pomeron exchange. As the Pomeron is believed to have
large gluonic content, one might expect that $X^0$ is a state dominated by
gluonic degrees of freedom (Close, 1997b).

This technique has been exploited extensively at CERN, in the
WA76 (Armstrong, 1989a, 1989b, 1991a, 1991b, 1992),
WA91 (Abatzis, 1994),
and WA102 experiments (Barberis, 1997a, 1997b, 1997c, 1998).

\subsection{Results from $e^+e^-$ Storage Rings}

High luminosity (${\cal L}\geq10^{32}/{\rm cm}^2\cdot{\rm sec}$)
$e^+e^-$ storage rings have been in operation for close to three decades.
Known primarily for their contributions to heavy quark spectroscopy,
they have shed valuable light on the light quark mesons in a variety
of ways.  These include direct production and spectroscopy of isovector
and isoscalar vector mesons (i.e. $\rho$, $\omega$, and $\phi$ states),
states produced in the radiative decay of the $J/\psi$, and indirect
production of various mesons in ``two-photon'' collisions.

\subsubsection{Vector Meson Spectroscopy}

The $e^+e^-$ annihilation process is mediated by a single virtual photon with
the quantum numbers $J^{PC}=1^{--}$.  These reactions therefore produce
vector meson resonances, and the isospin and other dynamic features are
studied through the appropriate final states.  By varying the $e^\pm$ beam
energy, experiments scan through the center of mass energy and trace out the
resonance shape, modified by interferences with overlapping states.

The reaction $e^+e^-\to\pi^+\pi^-$ has been carefully studied in the region
up to around 2~GeV center of mass energy (Barkov, 1985; Bisello, 1989).
These data have clearly established three $\rho$ resonances (Bisello, 1989)
and have also been used to precisely study $\rho/\omega$ mixing.
Other reactions have also been studied by scanning over center of mass
energies in this mass region, mainly by the DM2 collaboration at ORSAY
(Castro, 1994).

\subsubsection{Two-Photon Collisions}
\label{secIII:gg}

Typically, $e^+e^-$ colliders are used to acquire large amounts of data at
high energy resonances, such as $c\bar{c}$ or $b\bar{b}$ states, or (in the
case of LEP) at the $Z^0$ pole.
One very fruitful source of data on meson spectroscopy (as well as other
physics) in high energy $e^+e^-$ collisions is
the reaction $e^+e^-\to e^+e^- X$ where the state $X$ is produced by the 
collision of two photons radiated from the beam electron and positron.
This is rather analogous to the central production process
(Sec.~\ref{secIII:central}, Fig.~\ref{figIII:central}) where the photons
replace the less well understood Pomeron.
The field of ``two-photon'' physics has been rather extensively reviewed
(Morgan, Pennington, and Whalley, 1994; Cooper, 1988).
Data continues to be acquired and analyzed at operating $e^+e^-$ storage
ring facilities.

Some particular features of meson spectroscopy in two-photon collisions are
immediately apparent.  First, it is clear that only self-conjugate, $C=+1$
meson states $X$ will be formed in the collision.  Second, to the extent
that the photons couple directly to the $q$ and $\bar{q}$ that are formed,
the production rate will be proportional to the fourth power of the
quark charge.  Thus, $u$ (and $c$) quarks will be preferred, relative to
$d$ and $s$ quarks, and this has been used to determine the singlet/octet
mixing in the $\eta$ and $\eta^\prime$ (Cooper, 1988).  Also, if a state
is dominated by gluonic degrees of freedom (a ``glueball''), then there is
no valence charge to couple to photons, so we expect glueballs to {\em not}
be produced in these reactions.  This was discussed in 
Sec.~\ref{secII:glueballs} and quantified in Eq.~\ref{eq:sticky}.
Thirdly, two-photon reactions are a
powerful tool for spectroscopy in a way that is directly related to the
way the scattered $e^\pm$ are detected.

Spectroscopic data from two-photon collisions is generally separated into
``untagged'' and ``tagged'' samples.  The virtual photon spectrum is sharply
peaked in the forward direction, since the photon propogator is essentially
proportional to $1/q^2$ where $q$ is the photon four-momentum.  Consequently,
if the incident $e^\pm$ is scattered through a large enough angle to be
``tagged'' by the detector, the exchanged photon will have large enough
$q^2$ to be strongly ``virtual''.  That is, it will have a significant
component of longitudinal polarization.  On the other hand, if neither
the electron nor positron is tagged, one can safely assume that the
exchanged photons are essentially ``real''.

This leads to powerful selection rules (Yang, 1950) on the quantum numbers
of the meson formed in the collision.  In particular, for real ($q^2=0$)
photons, all spin $1$ states and odd spin states with negative parity are
forbidden.  Therefore, only states with
$J^{PC}=0^{\pm+},2^{\pm+},3^{++},\cdots$ are produced in untagged events,
and states with other quantum numbers should show a very strong dependence
on $q^2$ for tagged events.

\subsubsection{Radiative $J/\psi$ Decays}
\label{secIII:Jpsi}

All decays of the form $J/\psi\to\gamma X$ (except $J/\psi\to\gamma\eta_c$)
involve the annihilation of the $c\bar{c}$ pair into a photon and a hadronic
state of arbitrary mass.  To first order in perturbative QCD, this proceeds
through $J/\psi\to\gamma gg$ so one might expect the hadronic state to couple
strongly to two gluons.  Consequently, radiative $J/\psi$ decay has long been
regarded as a fertile hunting ground for glueballs.  In this manner, at least,
it is quite complementary to two-photon production.  Here again, the state $X$
must be self conjugate with $C=+1$.

The branching ratio for radiative $J/\psi$ decay is typically between
a few $10^{-4}$ and a few $10^{-3}$.  The best experiments acquire on the
order of several million produced $J/\psi$, so only a few thousand accepted
events can be expected for each of these states. Consequently, the statistical
power is meager and complete partial wave analyses are difficult.

This subject has not been reviewed for some time
(Hitlin and Toki, 1988; K\"{o}nigsman, 1986). Since then, however, new results
have been presented from the DM2 collaboration at Orsay
(Augustin, 1988, 1990)
and the Mark~III experiment at SPEAR (Dunwoodie, 1997; Burchell, 1991),
and new data is being collected and analyzed by the
BES collaboration at Beijing (Bai, 1996a, 1996b, 1998).

% ======================================================================
%\include{rev-v6-4}
% ======================================================================
\newpage
\section{THE QUARK MODEL: COMPARISON WITH EXPERIMENT}
\label{sec:COMPARISON}

\subsection{Heavy Quarkonia}

\begin{figure}[hbt]
\centerline{\epsfig{file=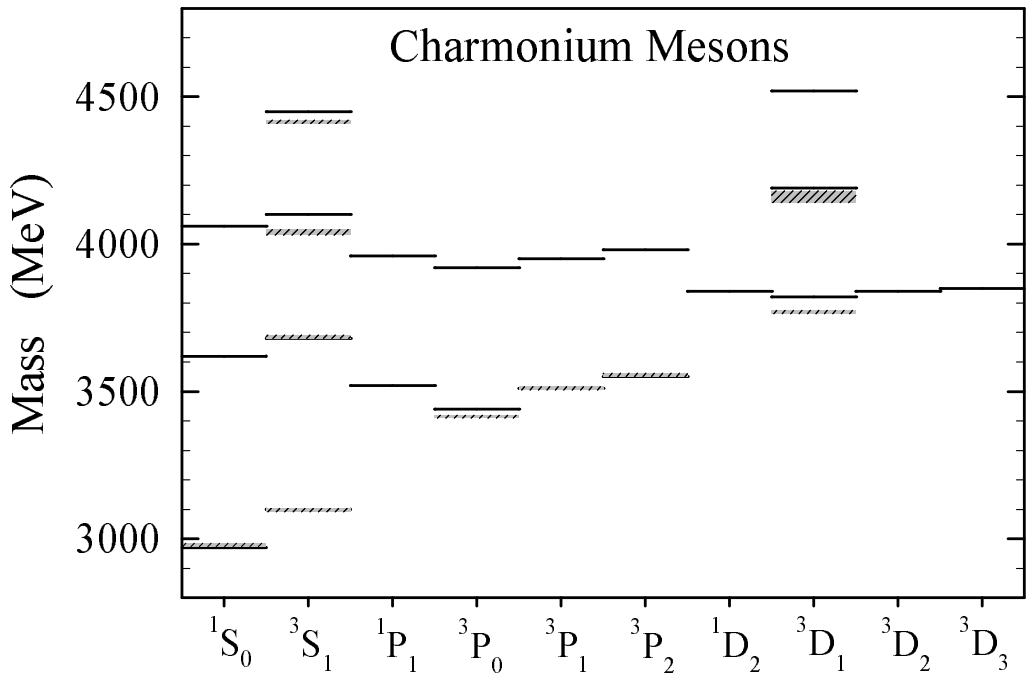,width=14.0cm,clip=}}
\caption[]{
The $c\bar{c}$ spectrum.  The solid lines are the quark model 
predictions and the shaded regions are the experimental states.  The 
size of the shaded regions approximates the experimental uncertainty.
Note that the interpretation of
the $\psi(4040)$ and $\psi(4160)$ as a single resonance is unclear because
of the substantial threshold effects in this energy region.  Although they
are included as $^3S_1$ states their classification is ambiguous.}
\label{fig:charmonium}
\end{figure}

It is useful to start with heavy quarkonium where there is theoretical 
justification for using potential models to calculate their
spectra and where there is some validity in identifying the static 
quark potential one obtains from lattice QCD calculations with the
phenomenological potential obtained empirically 
from heavy quarkonia spectra.  What is surprising is that the general 
spectroscopic features evident in the heavy quarkonia spectra persist 
to light meson spectroscopy where the quark model is on shakier grounds.
In figure \ref{fig:bb-mass} we compared quark model predictions
for the $b\bar{b}$ system to experiment and found the agreement to be good.
In figure \ref{fig:charmonium} 
we show a similar comparison for the $c\bar{c}$ spectrum, also with 
good agreement.
The basic experimentally observed level structure consists of a tower of $1^-$
radial excitations with only one P-wave orbitally excited multiplet for
the $c\bar{c}$ mesons and two for the $b\bar{b}$ mesons.  This is
because the $1^-$ states are directly produced in $e^+e^-$ collisions while
the higher orbital states require decays from the produced $1^-$ states
and are therefore much more difficult to produce and observe.  
There have been suggestions that D-wave quarkonium can be produced 
via gluon fragmentation at hadron colliders (Qiao 1997a), in $Z^0$ 
decays (Qiao 1997b), in B decays (Yuan 1997, Ko 1997) and in fixed 
target experiments (Yuan 1998).  There is some evidence for the D-wave 
$2^{--}$ charmonium state with Mass $3.836\pm 0.013$~GeV in Fermilab 
experiment E705 (Antoniazzi 1994) although this result is questioned 
by other experiments (Gribushin 1996).
It is also possible
that with the high statistics available at a B-factory that the $b\bar{b}$
D-waves might be observed via cascade radiative transitions from 
$\Upsilon (3S) \to 2P\gamma \to 1D \gamma\gamma$ (Kwong 1988).
For both the $b\bar{b}$ and $c\bar{c}$ spectra, the 
states which differ measurably from the predicted masses are the $1^{--}$
states near open bottom and charm threshold, respectively,
where the neglect of coupling to decay
channels may not have been justified (Eichten 1975, 1978, 1980).  
Note that there are classification 
ambiguities in a few cases for the high mass $1^{--}$ resonances.  

\subsection{Mesons With Light Quarks}

Given the successful description of heavy quarkonia by the quark 
potential model we proceed to the light quark mesons.
Following the argument of section \ref{sec:THEORY}
that the basic structure in heavy and light systems are 
qualitatively identical, 
we use the quark model to interpret
the spectra of mesons with light quark content.
As already noted, studies of mesons with light quarks 
complement those of the heavy quarkonium in that they probe a different 
piece of the $q\bar{q}$ potential which allows the study of the strength and
Lorentz structure of the long-range confining part of the potential.
In addition, the hadroproduction mechanism is sufficiently different 
from production in colliding $e^+e^-$ machines that the experimentaly
accessable excitations are nearly orthogonal.

In our survey of mesons with light quarks we begin with a general survey
of these states to establish the global validity of the quark model 
predictions.  With this backdrop, in section \ref{sec:PUZZLES} we will
focus on, and study in detail, the states which
are either poorly understood or pose a problem for the quark model.

\subsubsection{Mesons With One Light Quark and One Heavy Quark}

\begin{figure}
\centerline{\epsfig{file=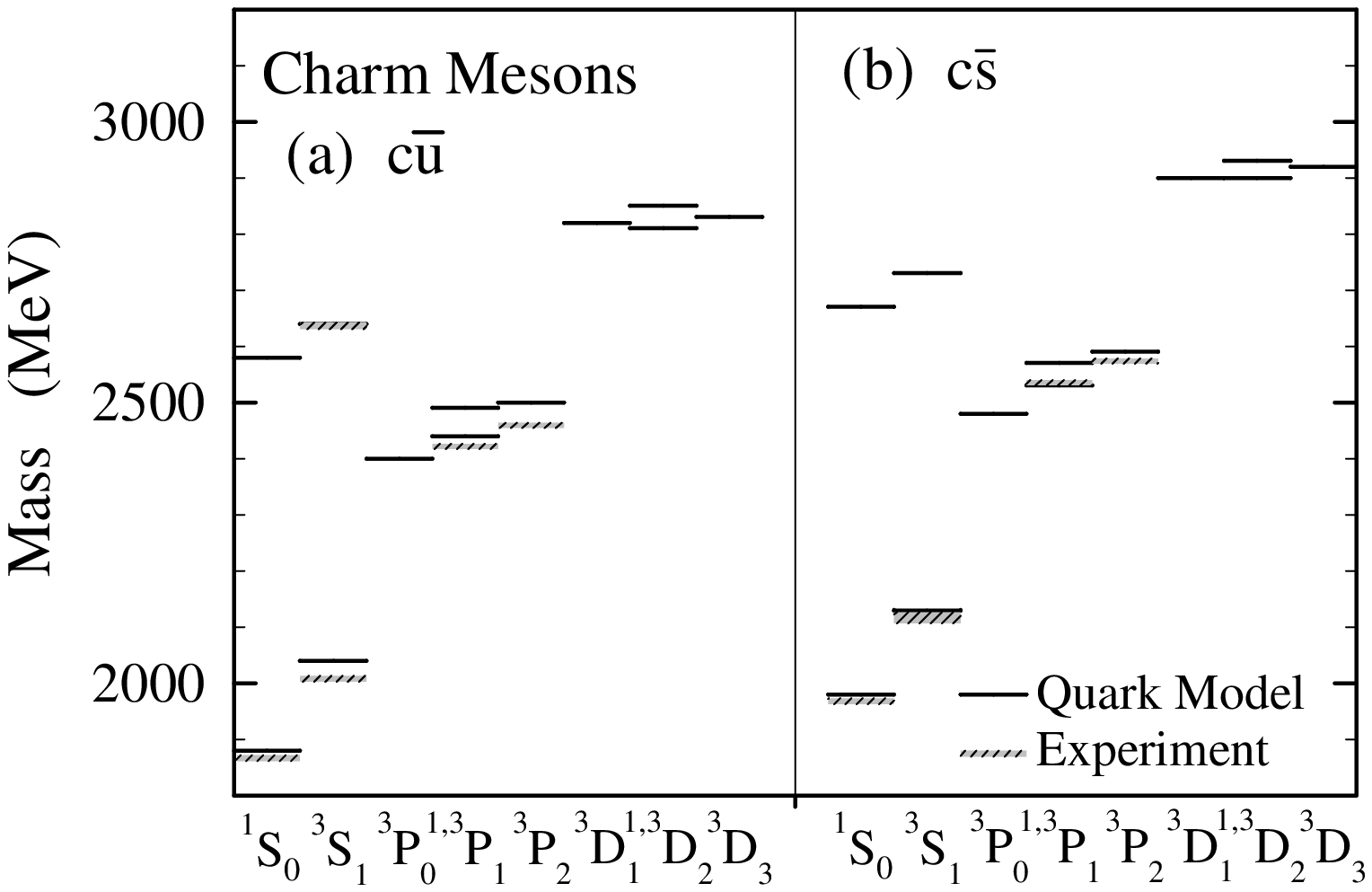,width=12.0cm,clip=}}
\centerline{\epsfig{file=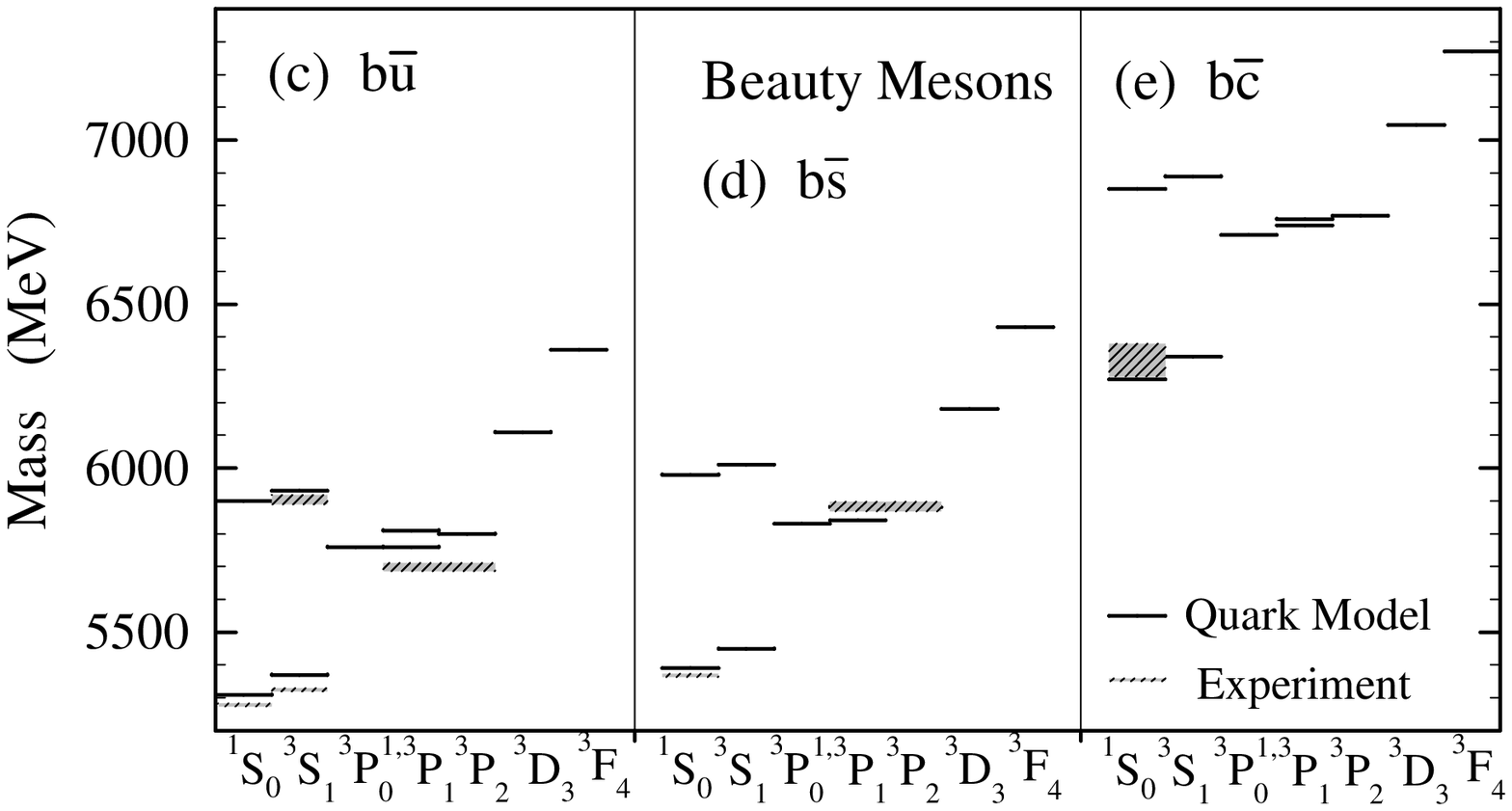,width=12.0cm,clip=}}
\caption[]{
Spectra for (a) $c\bar{u}$, (b) $c\bar{s}$, 
(c) $b\bar{u}$, (d) $b\bar{s}$, and (e) $b\bar{c}$.
The solid lines are the quark model predictions (Godfrey 1985a) and the 
shaded regions are the experimental measurements with the size 
representing the approximate experimental uncertainty.  The 
experimental results shown for the $^{1,3}P_1$ and $^3P_2$ $b\bar{u}$ 
and $b\bar{s}$ represent broad bumps interpreted as superpostions of 
more than one state assumed to be the $B_1$ ($B_{s1}$) and $B_2$ 
($B_{s2}$).  See the text and Fig~\ref{fig:opal}.}
\label{fig:charmed}
\end{figure}

We start with mesons
which contain one heavy quark such as the charmed and beauty mesons
(Rosner 1986, Godfrey 1991).
These systems are an interesting starting point
because, as pointed out long ago 
by De Rujula, Georgi, and Glashow (De Rujula 1976), 
as the heavy quark's mass increases,
its motion decreases, so the meson's properties will increasingly be governed
by the dynamics of the light quark and will approach a universal limit.  
As such, these states become the hydrogen atoms of hadron physics.  
Mesons with one heavy quark provide a spectroscopy as rich as 
charmonium but because the relevant scales are governed by the light 
quark, they probe different regimes.  They bridge the gap 
between heavy quarkonium and light hadrons providing an intermediate 
step on the way to studying the 
more complicated light quark sector in search 
of exotica like glueballs and hybrids. 
A growing number of excited charmed and beauty mesons have been observed 
by the ARGUS, CLEO, Fermilab E691 and E687, and more recently the LEP 
and CDF collaborations.  
The experimental situation and quark model predictions 
for the $c\bar{u}$, $c\bar{s}$, $b\bar{u}$,
$b\bar{s}$, and $b\bar{c}$ states are summarized in fig. 
\ref{fig:charmed}. 
See also Kwong and Rosner (Kwong 1991) and Eichten and Quigg (Eichten 1994).

For mesons composed of  an unequal mass quark and antiquark,
charge conjugation
parity is no longer a good quantum number so the triplet and singlet
states of the same total angular momentum 
can mix via the spin orbit interaction or some other mechanism
(Lipkin, 1977).
For example, the physical $J=$1 states are linear
combinations of $^3P_1$ and $^1P_1$ with mixing angle $\theta$
which has an important effect on the meson decay properties.
The OZI allowed decays for P-wave mesons 
can be described by two independent amplitudes, S-wave and D-wave.
One state is degenerate with 
the $^3P_0$ state and the other is degenerate with the $^3P_2$ 
state.  Furthermore,
the state degenerate with the $^3P_0$ state decays into
final states in a relative S-wave, 
the same as the $^3P_0$ state decay, while the state degenerate 
with the $^3P_2$ decays into final states in a relative
D-wave, the same as the $^3P_2$ state decay.  
Thus, in the heavy quark limit 
the P-wave mesons form two degenerate doublets.

These patterns in spectroscopy and decays can be extended to general 
principles and  this recognition that 
the heavy quark limit results in a new symmetry of QCD 
has led to considerable progress in our understanding of QCD through 
the study of mesons containing a single heavy quark
(Isgur 1989b; 1990; 1991, Voloshin 1987, see also Neubert 
1994, for a recent review).  
This symmetry arises because once a quark becomes 
sufficiently heavy its mass becomes irrelevant to the nonperturbative 
dynamics of the light degrees of freedom of QCD and the heavy quark 
acts as a static source of Chromoelectric field as far as 
the light degrees of freedom are concerned.
Thus, heavy hadron spectroscopy differs from that of hadrons containing 
only light quarks because we may separately specify the spin quantum 
number of the light degrees of freedom and that of the heavy quark. 
That is, 
$\vec{S}_Q$ and $\vec{j}_l = \vec{S}_q + \vec{L}$ are separately 
conserved so that each energy level in the excitation spectrum is 
composed of degenerate pairs of states 
$\vec{J} =\vec{j}_q + \vec{S}_Q = \vec{j}_q \pm 1/2$.  
This represents 
a new symmetry in the QCD spectrum in the heavy quark limit
and leads to relations between hadrons containing a single heavy quark.
The significance of these results cannot be overstated
as they follow rigorously from QCD in the heavy quark limit.

The heavy quark flavour symmetry also applies to the complete set of 
$n$-point functions of the theory including all strong decay amplitudes 
arising from the emission of light quanta like $\pi$, $\eta$, $\rho$, 
$\pi \pi$, etc., are independent of heavy quark flavour so that two 
states with spins $s_\pm$ must have the same total widths.  The heavy 
quark symmetry leads to a number of predictions.  
The two $D_1$'s both have $J^P= 1^+$ and are only distinguished by $J_l$ 
which is a good quantum number in the limit $m_c\to \infty$.
Since strong decays are entirely transitions of the light quark 
degrees of freedom the decays from both members of a doublet with 
given $J_\ell$ to the members of another doublet with $J_\ell^P$ are 
all essentially a single process.  This leads to the simple prediction 
that the two excited states should have exactly the same widths.
$D_0^* \to D\pi$ and $D_1 \to D^* \pi$ decay via S-wave and can 
be quite broad and therefore difficult to identify experimentally.
In contrast
$D_1 \to D^* \pi$ and $D_2^* \to D^* \pi, \; D\pi$ proceed via D-wave 
and are much narrower.  
These states are identified as the $D_1(2420)$ and $D_2(2460)$.  
These are the same conclusions obtained from the quark model.
There are numerous other predictions resulting from HQET that 
are relevant to weak decays and we encourage the interested reader to 
read one of the more specialized reviews on the important subject of the 
Heavy Quark Effective Theory (Neubert 1994).

\begin{figure}
\centerline{\epsfig{file=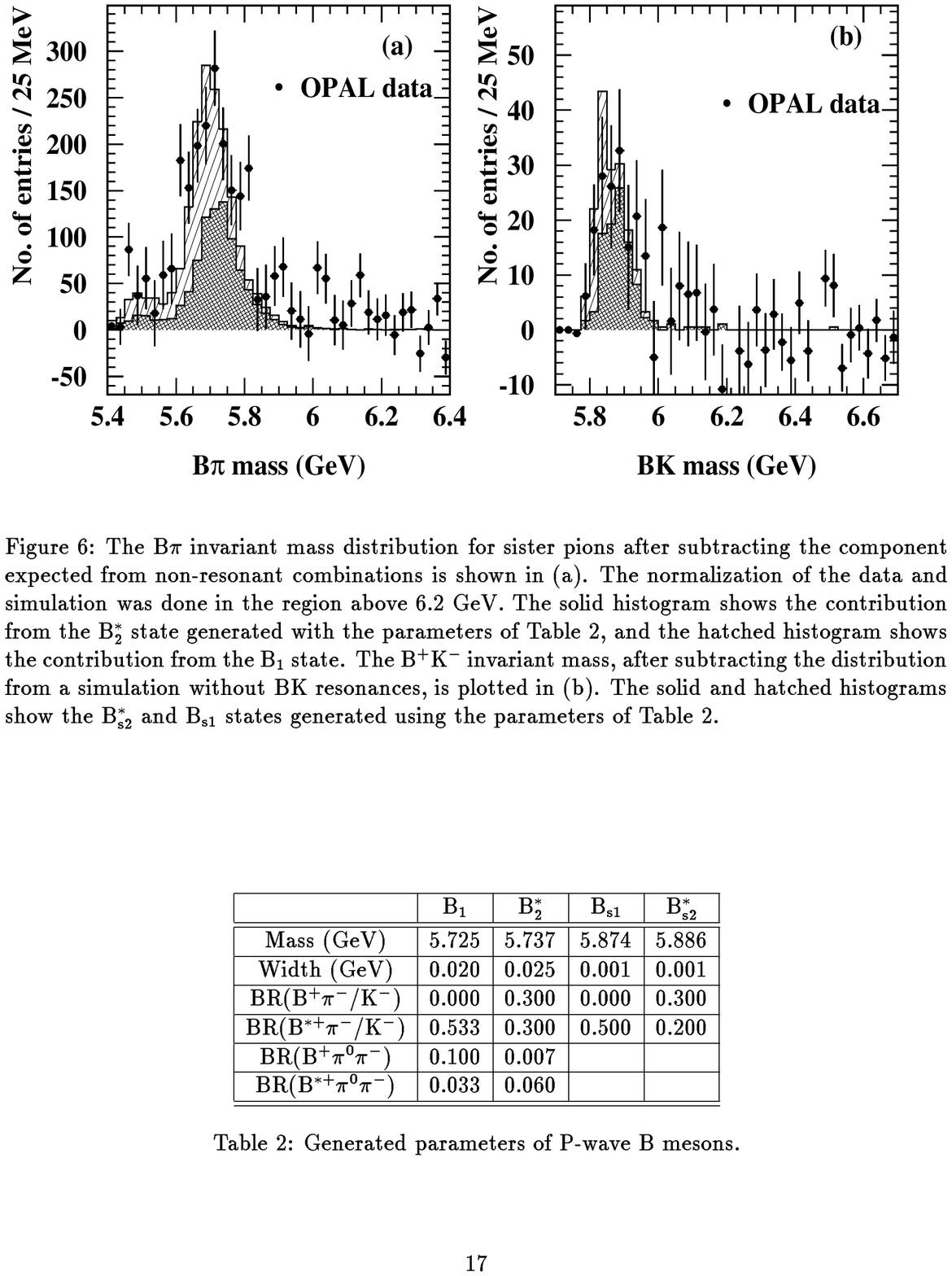,width=14.0cm,clip=}}
\caption[]{The $B\pi$ and $BK$ invariant mass distributions.  The solid 
histograms show Monte Carlo results for the $B_2^*$ and $B_{s2}^*$ 
states respectively and the hatched histograms show Monte Carlo 
results for  $B_1$ and $B_{s1}$ (Akers 1995).}
\label{fig:opal}
\end{figure}

While the P-wave charmed mesons have been known for some time,
the OPAL (Akers 1995), ALEPH (Buskulic 1996), and DELPHI (Abreu 1995)
collaborations at LEP have recently
reported the discovery of P-wave beauty mesons. 
The OPAL results are shown in Fig. \ref{fig:opal}.
Broad bumps are seen in $B\pi$ 
($M=5.68\pm 0.011$~GeV, $\Gamma=116\pm24$~MeV) and 
$BK$ ($M=5.853\pm 0.15$~GeV, $\Gamma=47\pm22$~MeV).  The $B\pi$ 
results are consistent with similar results by ALEPH and DELPHI. 
In both cases the 
widths are larger than the detector resolution of 40~MeV and the bumps 
are interpreted as superpositions of several states and/or decay 
modes.  In $B\pi$ the bump is assumed to be 
the $B_1$ and $B_2$ superimposed and in 
$BK$ the bump is assumed to be the $B_{2s}$ and $B_{1s}$ superimposed.   

The LEP collaborations have reported several candidates for the $B_c$ 
state.  The mean value for its mass averaged over the $\psi\pi$ decay 
mode is $m_{B_c}=6.33\pm 0.05$~GeV (Ackerstaff 1997, Abreu 1997).  
More recently CDF has also reported an observation of $B_c$ (Singh 1998).

The DELPHI collaboration (Abreu 1998, Ehret 1998) has reported evidence
for radial excitations of the $D_r^*$ and $B_r^*$.
From the invariant mass distribution $M(D^*\pi\pi)$ 
DELPHI obtains the mass measurement of $M(D_r^*) =2637\pm 2 \pm 6$~MeV 
which is in good agreement with the quark model prediction (Godfrey 1985a).
However, this state has not been confirmed by the OPAL collaboration 
(Krieger 1998)
or the CLEO collaboration (Shipsey 1998).
DELPHI also reports evidence for the $B^{\star\prime}$
from $Q(B^{(*)}\pi^+\pi^-)$ with
$M_{B^*_r}=5904\pm 4 \pm10$ which is also in good agreement with the 
quark model.
Although these results are preliminary they are promising and
together with the $B^{**}$ observations show the potential of high 
energy colliders for contributing to our understanding of hadron 
spectroscopy.

%A growing number excited mesons with one light and one heavy quark
%have been observed.  These observations agree with quark model predictions
%but more importantly they provide a test 
%of the Heavy Quark Effective Theory which provides a rigorous 
%calculational framework for QCD. 
%Further observations will give important information on the underlying
%dynamics.

\subsubsection{The Strange Mesons} 

An important ingredient in studying light meson spectroscopy is the study of
the level structure of the anticipated $q\bar{q}$ states. The strange mesons
(Aston 1986a, 1986b, 1987) 
are a good place to start for a number of reasons;
from an experimental perspective there is a reasonable amount of data
on strange mesons to test the model.  From a 
theoretical perspective strange mesons exhibit explicit flavour and
therefore don't have the additional 
complication of annihilation mixing which makes the isoscalar mesons much
more difficult to unravel.  It also eliminates the possibility of 
misidentifying a new meson as a glueball or $K\bar{K}$ molecule.

\begin{figure}[hbt]
\centerline{\epsfig{file=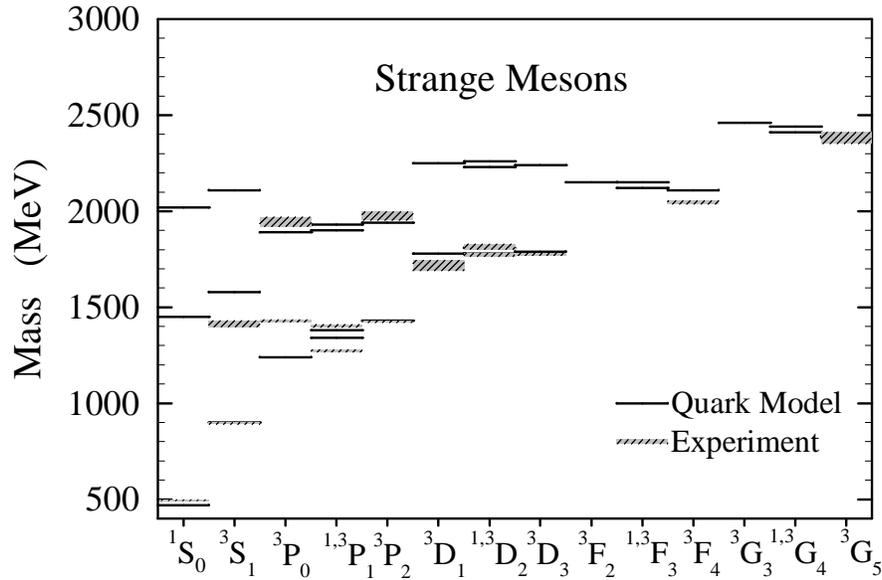,width=12.0cm,clip=}}
\caption[]{Level diagram for the strange mesons.  The quark model predictions 
(Godfrey 1985a) are given by the solid lines and the experimental measurements
are given by the shaded regions with the size representing the approximate
experimental uncertainty.}
\label{fig:strange}
\end{figure}

\begin{figure}
\centerline{\epsfig{file=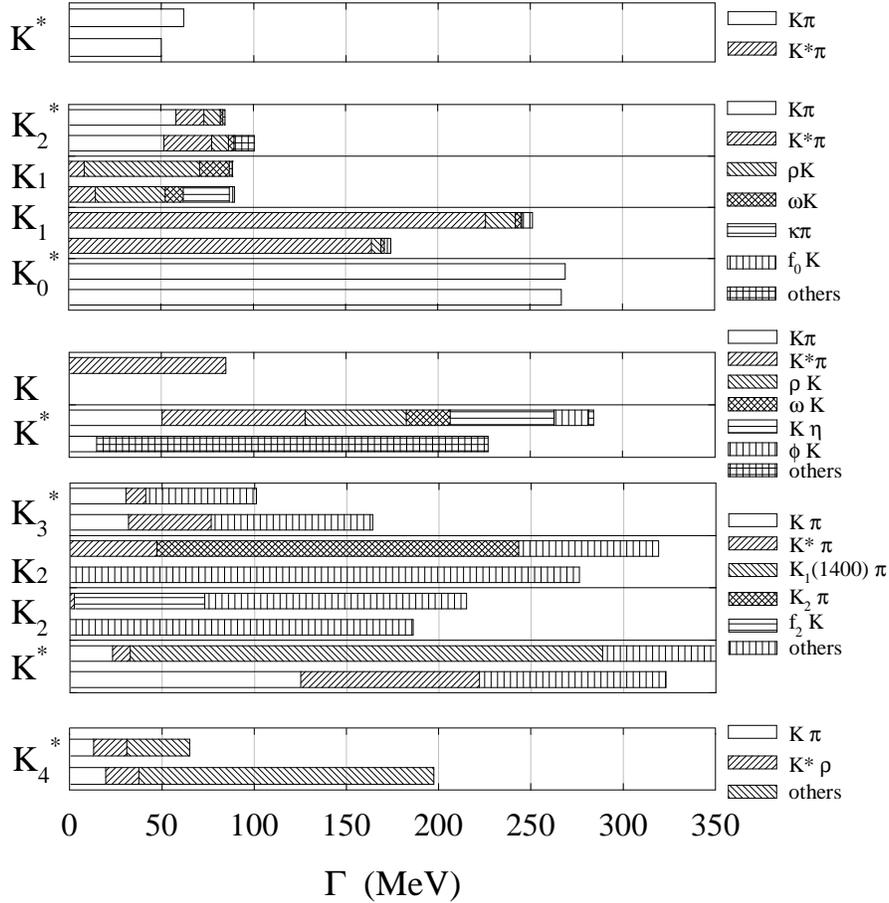,width=12.0cm,clip=}}
\caption[]{Strange meson strong decays. For each state
the quark model predictions are given by the upper bar 
(from Kokoski and Isgur (1987)) and the experimental results are given 
by the lower bar 
(from the Review of Particle Physics (Caso 1998) unless otherwise noted).
Some of the more important decay modes are indicated by the hatching.  
When only the total experimental width or at best only some of the 
partial widths have been measured we denote the unknown 
partial widths as {\em other}. }
\label{fig:strange-dec}
\end{figure}

The strange meson spectrum is shown in Fig. \ref{fig:strange} and the 
strong decay widths in Fig. \ref{fig:strange-dec}.
Much of the data comes from the
Large Aperture Superconducting Solenoid collaboration
(LASS) at SLAC, a high statistics study of  $K^-p$ interactions
at 11 GeV/c using the LASS  detector.
First note the qualitative similarity
to the $b\bar{b}$ (Fig.~\ref{fig:bb-mass}) 
and $c\bar{c}$ (Fig. \ref{fig:charmonium}) spectra.   
An important difference is that in the strange meson spectrum,
there are few candidates for radial excitations while the complete
leading orbitally excited $K^*$ series 
are observed up to $J^P=5^-$.  
A substantial number of the expected underlying
states are also observed in $K^-\pi^+$, $\bar{K}^0_s \pi^+ \pi^-$
and $K\eta$ final states. This reflects the importance of the production
mechanism in determining features of the spectrum.  
Another important difference between the heavy quarkonia spectra and
the light quark meson spectra is the relative 
importance of the electromagnetic and hadronic transitions in the two systems.
In light meson systems, OZI allowed decays are kinematically allowed 
and dominate while in heavy quarkonia the lower mass states are below 
the threshold for OZI allowed decays so that electromagnetic 
transitions dominate between these states.  The important strong 
decays are shown in Fig.~\ref{fig:strange-dec} for both the quark 
model $^3P_0$-model predictions (Kokoski 1987) and the measured widths.  
Qualitatively, the predictions are in reasonable agreement 
with experiment.  If a state is 
expected to have a broad width it generally does and the pattern of 
partial widths is approximately reproduced by experiment.  Thus, 
although quantitatively the predictions could be better, the
predicted decay patterns should provide useful information when trying to 
decide on the nature of a newly found meson.

In $K^- p \to K^- \pi^+ n$, a partial wave analysis reveals the
natural spin-parity states $J^P = 1^-\; K^*(892)$, $2^+\; K_2^* (1430)$,
$3^-\; K_3^* (1780)$, $4^+ \; K_4^*(2060)$, and $5^- \; K_5^* (2380)$
(Aston, 1986a).
The agreement is good for the masses of the leading orbital
excitations between the quark model and experiment supporting the picture
of linear confinement.
 
Partial waves of the reaction $K^-p \to \bar{K}^0 \pi^+ \pi^- n$ 
indicate structure in the $1^-$ wave around
1.4 and 1.8 GeV and in the $2^+$ wave around 2.0 GeV.
The individual $K^*$ and $\rho$ contributions
reveal 2 $1^-$ Breit-Wigner resonances;  a higher state
with $M=1717\pm 27$ MeV and $\Gamma=322\pm 110$ couples to both
channels while the lower mass state with $M= 1414\pm 15$ and 
$\Gamma =232\pm 21$ is nearly decoupled from the $K\rho$ channel 
(Aston 1987).
It is simplest to associate the higher state with the $1^3D_1$ state based
on the small triplet splitting and the lower state would be mostly the
first radial excitation of the $K^*(892)$. However, the
$2^3S_1\; K^* (1410)$ lies much lower in mass than quark model
predictions and its small coupling to $K\pi$ indicates a breakdown in the 
simple SU(3) model of decay rates.  This is likely due to mixing between
the two states via decay channels which would push the lower state down
in mass and the upper state higher.  This mixing would also
cause one of the states to couple strongly to 
one decay channel and the other state to another decay channel.

There is also evidence for 2 structures in the S-wave (Aston 1988e).
The first, 
with mass around\footnote{The mass and width quoted in Table 2 of
(Aston 1988e) are in fact incorrect due to a simple recording error.
We quote the correct values here, as communicated to us by W.~Dunwoodie.}
$M(K_0^*(1430))=1412\pm 6$~MeV and $\Gamma(K_0^*(1430))=294\pm 23$~MeV is
classified as the $^3P_0$ partner of the $K_2^*(1430)$.  Although this 
is $\sim 170$~MeV higher than the quark model prediction the $1^3P_0$ 
state is particularly sensitive to the details of the model so 
not too much should be read into this.
There is a second S-wave structure at around 1.9 GeV
with parameters $M= 1945\pm 22$~ MeV and $\Gamma \sim 201\pm 86$~ MeV.  
This
structure can only be classified as a radial excitation of the $0^+$
member of the L=1 multiplet, most probably the $2^3P_0$ state.  The $2^+$
also demonstrates resonance behavior in this same mass region with
$M=1973\pm 26$~MeV and $\Gamma=373\pm 68$~MeV, most probably the $2^3P_2$.

One can probe the internal dynamics of these states by studying their decays.
For example, SU(3) predicts that the $K\eta$ branching ratio will be
very small
from even spin $K^*$ states and large from odd spin states (Lipkin, 1981).  
The branching ratios are related to the ratios:
\begin{eqnarray}
 R_2  & = & {{ \Gamma(K_2^* \to K\eta)}\over {\Gamma (K_2^* \to K\pi)}}
      = {1\over 9} (\cos\theta_P + 2\sqrt{2}\sin\theta_P )^2
 	( { {q_{K\eta}} \over {q_{K\pi}} })^5  \\
R_3  & = & {{ \Gamma(K_3^* \to K\eta)}\over {\Gamma (K_3^* \to K\pi)}}
      = (\cos\theta_P )^2 ( { {q_{K\eta}} \over {q_{K\pi}} })^7
\end{eqnarray} 
where $\theta_P$ is the SU(3) singlet-octet mixing angle and has a value
of $\theta_P \sim -20^\circ$.  $R_2$ suffers a significant suppression
due to the cancellation between the two terms.  
This was studied
by the LASS collaboration in the reaction $K^-p \to K^- \eta p$ who found 
(Aston, 1988b)
$$BR(K_3^* \to K\eta) = 9.4\pm 3.4\% $$
$$BR(K_2^* \to K\eta) < 0.45\% \; (95\% \; C.L.) $$
in agreement with $SU(3)$.  For the $K\eta'$ channel the situation is reversed
with the even spin $K^*$'s expected to couple preferentially to $K\eta'$
and the odd spin $K^*$'s couplings to $K\eta'$  are expected to be suppressed.
The relative phases of the different decay amplitudes also agree with the
quark model predictions.

\subsubsection{The strangeonium mesons}
\label{secIV:ss}

The strangeonium ($s\bar{s}$) states provide an intermediate mass between the
heavier systems where the quark model is approximately valid and the lighter
mesons where it is on less firm foundation. For this reason strangeonium 
provides important 
input for hadron spectroscopy.  It is also important to understand these states
since a number of exotic candidates have been observed in final states 
where strangeonium might be expected.  

\begin{figure}
\centerline{\epsfig{file=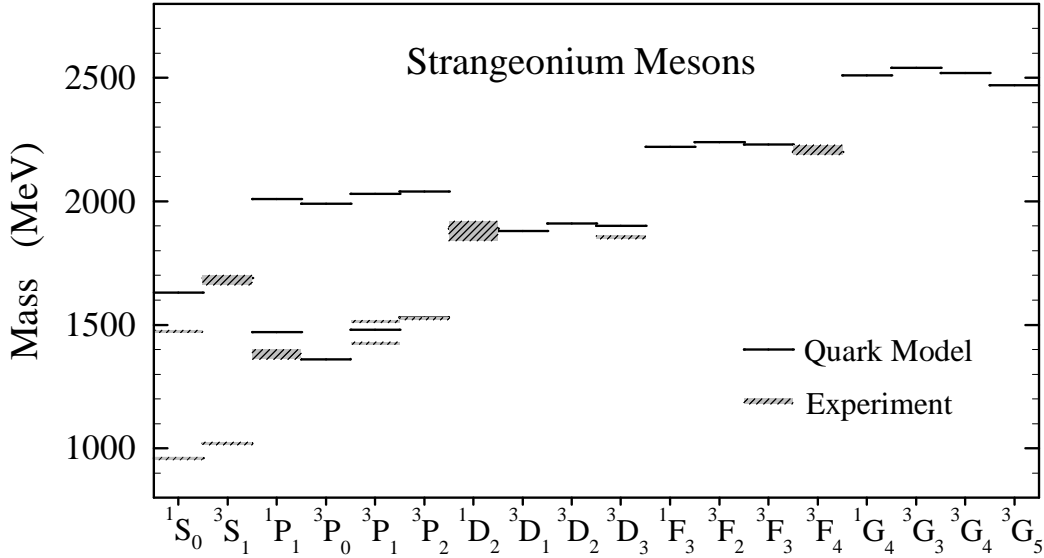,width=14.0cm,clip=}}
\caption[]{Level diagram for the strangeonium mesons. 
Note that there 
are 2 candidate $1^3P_1$ states, the $\eta(1470)$ is not unambiguously 
identified as the $2^1S_0$ state, and the $f_4(2220)$ is an 
unconfirmed report by the LASS collaboration (Aston 1988d).
See Fig.~\ref{fig:strange} for further details.}
\label{fig:strangeonium}
\end{figure}

\begin{figure}
\centerline{\epsfig{file=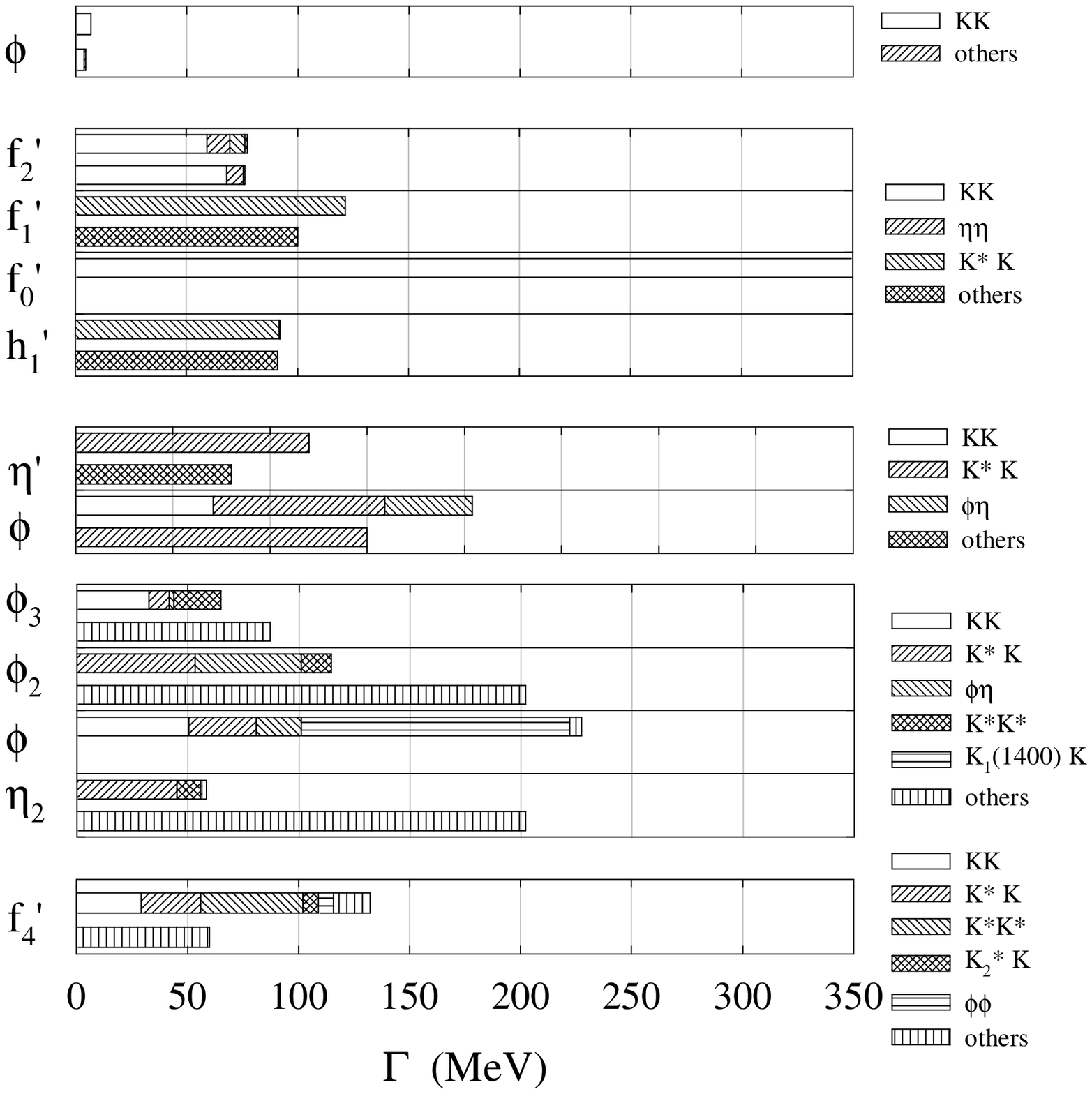,width=14.0cm,clip=}}
\caption[]{Strangeonium meson strong decays. For each state the 
the quark model predictions are given by the upper bar 
(Kokoski 1987, Blundell 1996, Barnes 1997) 
and the experimental results are given by the lower bar 
(from the Review of Particle Physics, Caso 1998).  See 
Fig.~\ref{fig:strange-dec} for further details.}
\label{fig:ss-dec}
\end{figure}

As in the case of the strange mesons, much of the data on the $s\bar{s}$
states have come from the high statistics study ($\sim$ 113 million triggers) 
of strangeonium mesons produced in the 
LASS detector by an 11 GeV/c $K^-$ beam.  The channels of interest are
dominated by hypercharge exchange reactions such as $K^-p\to K^-K^+ \Lambda$,
$K^-p\to K^0_sK^\pm \pi^\mp \Lambda$, and $K^-p\to K^0_sK^0_s \Lambda$,
which strongly favour the production of
$s\bar{s}$ mesons over glueballs.  The study
of strangeonium in hypercharge exchange reactions can provide revealing
comparisons with the same final states produced in gluon rich channels
such as $J/\psi$ radiative decays.  The level diagram for the strangeonium
spectrum is given in Fig. \ref{fig:strangeonium}
and their decay widths are shown 
graphically in Fig.~\ref{fig:ss-dec}.
The $s\bar{s}$ spectrum is similar to that of the strange meson spectrum.  
Except for the ground
state pseudoscalar mesons, the observed states fit into $SU(3)$ multiplets
consistent with magic mixing (pure $s\bar{s}$)
and agree well with quark  model predictions. 
The observed leading states lie on an essentially linear orbital ladder
that extends up through the $4^+\; f_4'$ and there are good candidates for 
the $^3P_0$ and $^3P_1$ partners of the $f_2'(1525)$.  
The couplings of natural parity states agree well with the SU(3) predictions
and the phases of the decay amplitudes are consistent with the quark model
predictions.
Overall, the parameters and decay transitions agree well with the predictions
of the quark model.  Some of the open issues in $s\bar{s}$ 
spectroscopy are the $\eta(1440)$, $f_J(1710)$, and the $f_J(2220)$ which 
we discuss below and in Section~\ref{sec:PUZZLES}.

\begin{figure}
\centerline{\epsfig{file=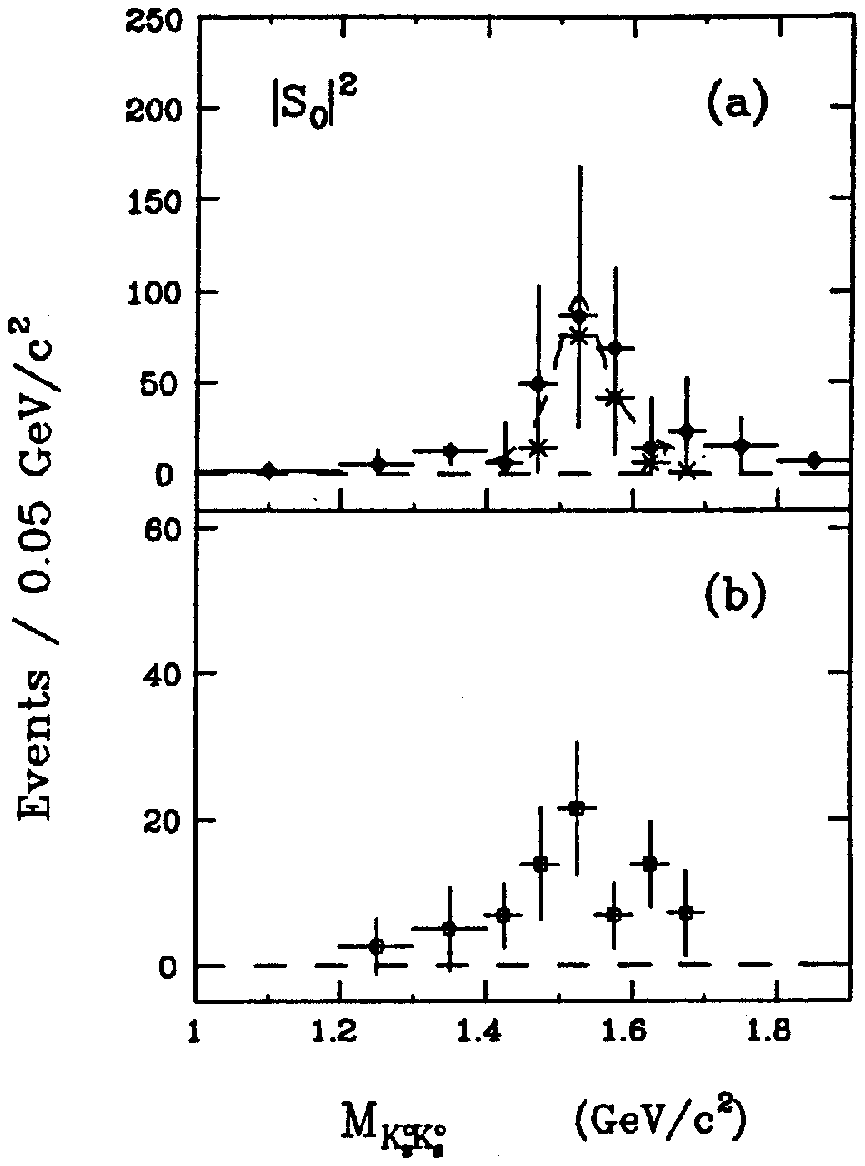,height=4.0in,clip=}}
\caption[]{S-wave intensity distribution for (a) the reaction
$K^-p\to K^0_S K^0_S \Lambda_{seen}$ and 
(b) $K^-p\to K^- K^+ \Lambda_{seen}$. (From Aston 1988a.)}
\label{fig:s-wave}
\end{figure}

An amplitude analysis of the reactions $K^-p \to K^0_S K^0_S\Lambda_{seen}$
and $K^-p \to K^-K^+\Lambda_{seen}$
indicates S-wave structure around the $f_2'(1525)$ mass, 
Fig.~\ref{fig:s-wave}, (Aston, 1988a)
suggesting the existence of a $0^+$ resonance which is most naturally
interpreted as the $^3P_0$ partner of the $f_2'(1525)$.
The approximate mass degeneracy of these states would imply that the 
spin-orbit interaction in the strangeonium sector is weaker that 
predicted which is consistent with the strange meson sector.
This implies that the 
the $f_0(975)$, normally interpreted as the $^3P_0\;(s\bar{s})$ state, is
not a conventional $q\bar{q}$ state.  There have been numerous sightings 
of another scalar meson with mass $\sim 1500$~MeV but with  
different properties than the state possibly observed by LASS in 
$K^-p \to K \bar{K}\Lambda_{seen}$.  We conclude that they are 
different states with the latter being discussed in detail in the 
following section.

\begin{figure}
\centerline{\epsfig{file=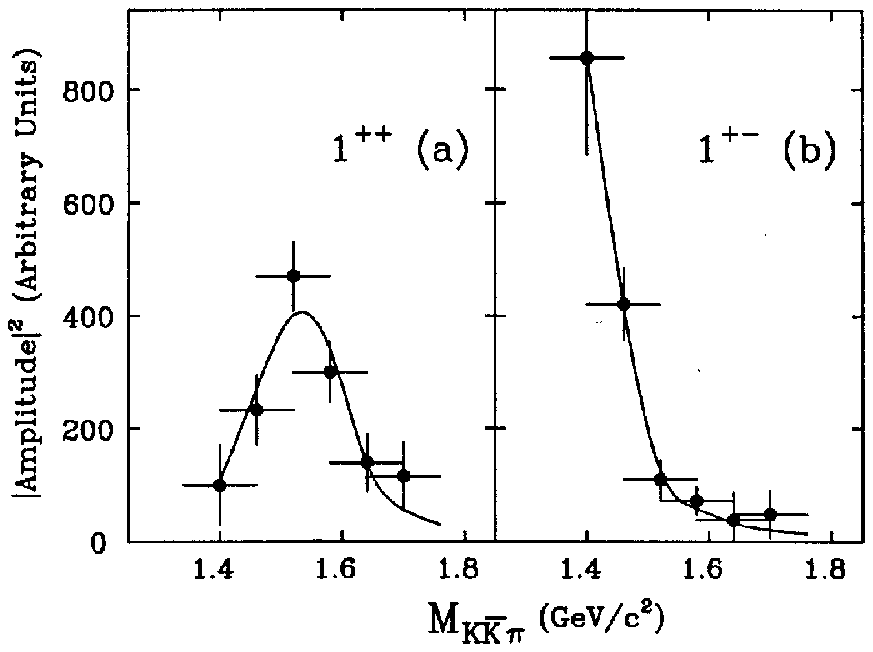,width=12.0cm,clip=}}
\caption[]{The mass dependence of the total production amplitude
intensities for the axial vector G-parity eigenstates.  The curves 
correspond to the Breit-Wigner fits described in the text. (From Aston 1988c.)}
\label{fig:ss-axials}
\end{figure}

There are two candidates for the $1^{++}$ $s\bar{s}$ state with masses 
$\sim 1510$~MeV and $\sim 1420$~MeV and 
disentangling them is tied up with the $E/\iota$ region to be discussed 
in Section~\ref{secV:Eiota}.
In the $K\bar{K}^* +c.c.$ modes, a partial wave analysis reveals 
structure around 1.5 GeV which is dominated by the $1^+\; K^*$ wave 
and around 1.85 GeV in the $2^-$ and $3^-$ waves.  The $1^+$ waves can be
combined to form eigenstates of G-parity as shown in 
Fig.~\ref{fig:ss-axials}(a) and (b). (Aston, 1988c)
These distributions are well described by Breit-Wigner curves as shown, and,
assuming I=0, represents good evidence of two $s\bar{s}$ 
axial vector meson states; $J^{PC}=1^{++}$ with $M\simeq 1530$ MeV
and $J^{PC}=1^{+-}$ with $M\simeq 1380$ MeV.  These states are good candidates
for the mainly $s\bar{s}$ members of the respective nonets since they
are strongly produced in the $Kp$ hypercharge exchange reaction.  
The $1^{+-}$ state has also been reported by E. King {\it et al.} (King, 1991) 
in the partial-wave anaysis of the $K^+K_S\pi^-$ system from the 
$K^-p$ interactions at 8 GeV/c (BNL-771 experiment)
and the Crystal Barrel Collaboration (Abele 1997b).  
The $1^+$ state, observed at around 1.5 GeV by LASS,
would be identified with the $f_1'(1530)$ claimed by Gavillet {\it et 
al.}. (Gavillet, 1982) although there is no evidence for this state by 
Fermilab E690 who see the $f_1(1420)$ (Berisso 1998).
These results, taken together with the $f_0(1530)$, 
indicate that the tensor and spin-orbit mass
splittings appear to be small.  
LASS shows no evidence for significant production of the $f_1(1420)$.
If this is the case, the $q\bar{q}$ interpretation of the
$f_1(1420)$, which has generally been taken as the $1^{++}$ strangeonium
state, comes into question indicating that the $f_1(1420)$ must be something
else, a $K\bar{K}^*$ molecule perhaps.  An alternative view advocated 
by the PDG (Caso, 1998)
assigns the $f_1(1420)$ as the $s\bar{s} \; 1^{++}$ state and 
concludes that the $f_1(1510)$ is not well established.

There is another $s\bar{s}$ candidate, the $f_J(1710)$, with width 
$\Gamma =150$ MeV,  seen in the invariant mass
distributions of $K^0_s\bar{K}^0_s$, $K\bar{K}$, and $\eta\eta$
pairs in $J/\psi$ radiative decay. (Baltrusaitis, 1987)
There is no evidence for such a state by
the LASS collaboration
indicating that the $f_J(1710)$ is not a conventional strangeonium state.
This state will be discussed in detail in  section~\ref{sec:PUZZLES}.

The $f_J(2220)$ was also first observed in a similar decay by the MARK III 
collaboration ($M=2231\pm 8$ GeV, $\Gamma=21\pm 17$ MeV)
(Baltrusaitis 1986a, Einsweiler 1984).
The LASS group sees a similar object in the reaction $K^-p \to K^+ K^- 
\Lambda_{seen}$ G-wave amplitude
which is evidence for a $4^{++}$ state (Aston, 1988d). 
This would be a member of the $4^{++}$ 
($f_4(2030)$, $a_4 (2040)$ (Cleland, 1982),
and $K_4^* (2060)$) (Aston, 1988e)
nonet predicted by the quark model. 
The LASS analysis yields mass and width values of $2209^{+17}_{-15}$ and
$60^{+107}_{-57}$ MeV/c$^2$ for this $J^{PC}=4^{++}$ state.
There is also evidence for this state
by the GAMS collaboration in $\pi ^-p \to
\eta\eta'$ (Alde 1986).  This implies
that the $f_J(2220)$, which has been conjectured to be an exotic hadron
of some sort (Chanowitz 1983a), 
is instead the $s\bar{s}$ member of the quark model
$^3F_4$ gound state nonet (Godfrey 1984, Blundell 1996).  On the 
other hand the BES collaboration (Bai, 1996a)
finds a decay width too narrow to 
be easily accomodated as the $1^3F_4 \; s\bar{s}$ state. More 
importantly they find that the decays are approximately flavour 
symmetric which supports a glueball interpretation.  One 
intrepretation which could accomodate the contradictory experimental 
evidence is that there are in fact two resonances;  the first is a 
broader conventional $s\bar{s}$ state and the second is a narrow 
glueball which is not observed in hadronic reactions.

To summarize, the $s\bar{s}$ spectrum agrees remarkably well with the 
predictions of the constituent quark model.  With this clarity, a
number of important puzzles have been revealed.  For example, it now
seems clear that there are too many low mass $0^{++}$ states and the
$f_1(1420)/\eta(1440)$, $f_J(1710)$, and $f_J(2220)$
regions contain intriguing hints of
non-quark model physics.  These puzzles will be explored in
section~\ref{sec:PUZZLES}.

\subsubsection{The Isovector Mesons}
\label{secIV:iv}

\begin{figure}[hbt]
\centerline{\epsfig{file=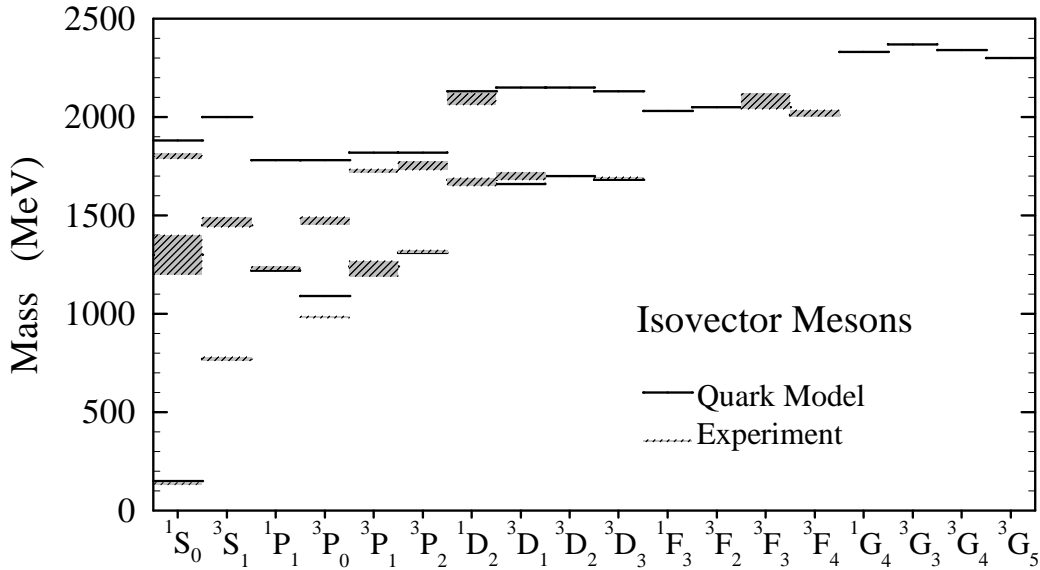,width=14.0cm,clip=}}
\caption[]{
Level diagram for the isovector mesons.
See Fig.~\ref{fig:strange} for further details.}
\label{fig:iv-mass}
\end{figure}

\begin{figure}
\centerline{\epsfig{file=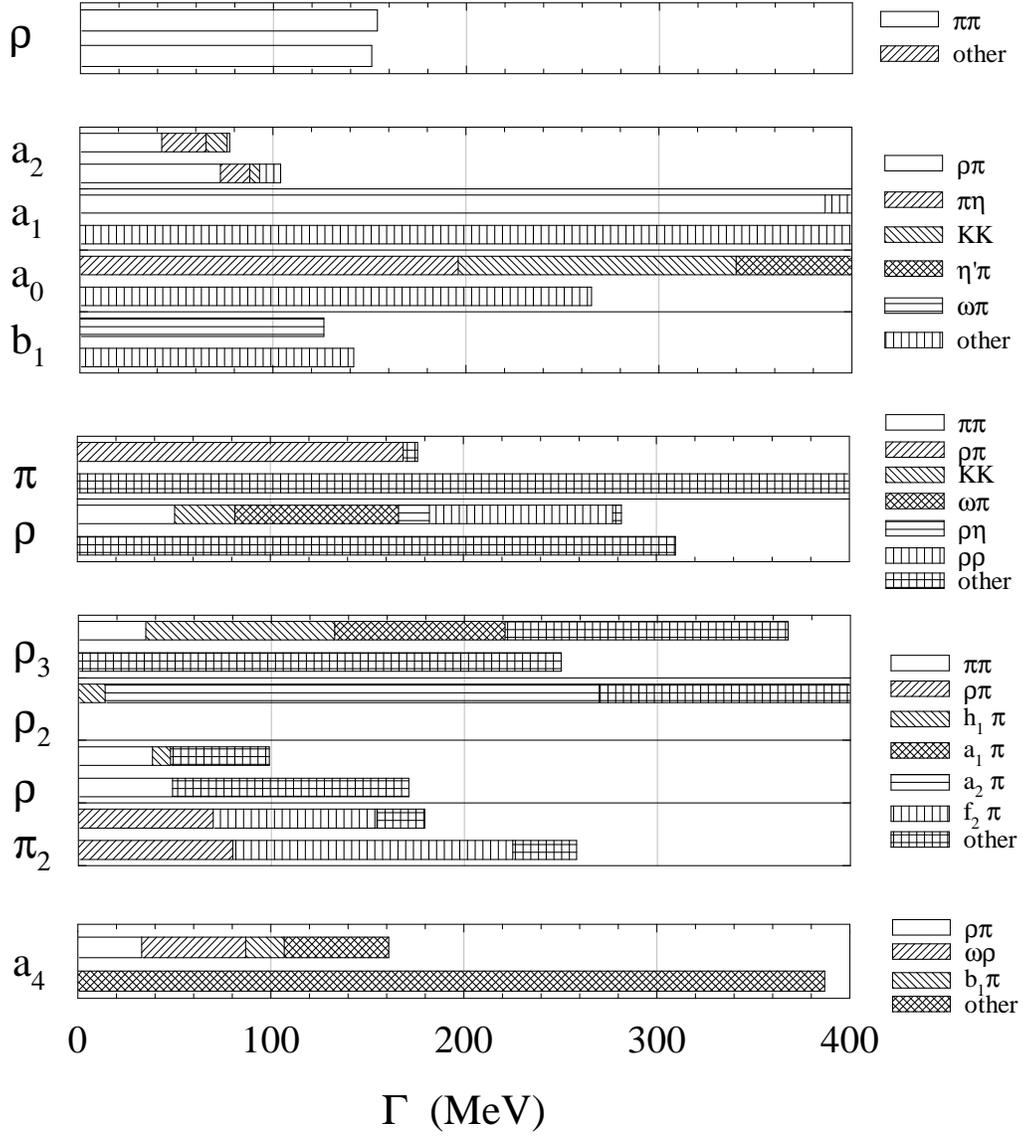,width=14.0cm,clip=}}
\caption[]{Isovector meson strong decays.  See 
Fig.~\ref{fig:strange-dec} for further details.}
\label{fig:iv-dec}
\end{figure}

Isovector mesons are made of the light up
and down quarks so they have the complications that arise from
relativistic effects but do not have the additional complication
of annihilation mixing which contribute to their isoscalar partners.  
The isovector meson spectrum is shown in Fig.~\ref{fig:iv-mass}
and the decay widths are given in Fig.~\ref{fig:iv-dec}.
For the most part there is good agreement 
between experiment and the quark model predictions.  
In the isovector sector, the orbitally excited states extend up as far as 
the L=5 $J^{PC}=6^{++}$ and are consistent with the quark model predictions, 
although the higher mass states need confirmation. 
The multiplet splittings for
the P and D wave mesons for the most part behave as expected.  
Given the general good agreement, it is the discrepancies which draw
attention to themselves and require further discussion.

Although the $a_0(980)$ scalar meson $J^{PC}=0^{++}$ lies about 100 MeV
below the quark model prediction for the $^3P_0$ state expected in this
mass region it is the measured width of $\sim 50$ MeV that is much
more difficult to reconcile with the quark model prediction of $\sim 500$
MeV.  This large discrepancy has led to numerous conjectures that the
$a_0(980)$ is not a $q\bar{q}$ state but rather, a more complicated
4-quark object (Jaffe, 1977a, 1977b, 1978), 
most probably a $K\bar{K}$ molecule (Weinstein and Isgur, 1982, 1983).  
The observation of the $a_0(1450)$  in 
$p\bar{p}$ annhilation by the Crystal Barrel Collaboration 
(Amsler 1994a, 1995) and the OBELIX Collaboration (Vecci 1998) is 
naturally assigned to be the isovector member of $1^3P_0$ nonet
reinforcing the interpretation that the $a_0(980)$ is a non-$q\bar{q}$ 
state. We will return to the scalar meson sector in the next section.

%{\bf Jim I'm not sure we need to keep the following paragraph:}
%The $\pi_2$ has been observed in two photon production experiments
%(Antreasyan, 1990; Behrend, 1990).  
%An analysis of the $2\pi^0$ subsystem shows evidence for the decay sequence
%$\pi_2 \to \pi^0 f_2(1270)$.  The two photon width is
%$\Gamma_{\gamma\gamma} =1.45 \pm 0.36$ keV (Antreasyan, 1990).  
%An analysis by the CELLO collaboration (Behrend, 1990) analysed the
%reaction $\gamma\gamma \to \pi_2 \to \pi^+\pi^-\pi^0$ which is complicated
%by the fact that in addition to a strong $a_2(1320)$ signal two intermediate
%states contribute and interfere; $\pi^0 f_2(1270)$ and $\pi^\pm \rho^\mp$.
%Depending on the relative phase of the interference they obtain
%$\Gamma_{\gamma\gamma}(\pi_2) =0.8 \pm 0.3$ for constructive interference,
%which is favoured by the data, and $\Gamma_{\gamma\gamma}(\pi_2) =1.3 \pm 0.4$
%for incoherent interference which is ruled out.  More recently, the L3 
%Collaboration (M. Acciarri 1997 L3 Collaboration Phys. Lett. {\bf 
%B413}, 147) has obtained the 90\% C.L. limit of $\Gamma(\pi_2)\cdot 
%BR(\pi^+\pi^-\pi^0) <0.072$ which contradicts the earlier results.

There is growing evidence for members of the radially excited L=1 
multiplet.  Both the CLEO (Kass, 1998) and DELPHI (Andreazza, 1998)
collaborations observed a 
signal in $\tau \to a_1' \nu_\tau \to \pi^-\pi^+\pi^- \nu_\tau$ with 
$M_{a_1}\simeq 1750$~MeV and $\Gamma_{a_1}\simeq 300$~MeV.  
Other observations of this state have been reported
by BNL E818 (Lee 1994), BNL E852 (Ostrovidov 1998) and VES (Amelin, 1995).  
There is also evidence for the  $J=2$ 
partner of this state, the $a_2'$, 
by the L3 collaboration which
observed it in two photon production with $\Gamma_{\gamma\gamma}(a_2) 
\times BR(\pi^+\pi^-\pi^0) =0.29 \pm 0.04$~keV (Hou 1998) making it 
the first radial excitation reported in $\gamma\gamma$ collisions.
The measured resonance parameters are $M(a_2')=1752\pm 21$~MeV and 
$\Gamma(a_2')= 150\pm 115$~MeV.
BNL E852 also reports observing it in 
$\pi^- p\to \eta \pi^- p$ (Ostrovidov 1998) while the
Crystal Barrel Collaboration 
reports an observation of an $a_2$ in $p\bar{p}$ 
annihilation in the final states $a_2\to 
\eta\eta\pi^0, \; \eta 3\pi^0$ with a much lower mass although the 
evidence for this state is not particularly strong (Degener 1997).

The $\pi(1800)$ has been observed by the VES Collaboration (Amelin 
1995).  Despite the fact that its mass is consistent with the $3^1S_0$ 
state there is speculation that it may be a hybrid because of its weak 
$\rho\pi$ decay mode.  However, the $3^1S_0$ interpretation predicts 
that the main decay mode would be $\rho\omega$ (Barnes 1997).  VES has 
studied the $\rho\omega$ final state and does indeed find evidence for 
a large $\pi(1800)$ signal (Ryabchikov 1998)
which supports the $3^1S_0$ assigment.

Finally, we mention the excited vector mesons.  At first 
only one excited vector meson state was observed, the $\rho(1600)$, 
whose properties did not agree well with the quark model predictions.  
Godfrey and Isgur (Godfrey and Isgur, 1985a) surmised that the observed
state was a mixture of two broad overlaping resonances, the $2^3S_1$ and
the $1^3D_1$ which gave rise to the observed properties.  
Subsequently, Donnachie and Clegg (Donnachie 1987)
performed a full analysis of the data for the annihilation reactions
$e^+e^-\to\pi^+\pi^-, \; 2\pi^+ 2\pi^-,\; \pi^+\pi^-\pi^0\pi^0$
and the photoproduction reactions 
$\gamma p\to \pi^+\pi^-p, \; 2\pi^+ 2\pi^-p,\; \pi^+\pi^-\pi^0\pi^0p$
and came to the conclusion that a consistent picture required
two states, the $2^3S_1$ at $1465\pm25$ MeV with width $220\pm 25 $ MeV and 
the $1^3D_1$ at $1700\pm 25 $ MeV with width $220\pm 25 $ MeV.  
These results have been confirmed by recent
results in $\bar{p}p$ annihilation by the Crystal Barrel Collaboration
(Abele 1997c) and by the OBELIX Collaboration (Bertin 1997b, 1998).
The properties of these two states are in reasonable 
agreement and more recent results examining the $4\pi$ decays of these 
states support the quark model assignments (Thoma 1998).
There have been additional claims of a $\rho(1250)$ seen 
in $\omega\pi$ but not in $\pi\pi$.  Although this state has been 
claimed many times it has never been generally accepted and there are 
theoretical difficulties reconciling it with $e^+e^-$ data
so at present its status is unclear.

\subsubsection{The Non-strange Isoscalar Mesons}
\label{secIV:is}

\begin{figure}
\centerline{\epsfig{file=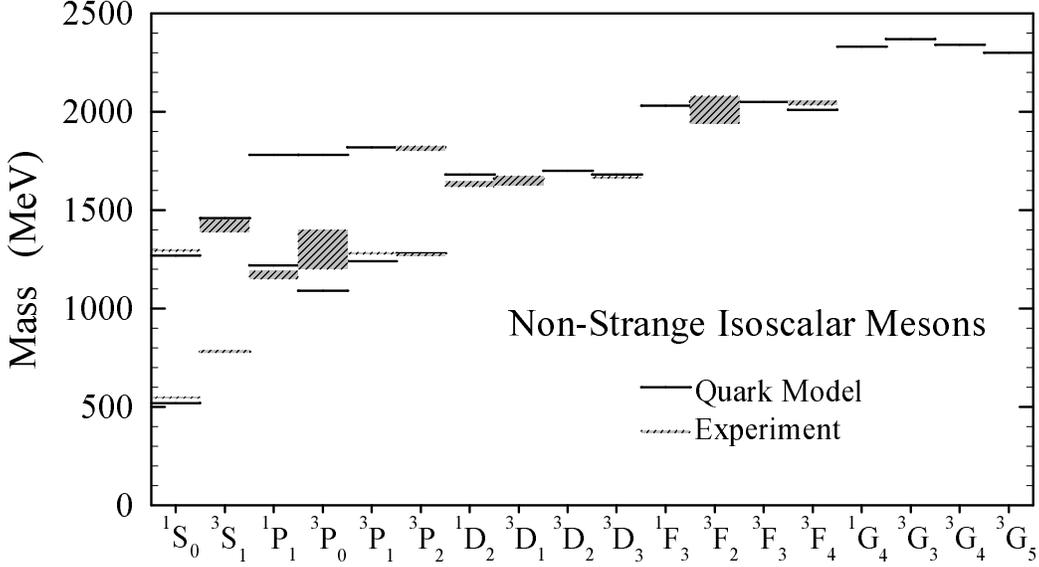,width=14.0cm,clip=}}
\caption[]{The non-strange isoscalar meson spectrum.
See Fig.~\ref{fig:strange} for further details.}
\label{fig:is-mass}
\end{figure}

\begin{figure}
\centerline{\epsfig{file=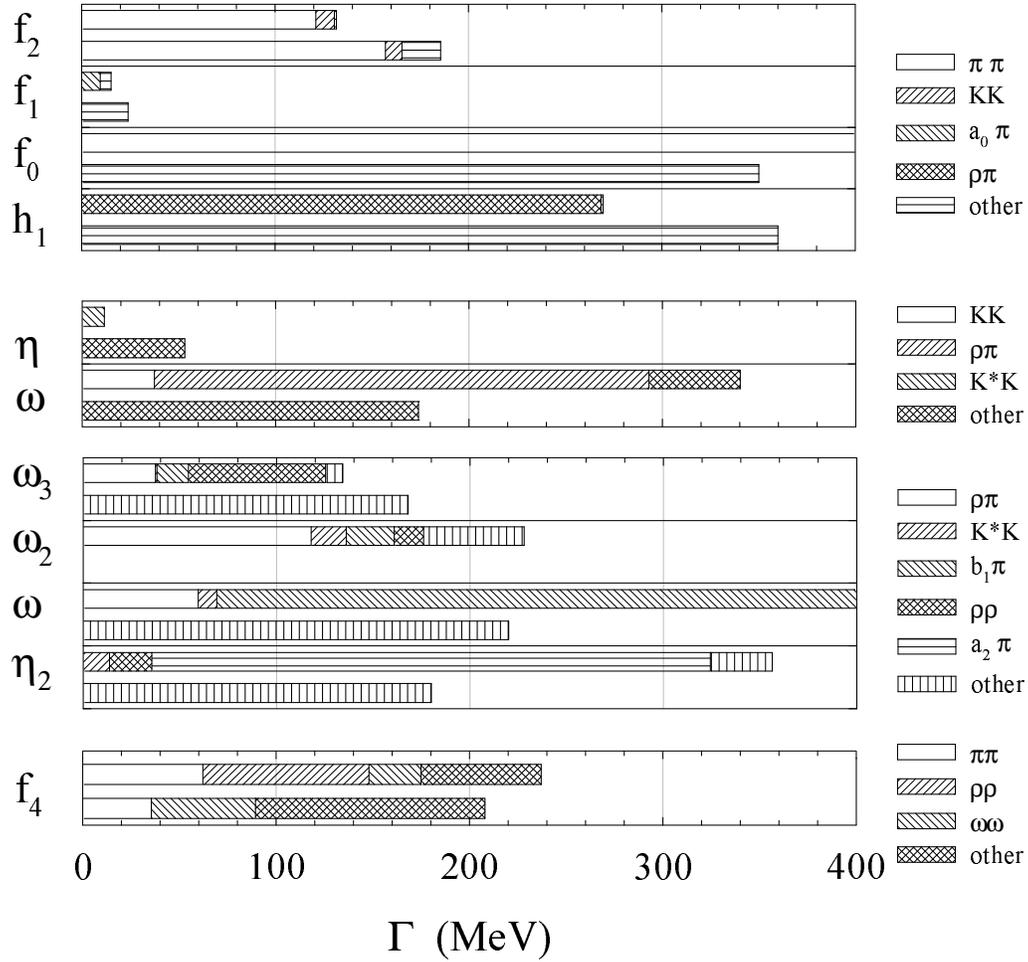,width=14.0cm,clip=}}
\caption[]{The non-strange isoscalar meson decays.
See Fig.~\ref{fig:strange-dec} for further details.}
\label{fig:is-dec}
\end{figure}

The non-strange isoscalar mesons 
are the last of the light quark meson families 
and are also the most problematic.
Experimentally, the isoscalars have been difficult to detect since
neutral mesons tend to be more difficult to detect than their charged
partners.  Phenomenologically, there is the problem of $n\bar{n}$ 
and $s\bar{s}$ mixing and, perhaps, the possibility of glueballs mixing
with ordinary isoscalars.  The spectrum and decay widths are shown in
Figs~\ref{fig:is-mass} and \ref{fig:is-dec} respectively.
%Recently, new experimental 
%information on $J^{PC}=2^{++}$, $0^{++}$, $0^{-+}$, and $1^{++}$
%has become available which might provide clues to the presence of gluonic
%excitations and multiquark systems.  This will be discussed in
%the following section on puzzles and problems.
%We point out their 
%existence in the appropriate location
%in Fig. \ref{fig:is-mass} of the isoscalar mesons but
%defer their discussion until the following section on puzzles and problems.
Although there is generally good agreement between the experimental
values and the quark model predictions, there are also numerous puzzles
which may point to the existence of ``exotic'' hadrons lying outside
the constituent quark model.  We will defer their 
discussion until the following section on puzzles and problems.

As in the case of the $s\bar{s}$ mesons the $^1P_1$ state $h_1(1170)$ and
the $^3P_1$ $f_1(1285)$ differ slightly from their 
respective quark model predictions, most likely indicating the need for 
further study of the annihilation mixing mechanism 
in the $^1P_1$ and $^3P_1$ sector of the model.  

The Crystal Barrel collaboration has reported two $\eta_2$ states in 
$p\bar{p}\to \eta 3\pi^0$ (Amsler 1996).  The lighter state has a 
total width of about $\Gamma \sim 180$~MeV and was seen in $a_2\pi$.  
This is consistent with the quark model expectations for the $I=0$ 
$^1D_2$ $n\bar{n}$ state which should appear near the 1D multiplet 
mass of $\sim 1670$~MeV with $\Gamma \sim 260$~MeV with $a_2\pi$ the 
dominant decay mode.  The 2nd state, $\eta_2(1875)$ is too high in 
mass to be a 1D $n\bar{n}$ state and the strong $f_2\eta$ mode argues 
against a mainly $s\bar{s}$ state.  This state has been reported 
previously by
the Crystal Ball Collaboration who measured the $\eta\pi^0\pi^0$ mass
spectrum in $\gamma\gamma\to 6\gamma$ (Karch, 1990) 
and found $M(\eta_2)=1876\pm35
\pm 50$ MeV, $\Gamma_{total}(\eta_2)=228\pm 90\pm 34$ MeV and
$\Gamma_{\gamma\gamma}(\eta_2) \cdot BR(\eta_2\to \eta\pi\pi)
=0.9\pm 0.2 \pm 0.3$ keV.  The angular distribution of the $\eta\pi^0$
subsystem gives $J^{PC}=2^{-+}$.
CELLO also observed an enhancement in the
cross section for $\gamma\gamma\to\eta\pi^+\pi^-$ yielding a mass
of $1850\pm 50$ MeV, $\Gamma_{total}=380\pm 50$ MeV, and 
$(2J_X +1)\cdot BR(X\to \eta\pi\pi) =15\pm 5$ keV (Feindt, 1990).  
The best single resonance assignment is $0^{-+}\; f_0\eta$ followed
by $2^{-+}\; a_2 \pi$.  This was the first new resonance to be discovered
in $\gamma\gamma$ collisions.  One possible explanation is that 
its higher than expected mass, like the $\pi_2$,
might be understood in terms of final state interactions.  A second 
possibility is that it is a hybrid candidate (Barnes 1998).  This 
could be tested by searching for the $a_2\pi$ decay mode which Close 
and Page (Close 1995a) predict to be dominant.

The excited pseudoscalar mesons, in particular the 
$E/\iota$ mass region, remains a puzzle. (See Sec.~\ref{secV:Eiota}.)
Although the quark model predicts two excited pseudoscalar mesons 
it is difficult to reconcile the properties of the 
$\eta(1295)$ and $\eta(1440)$ with the quark model predictions.  In addition,
the $\eta(1440)$ is now considered to be composed
of two separate resonances (PDG, Caso 1998). 
They are referred to as $\eta(1410)$ and 
$\eta(1490)$. One could identify the $\eta(1275)$  as the radial 
excitation of the $\eta$ and the $\eta(1490)$  as the mainly $s\bar{s}$
radial excitation of the $\eta'$.  This leaves the $\eta(1410)$ as an 
extra state.  There is some speculation that it is somehow related to 
the $f_1(1420)$.

The $0^{++}$ state, the $f_0(980)$, has problems with its quark 
model assignment similar to that of the $a_0(980)$ and is also
interpretated as a four quark state.   If the $f_0(1370)$ and the 
$f_0(1525)$ reported by LASS are identified with the quark model 
ground-state isocalars  the discrepancies between the observed and 
predicted masses would be due to an overestimate of the quark model 
spin-orbit splittings as in the isovector and strange meson sectors.

There is also some confusion in the $2^{++}$ sector which to be understood
will have to be studied in conjunction with similar puzzles in the $s\bar{s}$
$2^{++}$ mesons.  In particular, the $f_J(1710)$ has been mentioned
previously in that subsection.  In addition, there are three tensor mesons,
$f_2(2010)/g_T$, $f_2(2300)/g_T'$, and $f_2(2340)/g_T''$, 
produced in the reaction $\pi^- p\to \phi\phi n$  which do fit in with
the quark model prodictions.  Because they are produced
in an OZI forbidden process, it has been argued that they are strong
candidates for gluonium states. These are discussed further in Sec.~\ref{secV:gT}.

\subsubsection{Summary of light Mesons}

For the most part the observed meson properties are 
consistent with predictions of the quark model.  The 
experimental regularities fit the quark model well, with orbital excitations
on linear trajectories and triplet splittings at least qualitatively
described by the quark model.   To make further progress in
the isovector and non-strange isoscalar sectors a high statistics 
would be useful in finding higher mass states.

Given the generally good agreement between the quark model and experiment
it is the discrepancies which suggest interesting physics. 
With the rather complete picture of the low mass $q\bar{q}$ states 
described above
it is becoming increasingly clear that several states 
have no obvious home in the $q\bar{q}$ sector.  
%One of the most important conclusions that can be drawn from
%examining Figures 4.2 through 4.5 is this:  there appear to be
%extra states that seem to lie outside of the quarkonium picture
%of meson states.  
For example the low mass $0^{++}$ systems have been confusing for many years
and it now appears that there are too many such states.
Figures  \ref{fig:strangeonium} and \ref{fig:is-mass}
shows that two radial excitations
of the isoscalar pseudoscalars should occur in the mass region 1300-1600 MeV.
The $\eta(1295)$, $\eta(1410)$ and the $\eta(1470)$ cannot all 
fit into this picture and therefore at least one must appear
as a spurious state.  Two ground-state
isoscalar $1^{++}$ states occur at 1240 and 1480 MeV in the
quark model and are filled in by $f_1(1285)$ and $f_1'(1530)$ so that
the $f_1(1420)$ clearly appears as an extra state.
Similar observations apply to the $f_0(1500)$, 
the $f_J(1710)$, and the $f_2(2010)$, $f_2(2300)$ and the
$f_2(2350)$.
These states point to a need for a better understanding of hadronic structure,
perhaps by studying the relation between the $q\bar{q}$ meson properties
and experimental observations
or perhaps by enlarging the quarkonium picture
to include gluonic degrees of freedom and multiquark states.

In addition, there are other puzzles which do not fit in the $q\bar{q}$
spectroscopy at all and have not yet been discussed: for example, 
the state with exotic $J^{PC}$ quantum numbers,
$1^{-+}$ $\hat{\rho}(1405)$ seen  in $\pi^-p \to \pi^0\eta$
and structures in $\gamma\gamma \to VV$.

In the next section we will examine the properties and possible
interpretations of these anomalous meson resonances.

% ======================================================================
%\include{rev-v6-5}
% ======================================================================
\newpage
\section{PUZZLES AND POSSIBILITIES}
\label{sec:PUZZLES}

The quark model compares
very well with the light quark meson spectrum and meson decays.
Certainly, there are disagreements, but many of these can be ascribed to
the natural limitations of the model when compared to the inherent
complexity of QCD.

However, there is another class of disagreements which cannot be so ascribed.
These cases are more profound, and presumably point
to the features of QCD {\em not} contained in the quark model, and
may indeed point to the fundamental degrees of freedom needed to
fully describe hadron structure. These may help us identify the
necessary features which will one day have to be met by any purported solution
of the full theory.

This class of disagreement is identified by one (or more) of three aspects.
The most unequivocal of these is the establishment of states with ``exotic''
quantum numbers, that is, $J^{PC}$ which cannot be accomodated with only
$q\bar{q}$ degrees of freedom.  The second aspect is the overpopulation of
states in well-defined mass regions.  This is slightly ambiguous because
the ``expected'' mass of a state will depend on the dynamics of the model,
and we can never be entirely certain that the overpopulation doesn't arise
from an ``intruder'' state based on higher radial or orbital excitations.
Finally, we can identify puzzling states by their specific dynamical
characteristics, such as their mass and partial decay widths.  This is most
difficult, since we need to rely on specific quark model calculations to
claim a fundamental disagreement, yet there are some cases where this
disagreement is indeed quite profound.

\subsection{Exotic Quantum Numbers}

As discussed in Sec.~\ref{sec:THEORY} (see Table~\ref{tb:qn}), the quantum
numbers of a $q\bar{q}$ system must have either $P=C$ for all $J$ except
$J=0$ or $J^{PC}=0^{-+}, 1^{+-}, 2^{-+},\ldots$.  Specifically,
a state with any of the quantum numbers $J^{PC}=0^{--}, 0^{+-}, 1^{-+},\ldots$
would be manifest evidence for non-$q\bar{q}$ degrees of freedom.
Because of their unique role, these states have been sought after for quite
some time.  However, only recently has some clear evidence been obtained
experimentally.  There are a number of reasons for this, mainly because their
branching ratio to conventional final states are small. This means that one
needs to search very carefully amidst the forest of well-established states,
or to build and operate a dedicated experiment to search in more complicated
final states, or both.

The flux tube model
(Isgur, Kokoski, and Paton, 1985b; Isgur, and Paton, 1985a; Close and Page, 1995a)
predicts that exotic hybrid mesons preferentially decay to pairs of
$S$ and $P$ wave mesons while decays to two $S$-wave mesons
are suppressed on rather general grounds (Page, 1997b).
Examples of preferred final states would include $\pi b_1(1235)$,
$\pi f_1(1285)$, $\pi a_2(1320)$, and $KK_1(1270)$.
Essentially all of the $S+P$ decay modes are very
difficult experimentally, since they involve a large number of final state
particles which can arise from different quasi-twobody states.  For example,
$\pi^-f_1(1285)$ leads to $\pi^-\pi^+\pi^-\eta$ which might also be due to
such things as $\pi^-\eta(1295)$, $\eta a_2^-$, $\eta a_1^-$, $\rho^0a_2^-$,
and so forth.  Unless the particular reaction enhances {\em production} of
hybrid exotics (Isgur, Kokoski, and Paton, 1985b) it will likely be very hard
to clearly identify the exotic state amidst the debris of highly excited,
conventional $q\bar{q}$ mesons (Barnes, Close, Page, and Swanson, 1997).

Experimentally, however, it is reasonable to first look at less complex final
states, and this is how the field has progressed so far.
For example, $\eta\pi$ and
$\eta^\prime\pi$ final states are attractive since any resonant odd-$L$ partial
wave would be manifestly exotic.  Alternatively, states with relatively few
particles, all charged, are also attractive since photon detection (which costs
money and leads to poorer resolution and complex acceptance) is not needed and
the PWA is less complicated.

We take a historical approach to reviewing the experimental evidence for
mesons with exotic quantum numbers.  We are just now starting
to see clear evidence of such phenomena, and it will take some time to
sort out the misleading evidence of the past.

\subsubsection{$\eta\pi$ final states and the $\hat{\rho}(1400)$}

Any state decaying to $\eta\pi$ must have $J=L$ and $P=(-1)^L$ (since both the
$\eta$ and $\pi$ are spinless) and $C=+$.  Therefore, any resonant, odd-$L$
partial wave is good evidence for a meson state with exotic quantum
numbers.\footnote{We tacitly assume that $C$ is a good quantum number, even
for charged final states in which case $C$ refers to the neutral member of a
multiplet.  In principle, a non-exotic meson could decay to $\eta\pi^\pm$ if
isospin symmetry is violated, but this should be a negligible effect for the
cases discussed here.}

This is obviously attractive experimentally, and a number of experiments
have acquired data, mainly in peripheral production with $\pi^-$ beams.
As it happens, the $\eta\pi$ mass spectrum in $\pi^-N\to(\eta\pi)X$ reactions
is dominated by the $a_2(1320)$ and this is both an advantage and disadvantage.
Obviously, one is relegated to picking out an exotic state from a spectrum
dominated by a conventional $q\bar{q}$ meson.  However, the $D$-wave $a_2$
provides a convenient wave against which an odd-$L$ wave would interfere.
In the $\eta\pi$ rest frame, this must lead to a forward-backward asymmetry
in the decay, relative to the direction of the incoming $\pi^-$.  Although
such an asymmetry unambiguously implies the presence of an odd-$L$ wave, it
can only be shown to be resonant using a complete partial wave analysis.

Experiments searching for exotic resonance structure in $\eta\pi$ systems
are listed in table~\ref{tabV:etapi}.  Every experiment listed sees a clear
forward/backward asymmetry, and therefore strongly implies the need for an
odd-$L$ partial wave.  The asymmetry changes quickly in the region of the
$a_2(1320)$ signal which dominates the mass spectrum, although details of
the asymmetry differ from experiment to experiment.
Each experiment except the first one listed (Apel, 1981) suggest or
claim outright evidence
for an exotic $J^{PC}=1^{-+}$ resonance, although the only two consistent
results are from VES (Beladidze, 1993) and
E852 (Thompson, 1997).  Finally, a very recent result from the
Crystal Barrel at CERN (Abele, 1998a) is consistent with these two
peripheral experiments, yielding a mass of around 1400~MeV/$c^2$ and a width
between 300 and 400~MeV/$c^2$.  This is the state we call $\hat{\rho}(1400)$.
\begin{table}
\caption{Peripheral hadronic production experiments with 
$\eta\pi$ final states.}
\label{tabV:etapi}
\begin{center}
\begin{tabular}{lllrl}
\hline
           &            &          & $p_{\rm Beam}$  & \\
Experiment & Laboratory & Reaction & (GeV/$c$)       & Reference\\
\hline
NICE   & IHEP   & $\pi^-p\to\eta\pi^0 n$ &  40 & Apel, {\it et al.}, 1981\\
GAMS   & CERN   & $\pi^-p\to\eta\pi^0 n$ & 100 & Alde, {\it et al.}, 1988a\\
BENKEI & KEK    & $\pi^-p\to\eta\pi^- p$ & 6.3 & Aoyagi, {\it et al.}, 1993\\
VES    & IHEP   & $\pi^-N\to\eta\pi^- X$ &  37 & Beladidze, {\it et al.}, 1993\\
E852   & BNL/AGS& $\pi^-p\to\eta\pi^- p$ &  18 & Thompson, {\it et al.}, 1997\\
\hline
\end{tabular}
\end{center}
\end{table}

Both the NICE and GAMS experiments (see table~\ref{tabV:etapi})
work with all neutral final states,
detecting only the four photons from $\eta\to\gamma\gamma$ and
$\pi^0\to\gamma\gamma$.  The recoil neutron is undetected, and in principle
the target itself could be excited, decaying as $N^\star\to n\pi^0$ where
the recoil $\pi^0$ decays to low energy photons at large angles and are
therefore undetected.  The experimenters perform a constrained fit on the
four observed photons, and only accept events with a reasonably high probability
for the exclusive final state.

This is not a straightforward procedure, and therein lies one of the most
crucial potential problems in these $\eta\pi$ experiments.  If the detector
acceptance is not well understood, then after applying this correction, one
may end up with a forward/backward asymmetry that is due to {\em instrumental}
effects and not physics.  This is particularly difficult in the case of
photon detection.  This is less of a problem for BENKEI, VES, and E852, since
they examine the $\eta\pi^-$ final state.  Furthermore, VES detects the $\eta$
via $\eta\to\pi^+\pi^-\pi^0$ decay, and E852 obtains results using both $\eta$
decay modes.

The first positive result was claimed by GAMS (Alde, 1988a) and this
generated much interest in its theoretical interpretation
(Close and Lipkin, 1987; Iddir, 1988a, 1988b; Tuan, Ferbel, and Dalitz, 1988).
From the start, however, it was clear that there were some internal
inconsistencies in the experiment (Tuan, Ferbel, and Dalitz, 1988).  To be
brief, (Alde, 1988a) saw that both the $a_2(1320)$ and $\hat{\rho}(1400)$
were produced mainly in the unnatural parity exchange waves, namely the
$D_0$ and $P_0$ waves respectively.  This would imply $a_1(1260)$ exchange, as
the $a_1$ is the lightest particle with the appropriate quantum numbers.
However, as the $a_2(1320)$ is known to decay predominantly to $\rho\pi$,
one would have expected production through natural parity $\rho^+$
exchange with a $\pi^-$ beam.  Consequently, their result was met with
a good deal of skepticism.  In fact, a reanalysis of their data by two of
their collaborators (Prokoshkin and Sadovsky, 1995a, 1995b)
obtains different conclusions.

The KEK collaboration (Aoyagi, 1993), which studied the
$\eta\pi^-$ system at a much lower
beam energy, analyzed their asymmetry to find the $a_2(1320)$ produced
predominantly in the natural parity $D_+$ wave, consistent with $\rho^+$
exchange.  A $P$-wave resonance was suggested based on the shape of the
intensity distributions, but it was seen with about equal strength in both
the $P_0$ and $P_+$ waves.  Furthermore, the mass and width of the ``exotic''
resonance were completely consistent with those of the $a_2(1320)$, and
the relative phases of the $P_+$ and $D_+$ waves did not vary across the
mass region of interest.
These strongly suggest that the $P_0$ and $P_+$ waves were the result of
``leakage'' from the much stronger $D_+$ wave, caused by an imperfect
knowledge of the experimental acceptance.

VES obtained a large statistics data sample of both $\eta\pi^-$ and
$\eta^\prime\pi^-$ final states, using topologies with three charged pions
and two photons (Beladidze, 1993).  The beam was incident on a
beryllium target, however, and
the recoils were not identified.  Nevertheless, a clean signal was observed
for $a_2(1320)$ production, strictly in the $D_+$
wave.  A weak exotic $P_+$ wave was also observed, close to the mass of the
$a_2(1320)$ but much broader, and with a significant phase motion relative
to the $P_+$.  No claims were made as to the existence of an exotic meson,
but the data were certainly suggestive.

\begin{figure}
\begin{minipage}{3.0in}
\centerline{\epsfig{file=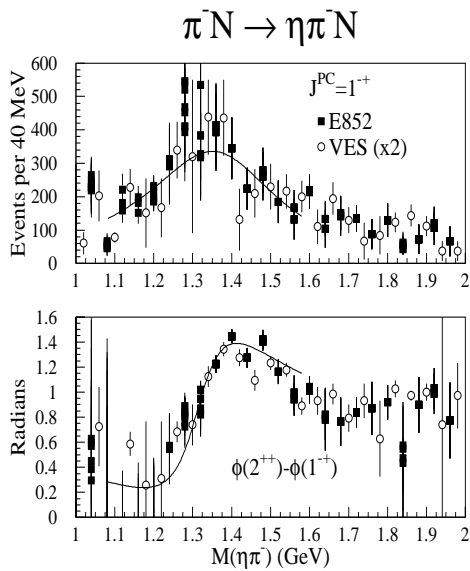,width=3.0in,height=3.5in}}
\end{minipage}
\begin{minipage}{3.4in}
\caption{A comparison of the exotic $J^{PC}=1^{-+}$ signal observed by
the E852 (Thompson, et al., 1997) and VES (Beladidze, et al., 1993)
experiments.  The E852 data explicitly shows the ambiguous solutions of the
partial wave analysis.
The VES intensity distribution is multiplied by a factor of two, and the
phase difference is offset by $9\pi/10$. 
The solid line is a simultaneous fit to the E852 data of Breit-Wigner forms
for the $a_2(1320)$ and $\hat{\rho}(1400)$, assuming appropriate
interference between the $D_+$ and $P_+$ partial waves.  The fit implies
a mass and width of $~1400$~MeV/$c^2$ and $~350$~MeV/$c^2$ respectively
for the $\hat{\rho}(1400)$.}
\label{figV:852ves}
\end{minipage}
\end{figure}

In E852, exclusivity of the final state was well established by tracking
the recoil proton, rejecting events with a $\pi^0$ associated with the
recoil, and requiring high probability with a constrained kinematic fit to
the full final state (Thompson, 1997).  Statistics were achieved that
were comparable to VES, and consistent $D_+$ and $P_+$ intensities and
phase motion were observed.  Figure~\ref{figV:852ves} compares the results
from the two experiments.  The E852 group fit the waves simultaneously to
two relativistic Breit-Wigner curves, and determine the mass and width of
the $\hat{\rho}$ to be approximately
$1370\pm50$~MeV/$c^2$ and $385\pm100$~MeV/$c^2$ respectively.  These fits are
shown in Fig.~\ref{figV:852ves}.

The recent Crystal Barrel result (Abele, 1998a) comes from $\bar{p}$
capture in liquid deuterium, the reaction being $\bar{p}d\to\eta\pi^-\pi^0p$
where the final state proton is a ``spectator''.  Resonances in either
$\eta\pi$ system recoil against the other pion, and $\pi^-\pi^0$ recoil against
the $\eta$.  The Dalitz plot is dominated by $\rho^-\to\pi^-\pi^0$, and there
is a clear $\eta\pi$ $P$-wave which interferes with it.  The Dalitz plot is
fit with a combination of resonances, and the $\chi^2$ is 1.29 per degree of
freedom when the $\hat{\rho}$ is included and 2.69 when it is removed.
They determine the mass and width of the $\hat{\rho}$ to be approximately
$1400\pm30$~MeV/$c^2$ and $310\pm70$~MeV/$c^2$ respectively, quite consistent
with the E852 result.

Curiously, the decay of an exotic $qqg$ hybrid meson to $\eta\pi$ 
is highly suppressed if one assumes
$SU(3)$ flavor symmetry and nonrelativistic degrees of freedom
(Close and Lipkin, 1987; 
Iddir, 1988a, 1998b; Tuan, Ferbel, and Dalitz, 1988; Page, 1997a).
Furthermore, the contribution expected from breaking these assumptions should
be small (Page, 1997a). If the $\hat{\rho}(1400)$ is confirmed as an
exotic meson, there is therefore the theoretical prejudice 
that it may be a $q\bar{q}q\bar{q}$ state.
A different suggestion (Donnachie and Page, 1998) is that the $\hat{\rho}(1400)$
is in fact an artifact of the $\hat{\rho}(1600)$ (see the next section)
which manifests itself through the $\rho\pi$ and $b_1(1235)\pi$ thresholds.

\subsubsection{The $\hat{\rho}(1600)$ in $\rho\pi$ and $\eta^\prime\pi$}

The E852 collaboration has recently put forth evidence for another
$J^{PC}=1^{-+}$ exotic meson (Adams, 1998), decaying in this case
to $\rho\pi$, in the reaction $\pi^-p\to\pi^-\pi^+\pi^-p$ at 18~GeV/$c$.
This state, the $\hat{\rho}(1600)$, has a mass and width of
$1593\pm8$~MeV/$c^2$ and $168\pm20$~MeV/$c^2$ respectively.  Tentative
evidence of the $\hat{\rho}(1600)$ was put forward by the VES collaboration
(Gouz, 1993), who also see evidence for a broad but resonant $P_+$ wave
near this mass in the $\eta^\prime\pi^-$ system (Beladidze, 1993).
Although the $\hat{\rho}(1600)$ data does not agree particularly well 
with recent lattice QCD results and 
phenomenological expectations for a hybrid meson it appears that it 
can still be accomodated as a hybrid meson (Page, 1997c).

The $\pi^-\pi^+\pi^-$ mass spectrum is dominated by the $a_1(1260)$, the
$a_2(1320)$, and the $\pi_2(1670)$.  (See Fig.~\ref{figIII:e852_3pi}.)
Each of these states is dominated by natural parity exchange and decay to
$\rho^0\pi^-$ or $f_2(1270)\pi^-$.  The $\hat{\rho}(1600)$ is seen clearly
in natural parity exchange as well, decaying to $\rho^0\pi^-$, and its
interference with the dominant waves shows clear phase motion, consistent
with resonant behavior, in all cases.  The peak intensity of the
$\hat{\rho}(1600)$ is about 5\%
of the nearby $\pi_2(1670)$.  There is
also indication of a peak in intensity in the unnatural parity waves,
but there is not enough intensity in other unnatural parity waves to
check the phase motion.

One important feature of the E852 result is that ``leakage'' has been checked
explicitly.  A data set was simulated, using the waves used
in the fit procedure but excluding the small $\hat{\rho}(1600)$.  Then, a full
partial wave analysis was performed, including the $1^{-+}$ waves, to see if
finite resolution and acceptance effects would lead to a spurious signal.
Indeed, there is a significant leakage of the large $a_1(1260)$ signal into
the $1^{-+}$ channel in the 1200-1300~MeV/$c^2$ region (Adams, et al, 1998),
but the region near 1600~MeV/$c^2$ is clean.

Simultaneous with their measurement of the $\eta\pi^-$ final state, the
VES collaboration (Beladidze, 1993) also studied the $\eta^\prime\pi^-$
system in the reaction $\pi^-N\to\eta^\prime\pi^-N$ with
$\eta^\prime\to\eta\pi^+\pi^-$.  The $\eta^\prime\pi^-$ mass spectrum shows
a clear peak at the $a_2(1320)$ (despite the limited phase space)
superimposed on a broad distribution peaked at $\sim1.6$~GeV/$c^2$.  In a
situation rather similar to the $\eta\pi^-$ system, the PWA finds the reaction
almost completely dominated by the natural parity exchange $P_+$ and $D_+$
waves.  The peak in the mass distribution near the $a_2(1320)$ is completely
absorbed into the $D_+$ wave, and a Breit-Wigner parameterization yields the
correct values for the mass and width.  (This analysis also yields the relative
branching ratio for $a_2\to\eta^\prime\pi$ to $a_2\to\eta\pi$ to be $\sim5\%$.)
The relative phase motion of the $P_+$ and $D_+$ waves also supports the
existence of the $\hat{\rho}(1400)$ seen in $\eta\pi^-$ and the matrix elements
they extract are quite comparable in this mass region.  The matrix elements,
however, are considerably stronger for $\eta^\prime\pi^-$ than for $\eta\pi^-$
over most of the accepted $\eta^\prime\pi$ masses, particularly in the
$\sim1.6$~GeV/$c^2$ region.  However, the structure appears to be quite broad
and the phase motion is not distinctive so it is difficult to associate this
with the $\hat{\rho}(1600)$ seen in $\rho\pi$ by E852 (Adams, 1998).

The E852 data indicates the $BR(\hat{\rho}(1600)\to\rho\pi)\approx20\%$
(Page, 1997c), while there is no indication of this state in the
$\eta\pi^-$ data (Beladidze, 1993; Thompson, 1997), and clearly
not all of the $\eta^\prime\pi^-$ signal (Beladidze, 1993) is
consistent with the $\hat{\rho}(1600)$.  It is therefore likely that its
dominant decay mode has yet to be observed, and (Page, 1997c) suggests that
this might be $b_1(1235)\pi$ or $f_1(1285)\pi$ since $b_1$ or $f_1$ exchange
would account for the unnatural parity exchange signal observed by E852.

\subsubsection{Searches for $S+P$ decays}

As noted previously, rather general symmetrization selection rules
(Page, 1997b) argue that hybrid mesons, exotic or otherwise, should not decay
predominantly to a pair of $S$-wave mesons.  Consequently, the cases discussed
in the previous two sections, namely $\eta\pi$, $\eta^\prime\pi$, and $\rho\pi$,
should not represent the dominant decay modes of the
$\hat{\rho}(1400)$ or $\hat{\rho}(1600)$, if those states are indeed hybrid,
exotic mesons.

Decays to $S+P$ pairs of mesons are rather
complex, however, and until now only a very limited number of experiments
have attempted to search for exotic mesons in these signatures.  In response
to the suggestion of (Isgur, Kokoski, and Paton, 1985b) that exotic hybrids
decaying to $b_1(1235)\pi$ should be produced with good signal to noise in
peripheral photoproduction, the $\Omega$ spectrometer group at CERN reexamined
their previous data (Atkinson, 1983) on the reaction
$\gamma p\to\omega\pi^+\pi^-p$.  This group showed that production of
$b_1(1235)\to\omega\pi$ was enhanced when
$1.6\leq{\rm Mass}(\omega\pi\pi)\leq2.0$~GeV/$c^2$.  Although the number of
events was limited, their reanalysis (Atkinson, 1987) showed what
appeared to be production of $\omega_3(1670)$ and a new state at
$\sim1.9$~GeV/$c^2$, decaying to $b_1(1235)\pi$.  This was based strictly on
the invariant mass distribution; no partial wave or angular distribution
analyses were demonstrated.

One would prefer to search for $S+P$ wave decays where the
$P$-wave meson is not too broad, and which decays mainly to experimentally
accessible final states.  Good candidates include
$b_1(1235)\to\omega\pi$ (with total width $\Gamma=142$~MeV/$c^2$),
$a_2(1320)\to\rho\pi$ ($\Gamma=107$~MeV/$c^2$), and
$f_1(1285)\to a_0(980)\pi$ ($\Gamma=25$~MeV/$c^2$).

At this time, results
have only been presented for $f_1(1285)\pi$ final state.  These are from
the VES collaboration (Gouz, 1993), who made use of the
$f_1\to\eta\pi^+\pi^-$ final state,
and BNL Experiment E818 (Lee, 1994) for which $f_1\to K^+\bar{K}_0\pi^-$.
In both cases, there are not very many events, and the VES result remains
preliminary.  However, E818 show that the intensity distribution and phase
motion of the $J^{PC}=1^{-+}$ signal are suggestive of the presence of an
exotic meson with mass around 1.9~GeV/$c^2$ and a $J^{PC}=1^{++}$ resonance
near 1.7~GeV/$c^2$.  As stated in (Lee, 1994), additional data are
required for a more complete understanding of the $J^{PC}=1^{-+}$ wave.

There have been two attempts to search for $S+P$ final states in high
energy photoproduction, since the original CERN data (Atkinson, 1987).
One of these, experiment E687 at Fermilab, performed an inclusive search
for $f_1(1285)\pi^\pm$ states (Danyo, 1995) in the reaction
$\gamma {\rm Be}\to f_1(1285)\pi^\pm X$ with
$f_1(1285)\to a_0(980)^\pm\pi^\mp$ and $a_0(980)^\pm\to K_SK^\pm$. A tagged
bremsstrahlung beamline provided photons with energies near $200$~GeV.  A peak
in the $f_1\pi$ mass distribution is clear above background. The
background was measured using combinations of $f_1$ and $\pi$ taken from
different events.  The fit yields a mass and width of $1748\pm12$~MeV/$c^2$
and $136\pm30$~MeV/$c^2$ respectively, not inconsistent with that suggested
by E818.  The angular distribution in the helicity frame is not inconsistent
with $J^{PC}=1^{-+}$, although statistics are quite poor.

The only other photoproduction experiment is a reanalysis by the
Tennessee group (Blackett, 1997) of data taken with the SLAC
Hybrid Photoproduction experiment (Abe, 1984).  Data was collected
for the reaction $\gamma p\to p\omega\pi^+\pi^-$ in a search for states
decaying to $b_1^\pm(1235)\pi^\mp$ with $b_1^\pm(1235)\to\omega\pi^\pm$.
However, most of the sample was in fact more consistent with the charge
exchange reaction $\gamma p\to\Delta^{++}b_1^-(1235)$ and no significant
structure was observed in the $b_1^\pm(1235)\pi^\mp$ mass spectrum when these
events were removed.

\subsection{The scalar mesons}

The scalar ($J^P=0^+$) mesons have long been a source of controversy
(Morgan, 1974).
As discussed in Sections~\ref{secIV:ss}, \ref{secIV:iv}, and \ref{secIV:is},
it is difficult to compare observed states to the theoretical predictions in
this sector.
For masses below 2~GeV/$c^2$ the two-body $S$-wave structure
is very complicated, and overlapping states interfere with each other
differently in different production and decay channels.  Interpreted in terms
of resonances, the quark model is clearly oversubscribed, and only very
recently has a semblance of order emerged.  This subject has already been 
touched on in Section~\ref{sec:COMPARISON} 
and a brief techical review 
is given by the Particle Data Group (Caso, 1998).

We will take a different approach here, and focus directly on the observed
states where the $J^{PC}=0^{++}$ assignment is well established. We note
that this discussion is linked to the $f_J(1710)$ and $f_J(2220)$, which are
discussed separately in Section~\ref{secV:fJ}.

The present bias is that the
quark model nonet is tentatively filled, with only the predominant $s\bar{s}$ state
in need of confirmation, and there are two manifestations of degrees of
freedom beyond the quark model.  These are the $f_0(1500)$, suggested as being
dominated by gluonic degrees of freedom (Amsler and Close, 1996), and the
$a_0(980)$ and $f_0(980)$ which are candidates for multi-quark states
(Jaffe, 1977a; Weinstein and Isgur, 1983; Rosner, 1983).  Still, there is
significant controversy (Janssen, 1995; T\"{o}rnqvist and Roos, 1996).

The strange members of the scalar nonet are the $K_0^\star(1430)$ states,
clearly established by the LASS collaboration (Aston, 1988e).
The $I=1$ member is most likely the $a_0(1450)$, observed (Amsler, et al, 1994a)
by the Crystal Barrel collaboration, who also measured its decay branches to
$\eta^\prime\pi$ (Abele, 1997a) and to
$K\bar{K}$ (Abele, 1998b).  The broad, isoscalar $\pi\pi$ $S$-wave
has within it three clear resonances, namely the $f_0(980)$, $f_0(1370)$,
and $f_0(1500)$. We focus our discussion on the isoscalar
states near $1500$~MeV/$c^2$, mainly the $f_0(1500)$, and on the enigmatic
$a_0(980)$ and $f_0(980)$ mesons.  See also (Amsler, 1998).

\subsubsection{The $f_0(1500)$}

The $f_0(1500)$ has been touted as a likely candidate for a ``glueball''
(Amsler and Close, 1995, 1996). The evidence
is circumstantial, based on its peculiar width and decay properties.  This
suggestion is supported by various theoretical calculations
(see Section~\ref{secII:glueballs} as well as Szczepaniak, 1996)
which indicate that the lightest glueball
should have $J^{PC}=0^{++}$ and a mass near 1500~MeV/$c^2$.
The scalar nonet is {\em not} overpopulated (discounting the $a_0(980)$
and $f_0(980)$) until the $s\bar{s}$ state is unambiguously identified.

Most of the data on the $f_0(1500)$ is from the Crystal Barrel collaboration,
who resolved two
scalar states in this mass region (Anisovich, 1994),
and who determine its decay branches to a number of final states, including
$\pi^0\pi^0$ and $\eta\eta$ (Amsler, 1995; Abele, 1996c),
$\eta\eta^\prime$ (Amsler, 1994b),
$K_LK_L$ (Abele, 1996b) and $4\pi^0$ (Abele, 1996a), all using
$\bar{p}p$ annihilation at rest in liquid hydrogen.
The $f_0(1500)$ has also been observed by the OBELIX experiment
(Bertin, 1998) in the $\bar{n}p\to\pi^+\pi^+\pi^-$ reaction for
neutrons in flight, and also in ``glue rich'' central production reactions
$pp\to p_f(\pi^+\pi^-\pi^+\pi^-)p_s$ and $pp\to p_f(\pi^+\pi^-)p_s$, most
recently by the WA91 (Antinori, 1995) and
WA102 (Barberis, 1997b) collaborations.  There is some evidence of
$J/\psi\to\gamma f_0(1500)$ with $f_0(1500)\to\pi^+\pi^-\pi^+\pi^-$
(Bugg, 1995)
and perhaps also with $f_0(1500)\to\pi^+\pi^-$ (Dunwoodie, 1997),
but $S$-wave structure in the $K^+K^-$ and $K_SK_S$ final states seems absent
in this mass region for $J/\psi$ radiative decay
(Bai, 1996a; Dunwoodie, 1997).
For a collective discussion of these data and the evidence that the $f_0(1500)$
is a glueball, see
(Amsler and Close, 1995;
 Amsler and Close, 1996;
 Close, Farrar, and Li, 1997c;
 Close, 1997a;
 Amsler, 1998).
No evidence has been reported in $\gamma\gamma$ collisions 
(as would be expected to
be the case if the $f_0(1500)$ is predominantly glue), but the $\pi\pi$ and
$K\bar{K}$ mass distributions are dominated by the $f_2(1270)$ and
$f_2^\prime(1525)$ in this mass region (Morgan, Pennington, and Whalley, 1994)
making a search for scalar structure quite difficult.

The mass and width of the $f_0(1500)$ are given (Caso, 1998) as
$1500\pm10$~MeV/$c^2$ and $112\pm10$~MeV/$c^2$ respectively.  The mass is
rather close to the nominal $I=1$ and $I=1/2$ members of the scalar nonet,
i.e. the $a_0(1450)$ and the $K_0^\star(1430)$.  However, it is decidedly
more narrow, with
$\Gamma[a_0(1450)]=265\pm13$~MeV/$c^2$ and
$\Gamma[K_0^\star(1450)]=287\pm20$~MeV/$c^2$, which are more in line with the
very broad $f_0(1370)$.  This comparison, and the penchant of the $f_0(1500)$
to be produced in ``glue-rich'' environments, hints to its special status as
something different that a standard $q\bar{q}$ meson.

Of course, if we take the $f_0(980)$ and $a_0(980)$ to be multiquark states,
then there should be {\em two} isoscalar $0^{++}$ mesons with mass typical of
of the $0^+$ nonet, and it is tempting to take these to be the $f_0(1370)$ and
$f_0(1500)$.  This is clearly not workable, however, even disregarding the
narrow width of the $f_0(1500)$, since neither can convincingly be associated
with the predominantly $s\bar{s}$ member.  This information is based on the
decay patterns of the $f_0(1370)$ and $f_0(1500)$.

It is difficult to assign specific branching ratios to the $f_0(1370)$ because
it is not only broad, but it also interferes strongly with the even broader
underlying $S$-wave $\pi\pi$ structure as well as with the $f_0(980)$
(Caso, 1998; Amsler, 1998).  It is nevertheless clear, however, that
this state couples mainly to pions and $K\bar{K}$ is suppressed (Amsler, 1998).
Given that its width is in line with the $a_0(1450)$ and $K_0^\star(1430)$,
we associate it with the $n\bar{n}\equiv[u\bar{u}+d\bar{d}]/\sqrt{2}$ member of
the nonet.

It is possible to derive branching ratios for the $f_0(1500)$ (Amsler, 1998).
From a coupled channel analysis (Amsler, 1995) of $\bar{p}$ annihilation
at rest in liquid hydrogen to $\pi^0\pi^0\pi^0$, $\eta\pi^0\pi^0$, and
$\eta\eta\pi^0$ final states, the Crystal Barrel collaboration determine the
following branching fraction for production and decay of $f_0(1500)$:
\begin{center}
\begin{tabular}{lcl}
$B[\bar{p}p\to f_0(1500), f_0(1500)\to\pi^0\pi^0$] & $=$ &
                          $(12.7\pm3.3)\times 10^{-4}$\\
$B[\bar{p}p\to f_0(1500), f_0(1500)\to\eta\eta$] & $=$ &
                          $(6.0\pm1.7)\times 10^{-4}$
\end{tabular}\\[0.25in]
\end{center}
An analysis (Amsler, 1994b) by the Crystal Barrel of the reaction
$\bar{p}p\to\pi^0\eta\eta^\prime$ gives
$$B[\bar{p}p\to f_0(1500), f_0(1500)\to\eta\eta^\prime]
   =(1.6\pm0.4)\times 10^{-4}$$
They also determine (Amsler, 1998; Table~11)
$$B[\bar{p}p\to f_0(1500), f_0(1500)\to K_L K_L]=(1.13\pm0.09)\times 10^{-4}$$
from data (Abele, 1996b) on the reaction
$\bar{p}p\to\pi^0K_LK_L$, using data from the reaction
$\bar{p}p\to K_LK^\pm\pi^\mp$ (Abele, 1996a)
to fix the contribution from $a_0(1450)\to K_LK_L$.

A simple phenomenological model makes it possible to derive couplings of the
$f_0(1500)$ to various pairs of pseudoscalar mesons (Amsler and Close, 1996).
The model incorporates $SU(3)$ flavor symmetry breaking and meson form factors,
and is tested on decays of the well-understood $2^{++}$ nonet.  Following this
model and incorporating two-body phase space, we arrive at the following
relative decay rates for $f_0(1500)$ into two-body, pseudoscalar meson pairs:
\begin{equation}
\pi\pi:KK:\eta\eta:\eta\eta^\prime=
  (5.1\pm2.0):(0.71\pm0.21):(\equiv1.0):(1.3\pm0.5)
\label{eqV:f0coup}
\end{equation}
where the value for $\pi\pi$ ($KK$) multiplies the branching ratio for
$\pi^0\pi^0$ ($K_LK_L$) by 3 (4) to account for charge combinations.
These are obviously {\em inconsistent} with the $f_0(1500)$ being the $s\bar{s}$
member of the nonet.  In fact, detailed considerations of isoscalar mixings
(Amsler and Close, 1996; Amsler, 1998) show that they are consistent with the
$f_0(1500)$ being the $n\bar{n}$ member, although we've already argued that
the $f_0(1370)$ is a better candidate.

Therefore, the circumstantial evidence for $f_0(1500)$ being the scalar
glueball is clear.  It is produced primarily in glue-rich environments;
its mass and width are consistent with theoretical predictions;
and it overpopulates the $q\bar{q}$ states in this mass region if we
assume it is not the $s\bar{s}$ state.
Direct evidence, however, is lacking.  If the 
$f_0(1500)$ were a pure glueball,
one would naively expect ``flavor blind'' decays to all available $SU(3)_f$
singlets, and as discussed in Section~\ref{secII:glueballs}
the couplings (\ref{eqV:f0coup}) would be $3:4:1:0$ (ignoring
single/octet mixing in the $\eta$ and $\eta^\prime$).  Most notable here is
the strongly suppressed $K\bar{K}$ coupling relative to $\pi\pi$ whereas the
naive prediction is that it should be comparable if not larger.

Amsler and Close, 1996, argue that this problem is linked to the other
outstanding problem in the isoscalar $0^+$ nonet, namely the missing $s\bar{s}$
state, i.e. the $f_0^\prime(\sim1600)$.  Using first order perturbation theory,
one finds that $f_0(1500)\to K\bar{K}$ can be strongly suppressed by mixing
between $f_0(1370)$, $f_0(1500)$, and the hypothetical $f_0^\prime(\sim1600)$.
In fact, they find that 
if pure glue is indeed flavor blind with respect to $s\bar{s}$ and
$n\bar{n}$, then $f_0(1500)\to K\bar{K}$ goes to zero if the $f_0(1500)$ lies
exactly between the other two states in mass.  Turning this analysis around,
the $f_0(1500)\to K\bar{K}$ branch above infers two possible values for the
mass of the $f_0^\prime$, namely 1600 or 1900~MeV/$c^2$ (Amsler, 1998).

{\em It is essential to confirm the $f_0^\prime$} and to measure its decay
properties in order to clearly establish the $f_0(1500)$ as the scalar glueball.
See Section~\ref{secIV:ss}.
There are two candidates at present.  One possibility is the tentative
identification (Aston, 1988a) of an $S$-wave resonance, produced and
decaying through $K\bar{K}$ in the reaction $K^-p\to K^0_SK^0_S\Lambda$,
directly underneath the dominant $f_2^\prime(1525)$.  Not only is this a weak
observation, however, it may in fact be an observation of the $f_0(1500)$
itself.  Another candidate is the $f_J(1710)$ (Sec.~\ref{secV:fJ}). Although
its spin assignment is somewhat uncertain and controversial, it shows some
characteristics of being a glue-dominated state itself.

\subsubsection{The $a_0(980)$ and $f_0(980)$}

These states have been known for a very long time (Morgan, 1974) but their
nature continues to generate controversy (Janssen, 1995).
With the establishment of the
$a_0(1450)$, $f_0(1370)$, and $f_0(1500)$ (Amsler, 1998), it is no longer
feasible to argue that the $a_0(980)$ and $f_0(980)$ are members of the
$q\bar{q}$ scalar nonet.  Their near degeneracy in mass, as well as their
proximity to the $K\bar{K}$ threshold and their propensity to decay to
$K\bar{K}$, strongly suggest they are $I=1$ and $I=0$ bound states of
$K\bar{K}$ (Weinstein and Isgur, 1983).

Because of their very peculiar decay properties, it is difficult to quantify
even the mass and width of these states (Caso, 1998).  For example,
the states are somewhere between 50 and 100~MeV/$c^2$ wide, and so their
nominal mass allows the width to straddle the $K\bar{K}$ threshold at
990~MeV/$c^2$.  The non-$K\bar{K}$ decays are fully dominated by
$f_0(980)\to\pi\pi$ and $a_0(980)\to\eta\pi$ which are not significantly suppressed
by their own kinematic thresholds.  A recent measurement by the E852
collaboration (Teige, 1998) gives the charged $a_0(980)$ mass and
width as $995.8\pm1.6$~MeV/$c^2$ and $62\pm6$~MeV/$c^2$ respectively, when
the $\eta\pi^\pm$ final state is fit to a relativistic Breit-Wigner, and
$1001.3\pm1.9$~MeV/$c^2$ and $70\pm5$~MeV/$c^2$ based on the coupled-channel
description developed by Flatt\'{e} (Flatt\'{e}, 1976).

The $\gamma\gamma$ decay widths (Caso, 1998) have been measured in
photon-photon collisions (Morgan, Pennington, and Whalley, 1994), but
theoretical estimates vary widely and it is difficult to make a definitive
statement (Barnes, 1985b; Antreasyan, 1986).

A novel measurement to elucidate the nature of these states was suggested by
(Close, Isgur, and Kumano, 1993.)  By determining the radiative decay rate
$\phi\to a_0(980)\gamma$ or $\phi\to f_0(980)\gamma$, one could infer the
$s\bar{s}$ content of the $a_0$ or $f_0$ wave function since the rate is
proportional to the overlap with the $\phi$, a well-known $s\bar{s}$ state.
They calculate that
$BR(\phi\to a_0\gamma)\approx BR(\phi\to f_0\gamma)\approx4\times10^{-5}$
if the $a_0$ and $f_0$ are indeed $K\bar{K}$ molecules, whereas the branching
ratio should be $10^{-6}$ or less for $q\bar{q}$ or other multiquark
configurations.
Very recent results from the SND collaboration running at the VEPP-2M storage
ring in Novosibirsk (Achasov, 1997a; Aulchenko, 1998) yield a
value $BR(\phi\to f_0\gamma)=(3.42\pm0.30\pm0.36)\times10^{-4}$,
much larger than expected for a $K\bar{K}$ molecule.  They also report that
$BR(\phi\to\eta\pi^0\gamma)\approx(1.3\pm0.5)\times10^{-4}$ but with no
suggestion of a peak at the $a_0(980)$.
A new experiment using $\phi$ photoproduction on hydrogen (Dzierba, 1994)
will take data on $\phi\to\gamma X$ in 1999.
It will take some time to sort out all
the new information, but it is reasonable to expect that important new
results will be available soon.

\subsection{Other possible glueballs: The $f_J(1710)$ and the $f_J(2220)$}

Focus on the $f_0(1500)$ as the lightest glueball comes partly because of the
wealth of information provided on this state by the Crystal Barrel experiment.
A number of branching ratios have been measured, many including complicated,
all neutral final states, and this has made detailed analyses possible.  These
analyses have lead to clear inconsistencies with what is expected from the
scalar $q\bar{q}$ nonet, and that has fueled the conjecture.

It is worth noting, however, that the search for glueballs began much
earlier, using $J/\psi$ radiative decays.  As discussed in
Sec.~\ref{secIII:Jpsi}, the annihilation of a $c\bar{c}$ pair into a photon
and two gluons is certainly expected in lowest order QCD, so one is naturally
lead to search for gluonic states in these decays.
This was discussed in some detail recently by
(Close, Farrar, and Li, 1997c; Page and Li, 1998).
Further information is provided by the coupling, or its upper limit, of the
candidate state to $\gamma\gamma$ (Sec.~\ref{secIII:gg}).  Pure glue can only
couple to photons through the creation of an intermidiate $q\bar{q}$ pair and
is therefore suppressed relative to quark model states.
On the other hand, central production in $pp$ collisions
(Sec.~\ref{secIII:central}) may be a rich source of gluonic states, based on
speculation that it proceeds through double Pomeron exchange
(Close, 1997a, Close and Kirk, 1997b).

\begin{figure}
\centerline{\epsfig{file=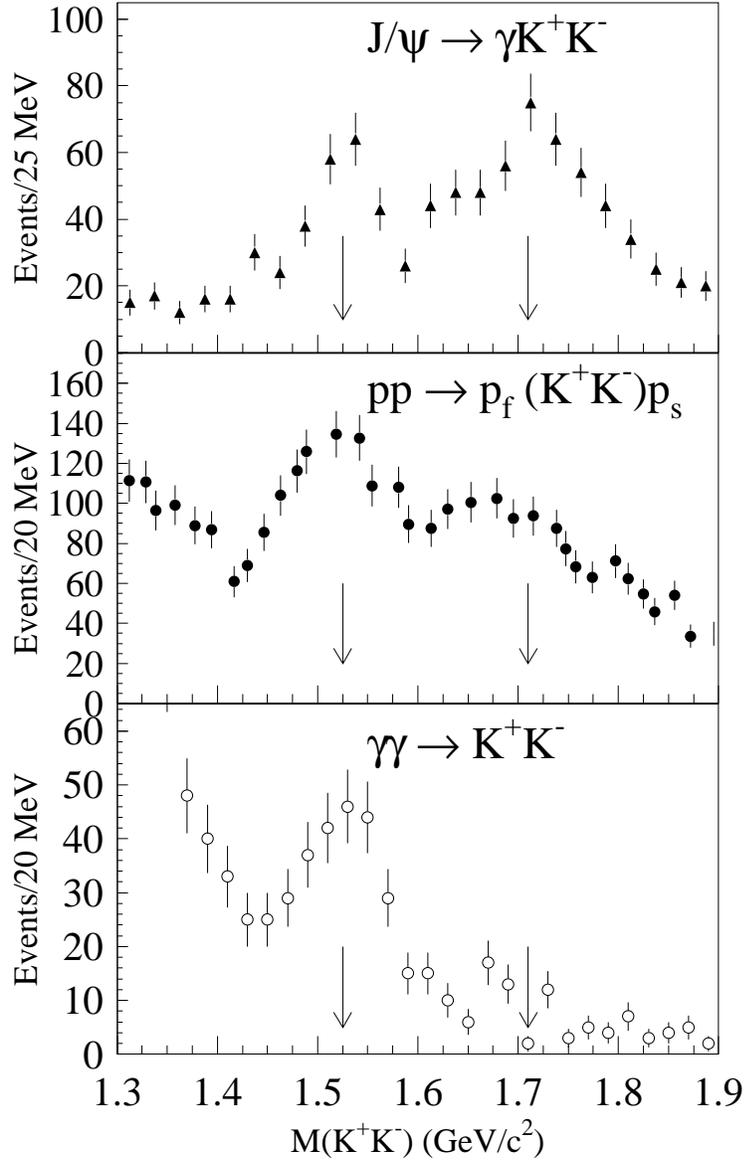}}
\caption{Possible glueball sensitivity in different reactions, for states
decaying to $K^+K^-$.  Shown are the $K^+K^-$ invariant mass distributions for
$J/\psi\to\gamma K^+K^-$ (Bai, et al., 1996b),
$pp\to p_f(K^+K^-)p_s$ (Armstrong, et al., 1991a), and
$\gamma\gamma\to K^+K^-$ (Albrecht, et al., 1990).
The arrows mark the positions of the $f_2^\prime(1525)$ and the
$f_J(1710)$.}
\label{figV:glueKK}
\end{figure}

These processes are shown in Fig.~\ref{figV:glueKK} for excited mesons decaying
to $K^+K^-$.
The figure shows the invariant mass of $K^+K^-$ pairs produced in
radiative $J/\psi$ decay, $J/\psi\to\gamma K^+K^-$ (Bai, 1996b);
central $pp$ collisions, $pp\to p_f(K^+K^-)p_s$ (Armstrong, 1991a); and
two-photon collisions, $\gamma\gamma\to K^+K^-$ (Albrecht, 1990).
For both radiative $J/\psi$ decay and for central $pp$ collisions, two
enhancements are clear, one near 1500~MeV/$c^2$ and the second near
1700~MeV/$c^2$.  The 1500~MeV/$c^2$ structure is consistently found to be
dominated by $J=2$ and is identified as the $f_2^\prime(1525)$, the mainly
$s\bar{s}$ member of the tensor nonet.  The 1700~MeV/$c^2$ structure contains
the $f_J(1710)$.  Note that this second structure is {\em not} seen in
two-photon collisions, and this has fueled speculation that the $f_J(1710)$
is a glueball.  The relative ratios of the abundances of $f_2^\prime(1525)$
and $f_J(1710)$ for radiative $J/\psi$ decay and two-photon collisions, can
be used to evaluate the ``stickiness'' (Chanowitz, 1984) of these two
enhancements.  
Recall from Section~\ref{secII:glueballs} (Eqn.~\ref{eq:sticky}) that 
this quantity is proportional to the ratio of squared matrix
elements for coupling of the state to $gg$ and $\gamma\gamma$. 
For a state of pure glue, there would be no coupling to
photons, and the state would have infinite stickiness.

Both $J/\psi\to\gamma X$ and $\gamma\gamma\to X$ can only produce states $X$
with $C=+1$.  Furthermore, in this section, our discussion is mainly limited
to decays to pairs of identical psuedoscalar particles, so $P=+1$ and the
total spin $J$ must be even.  There is in fact considerable controversy
regarding the total spin $J$ of the $f_J(1710)$ and $f_J(2220)$, hence the
indeterminate nomenclature.

\subsubsection{The $f_J(1710)$}
\label{secV:fJ}

The $f_J(1710)$ is the main competitor of the $f_0(1500)$ for status as the
lightest glueball, assuming that $J=0$.  Our best estimates for glueball
properties are from lattice gauge theory calculations, and although they all
agree that the lightest glueball should have $J^{PC}=0^{++}$, there is some
disagreement on the mass.  For example, two comprehensive studies find
$M(0^{++})=1550\pm50$~MeV/$c^2$ and $M(2^{++})=2270\pm100$~MeV/$c^2$
(Bali, 1993) and
$M(0^{++})=1740\pm71$~MeV/$c^2$ and $M(2^{++})=2359\pm128$~MeV/$c^2$
(Chen, 1994) for the lowest mass scalar and tensor glueballs.
In fact, one of these groups compares the measured properties of the
$f_J(1710)$ to their calculations, and directly argue that it must be the
lightest scalar glueball (Sexton, Vaccarino, and Weingarten, 1995a).
(See Sec.~\ref{secII:glueballs} for more details.)
In all cases, however, the tensor mass remains in
the region near 2.2~GeV/$c^2$.

It is also important to recall that in order to accomodate a scalar glueball
anywhere in the 1.5-1.7~GeV/$c^2$ region, one needs to identify the $s\bar{s}$
partner to the $n\bar{n}$ $f_0(1370)$.  If $J=0$, then the $f_J(1710)$ and
$f_0(1500)$ might well represent the glueball and the $s\bar{s}$ state, or more
likely each is a mixture of both.  
Recenty, the IBM group has computed mixing with quarkonia
(Weingarten, 1997; Lee and Weingarten, 1998a; Lee and Weingarten, 1998b)
and again claim good agreement with the $f_J(1710)$ as mainly the $0^{++}$
glueball, while establishing
that the $f_0(1500)$ is a good candidate for the mainly $s\bar{s}$ member
of the nonet. If $J=2$, however, it will be difficult to
assign a glueball status to the $f_J(1710)$ since that would be at odds with
all current lattice gauge calculations.

The Particle Data Group (Caso, 1998) estimates the mass and width of
the $f_J(1710)$ to be $1712\pm5$~MeV/$c^2$ and $133\pm14$~MeV/$c^2$,
respectively.
The differing experimental results for the spin of this state are clearly
intertwined with determining its other properties.  We will therefore go
through the experimental evidence for the $f_J(1710)$ and point out the various
important agreements and disagreements.  Controversy still remains, and there
is some suggestion that the $f_J(1710)$ is actually more than one state.

\paragraph{Radiative $J/\psi$ decay.}
This state was first observed by the Crystal Ball collaboration, in radiative
$J/\psi$ decay (Edwards, 1982b).  It was immediately recognized as a
glueball candidate. Called $\theta(1640)$, it was seen as
a peak in the $\eta\eta$ mass distribution of $39\pm11$ events over background.
The width was large ($\sim220$~MeV/$c^2$) and the $\theta\to\eta\eta$ angular
distribution favored $J=2$, however the analysis did not include the presence
of the $f_2^\prime(1525)$ which decays $\sim10\%$ of the time to $\eta\eta$.
Soon afterwards, however, a consistent peak was observed in $K^+K^-$ mass by
the Mark~II collaboration (Franklin, 1982) in the reaction
$J/\psi\to\gamma K^+K^-$ and this analysis did include the $f_2^\prime(1525)$,
again slightly favoring $J=2$.  The simultaneous observation of a state decaying
both to $\eta\eta$ and $K\bar{K}$ fueled speculation that this was a glueball.
A reanalysis of the Crystal Ball data which included the $f_2^\prime(1525)$
(Bloom and Peck, 1983; K\"{onigsman}, 1986) led to a larger mass,
near 1700~MeV/$c^2$, and a smaller width, but $J=2$ was still preferred.

The Mark~III collaboration followed up with higher statistics measurements of
the charged particle decays $\pi^+\pi^-$ and $K^+K^-$ in the reactions
$J/\psi\to\gamma K^+K^-$ and $J/\psi\to\gamma\pi^+\pi^-$
(Baltrusaitis, 1987).  The $f_J(1710)$ was observed in both modes.
The $K^+K^-$ mode was particularly clean, apparently obstructed only slightly
by the nearby $f_2^\prime(1525)$, and once again the angular distribution
preferred $J=2$.  Peaks were confirmed by the DM2 collaboration in the
$\pi^+\pi^-$ (Augustin, 1987) and $K^+K^-$ (Augustin, 1988)
channels, as well as in $K_SK_S$ (Augustin, 1988).

\begin{figure}
\centerline{
\epsfig{file=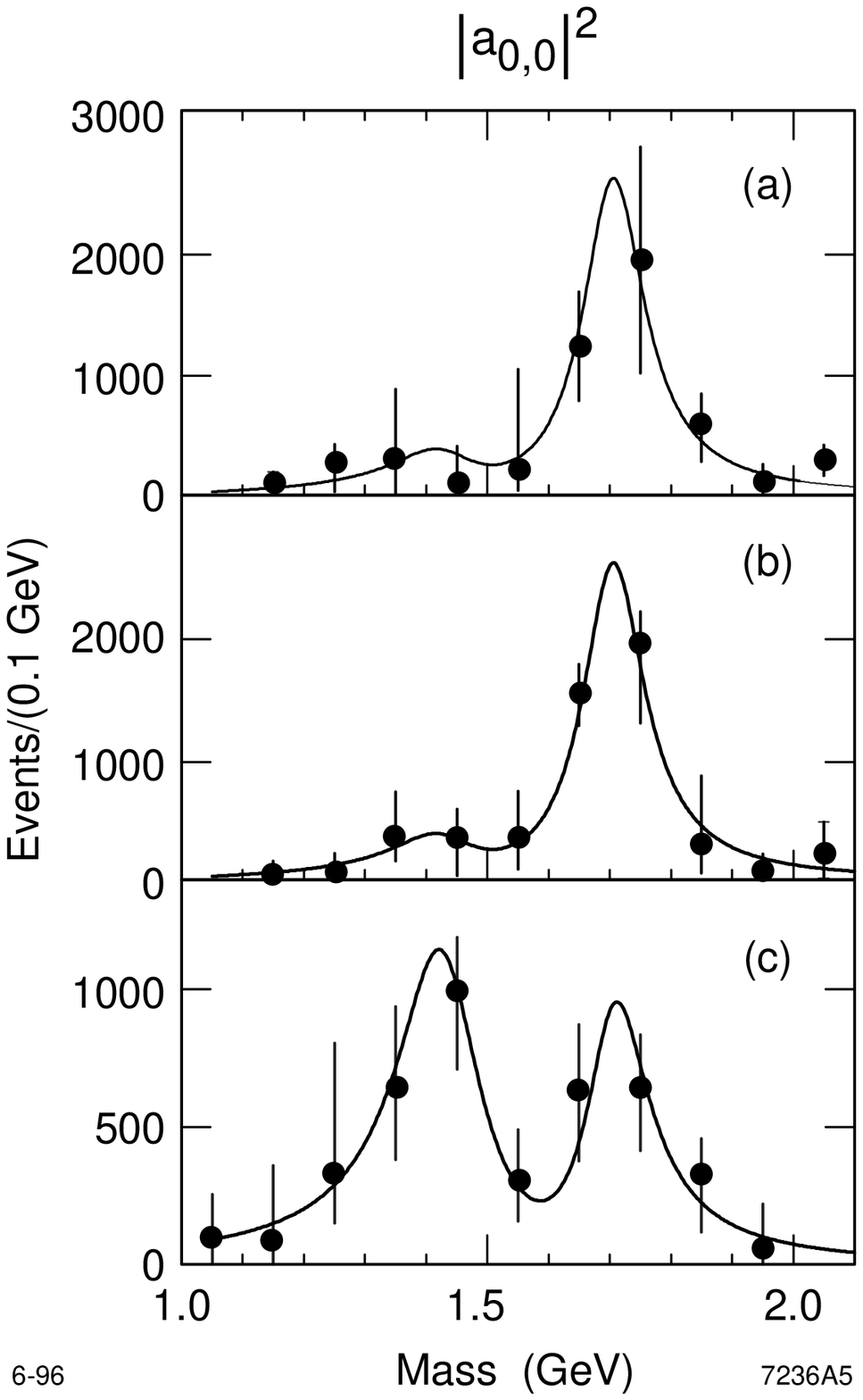,height=3.0in,width=3.0in}\hfill
\epsfig{file=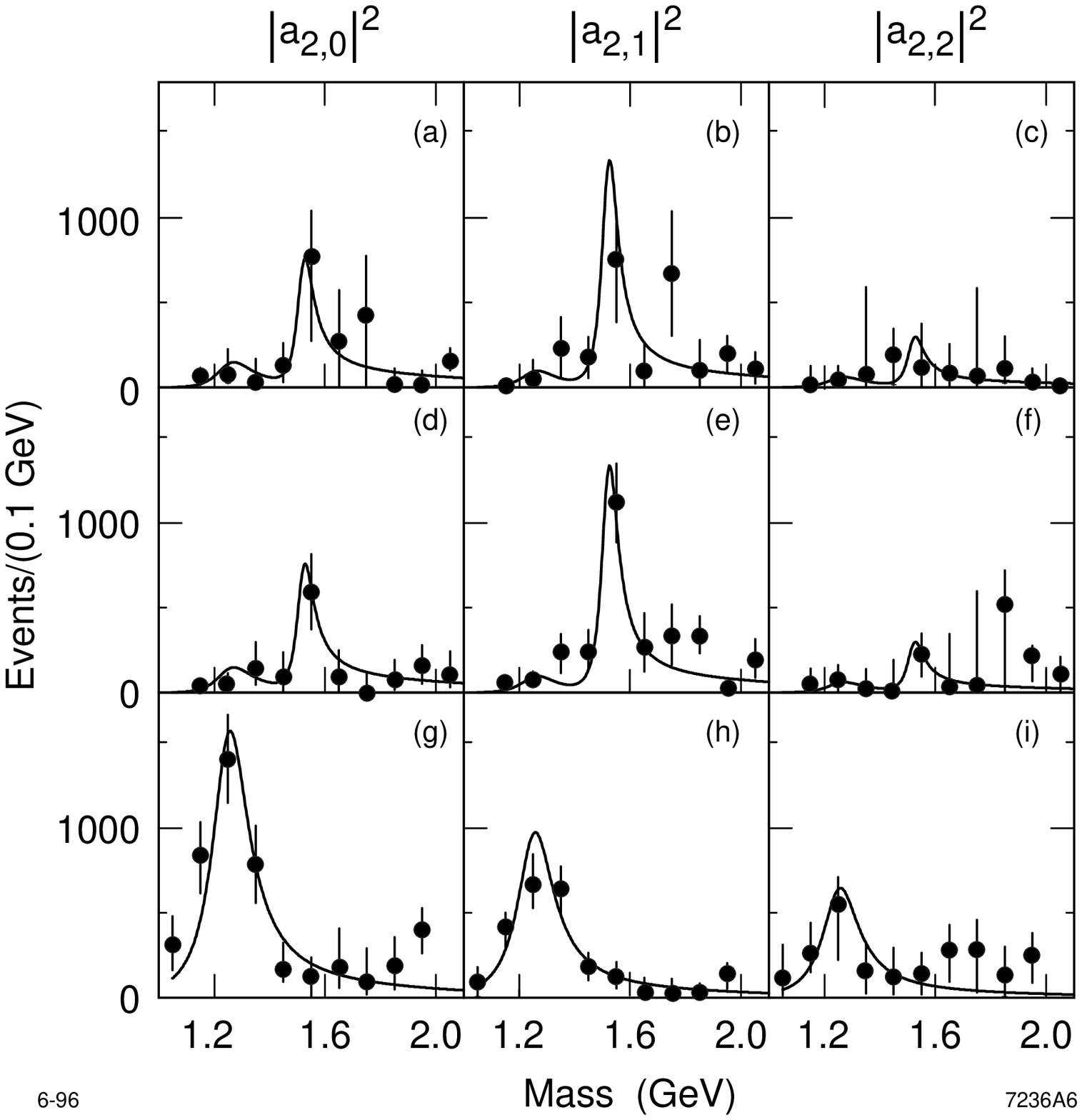,height=3.0in,width=3.0in}}
\caption{Partial wave analysis of $J/\psi$ radiative decay to two
pseudoscalar mesons, from the Mark~III collaboration.  Tha analysis
allows all partial wave components to vary independently for both
$D$-wave (left) and $S$-wave (right).  The top row is the analysis
of $J/\psi\to\gamma K_SK_S$, middle row for $J/\psi\to\gamma K^+K^-$,
and bottom row for $J/\psi\to\gamma \pi\pi$.  The structure near
$1700$~MeV/$c^2$ is clearly dominated by an $S$-wave state.}
\label{figV:fJwmd}
\end{figure}

The published Mark~III result suggesting $J=2$ (Baltrusaitis, 1987)
has actually been called into question in unpublished reports by the same
collaboration (Chen, 1990; Chen, 1991; Dunwoodie, 1997).  This is the result
of a separate analysis, using a somewhat larger event sample
($5.8\times10^6$ $J/\psi$ decays as compared to $2.7\times10^6$)
and also including $J/\psi\to\gamma K_SK_S$ decays.  The key difference in the
analyses, however, is that (Baltrusaitis, 1987) assumed that the
peaks at 1525~MeV/$c^2$ and 1700~MeV/$c^2$ (Fig.~\ref{figV:glueKK}) consisted
of pure resonances and compared observed and predicted angular distributions
for $J=0$ and $J=2$.  That is, interference effects in the amplitudes were
unaccounted for.  However an analysis of moments (see Dunwoodie, 1997) clearly
demands the presence of $S$-wave in the 1710~MeV/$c^2$ region for both
$\pi\pi$ and $K\bar{K}$ final states.  The full amplitude analysis,
shown in Fig.~\ref{figV:fJwmd},
finds that $K\bar{K}$ is nearly entirely $S$-wave
in the $f_J(1710)$ region (with a clear $D$-wave signal for the
$f_2^\prime(1525)$).  In $\pi^+\pi^-$ this analysis clearly sees the $f_2(1270)$
in $D$-wave, and confirms the $K\bar{K}$ $S$-wave result near 1700~MeV/$c^2$
while also identifying what appears to be a scalar with mass near 1400~MeV/$c^2$.
This result in particular would argue for the $f_J(1710)$ to consist mainly of
the scalar glueball, as its decays to $\pi\pi$ preclude identifying it as the
$s\bar{s}$ scalar nonet member.

Other analyses of $J/\psi$ radiative decay, leading to $4\pi$ decay modes of
the $f_J(1710)$, also did not support the original $J=2$ assignment.
Baltrusaitis, et al, 1986a, observed a strong enhancement of $\rho^0\rho^0$
and $\rho^+\rho^-$ masses in the 1.4 to 2.0~GeV/$c^2$ region, well above phase
space.  The angular analysis strongly argued that a single $J^{PC}=0^{-+}$
component dominated this region.  A recent reanalysis of this data
(Bugg, 1995) includes $\sigma\equiv(\pi\pi)_S$ components as well, and
finds this region populated with both $J^{PC}=2^{++}$ states decaying mainly
to $\rho\rho$, and $J^{PC}=0^{++}$ states decaying to $\sigma\sigma$.

Two body decays of the $f_J(1710)$ have been recently reexamined with new
data on $K^+K^-$ from the BES collaboration (Bai, 1996b).
BES finds the $f_J(1710)\to K^+K^-$ region dominated by
a $2^{++}$ state near 1700~MeV/$c^2$, but also resolves a $0^{++}$ state at
1780~MeV/$c^2$ with about half the strength of the $2^{++}$.  However, this
analysis is more tightly constrained than the Mark~III analysis by
(Chen, 1990; Chen, 1991; Dunwoodie, 1997).  In particular, the $D$-wave
amplitudes are forced to be relatively real, reducing the number of unknowns
which need to be determined from the fit.  Their analysis, however, shows that
there are indeed more than one overlapping $D$-wave state in this region,
and the assumption is therefore questionable.

We conclude that the $f_J(1710)$ region, as observed in two-body modes in
$J/\psi$ radiative decay, is most likely dominated by a single $J^{PC}=0^{++}$
state.  Branching ratios for different modes are given by
(Chen, 1990; Chen, 1991; Dunwoodie, 1997).
Their analysis (Fig.~\ref{figV:fJwmd}) gives
$$\frac{\Gamma\left[f_{J=0}(1710)\to \pi\pi  \right]}
       {\Gamma\left[f_{J=0}(1710)\to K\bar{K}\right]}=0.27^{+0.17}_{-0.12}$$

\paragraph{Central production in $pp$ collisions.}

Supposing that it proceeds through double Pomeron exchange, one might
suspect that glue-rich states are produced centrally in $pp$ collisions.
See Sec.~\ref{secIII:central}.  The $f_J(1710)$ has been studied in this
way at CERN by the WA76 (Armstrong, 1989a, 1991a)
and the WA102 (Barberis, 1997b).  Final states include
$\pi^+\pi^-$ (Armstrong, 1991a),
$K^+K^-$ and $K_SK_S$ (Armstrong, 1989a), and
$\pi^+\pi^-\pi^+\pi^-$ (Barberis, 1997b).
As seen in Fig.~\ref{figV:glueKK}, there is a clear enhancement of $K^+K^-$
in the region of the $f_J(1710)$.
The shape of the mass spectrum is quite sensitive to momentum transfer,
with the $f_J(1710)$ region enhanced for more peripheral reactions, i.e.
where Pomeron exchange is expected to dominate.  The $4\pi$ spectrum
shows a clear peak associated with the $f_1(1285)$, and other peaks at
$1440$ and $1920$~MeV/$c^2$.  Again, the shape depends very much on the
momentum transfer.

The angular distribution of the two body decays (Abatzis, 1994)
seems to prefer $J^P=2^+$.  The $4\pi$ system is analyzed assuming the
contribution of a number of isobars, including $\rho$, $f_2(1270)$,
$a_1(1260)$, $a_2(1320)$, and $(\pi\pi)_S$.  Both $1^{++}$ and $2^{++}$
structures are found throughout this region as well as other 
structures.

\paragraph{Peripheral hadronic reactions.}

The $f_J(1710)$ has generally been unobserved in peripheral hadronic reactions.
A measurement of the reaction
$\pi^-p\to K_SK_Sn$ with 22~GeV/$c$ pions, which included a systematic study
of the $2^{++}$ meson spectrum for states decaying to $\pi\pi$ and $K\bar{K}$
(Longacre, 1986), found no evidence for the $f_J(1710)$ except in
the $J/\psi$ radiative decay data.  A measurement of $K^-p\to K_SK_S\Lambda$
by the LASS collaboration (Aston, 1988a) sees a clear signal for the
$f_2^\prime(1525)$ with no evidence for any structure near the $f_J(1710)$.

Interestingly, however, a rather old measurement of $\pi^-p\to K_SK_Sn$ at BNL
(Etkin, 1982b) and a detailed analysis of the $K_SK_S$ $S$-wave
(Etkin, 1982c) reveals two states that are more or less consistent
with both the $f_0(1500)$ and $f_0(1710)$.  Produced in charge exchange with
a $\pi$ beam, it is reasonable to assume the states also couple to $\pi\pi$.
It might be plausible to argue that these two states are in fact significant
mixtures of the scalar glueball and the $s\bar{s}$ member of the $0^{++}$
nonet.  These states were confirmed in a subsequent experiment at Serpukhov
(Bolonkin, 1988).
In fact, a reanalysis of this data, in combination with the data from both LASS
and $J/\psi$ radiative decay (Lindenbaum and Longacre, 1992), shows clearly
that $J=0$ and derives branching ratios to $\pi\pi$, $K\bar{K}$, and
$\eta\eta$.

The GAMS collaboration (Alde, 1992) observes a state which may or may
not be  the $f_J(1710)$.  Called $X(1740)$, it is observed decaying to
$\eta\eta$
in the reaction $\pi^-p\to\eta\eta N^\star$, for 38~GeV/$c$.  The
$\eta\eta$ mass distribution shows a significant peak at $1744\pm15$~MeV/$c^2$
when the recoiling nucleon is accompanied by photons at large angle.  That is,
the signal is present for $\pi^-p\to\eta\eta N^\star$ with $N^\star\to n+\gamma$'s,
but not for $\pi^-p\to\eta\eta n$.  The peak is narrower than has been
observed for the $f_J(1710)$, with $\Gamma<80$~MeV/$c^2$.  No structure is
observed in the $\pi^0\pi^0$ or $\eta\eta^\prime$ mass spectra in the same
experiment.

\paragraph{Two-photon collisions.}

One expects glueballs to be absent in two-photon production. In studies of
$\gamma\gamma\to K\bar{K}$ (Althoff, 1985; Behrend, 1989c;
Albrecht, 1990) a clear signal for $f_2^\prime(1525)$ is evident, but
only upper limits are put on $\Gamma_{\gamma\gamma}$ for $f_J(1710)$.  The
analysis in this case is difficult, because overlap from the various
amplitudes producing $f_2^\prime(1525)$ must be taken into account.  This is
particularly true for $K^+K^-$ where the broad $a_2(1320)$ also contributes
to the sample.  A high
statistics measurement of $\gamma\gamma\to K_SK_S$ would be particularly
useful along with a complete partial wave analysis in the 1400 to
1800~MeV/$c^2$ region.

\subsubsection{The $f_J(2220)$}

The $f_J(2220)$, also known as $\xi(2220)$ or $\xi(2230)$, is a candidate for
the lightest tensor glueball.  However, this association is tenuous for a
number of reasons.  As listed by the Particle Data Group (Caso, 1998),
its mass and width are $2231\pm4$~MeV/$c^2$ and $23\pm8$~MeV/$c^2$ 
respectively.
The mass is close to that expected for the $2^{++}$ glueball from lattice
gauge calculations (Bali, 1993; Chen, 1994) but the width
is very small.  The state has been seen mainly in $J/\psi$ radiative decay,
with a number of decay channels, but never with a strong statistical
significance.

This state was first observed in
$J/\psi\to\gamma K^+K^-$ and $J/\psi\to\gamma K_SK_S$ by the
Mark~III collaboration (Baltrusaitis, 1986b), based on a sample of
$5.8\times10^6$ $J/\psi$ decays.  In both decays, the $K\bar{K}$ mass
distribution rises near 2~GeV/$c^2$ producing a broad enhancement at high
masses.  Superimposed on this enhancement is a narrow signal of $\sim3-4$
standard deviations in {\em each} channel, consistent with the mass and
width of one state, the $f_J(2220)$.  The state was unobserved in a number
of other two-body channels, and upper limits are quoted.  No attempt is made
to identify the spin.

The DM2 collaboration searched through a sample of $8.6\times10^6$ $J/\psi$
for radiative decays, to $\pi^+\pi^-$ (Augustin, 1987) and
$K^+K^-$ and $K_SK_S$ (Augustin, 1988) and do not see the $f_J(2220)$
in any of these three channels.  They quote a limit on the product branching
ratio, that is \mbox{$B[J/\psi\to\gamma f_J(2220); f_J(2220)\to K^+K^-]$},
incompatible with the value determined by Mark~III.  However, they observe
the same broad high mass enhancement in $K\bar{K}$ and suggest it may
represent a state at $2197\pm17$~MeV/$c^2$ with width $\sim200$~MeV/$c^2$.

\begin{figure}
\begin{minipage}{3.0in}
\centerline{\epsfig{file=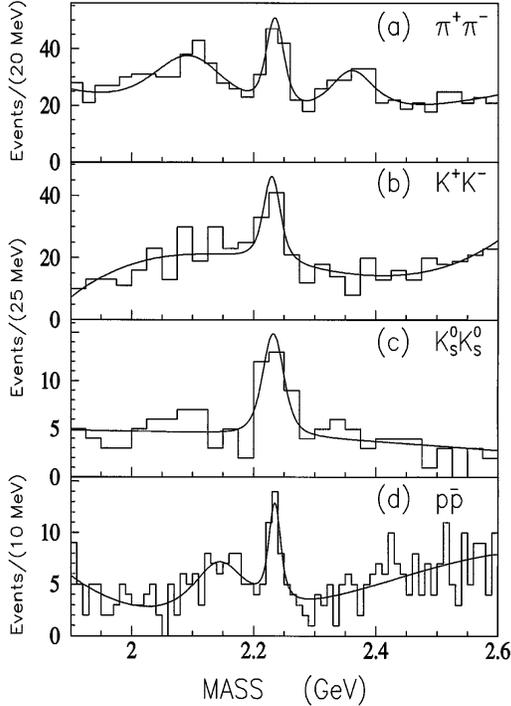,width=2.75in}}
\end{minipage}
\hfill
\begin{minipage}{3.0in}
\caption{Figure 2 from (Bai, et al., 1996a) showing various two-particle
mass distributions observed in $J/\psi$ radiative decay.  Each shows a
signal for the $f_J(2220)$.  The lines represent fits to a smooth background
and one or more Breit-Wigner resonance shapes, convoluted with the
appropriate Gaussian resolution function.}
\label{figV:xiBai}
\end{minipage}
\end{figure}

Recent measurements of $J/\psi$ radiative decay by BES, also with
$\sim8\times10^6$ $J/\psi$ events, claim observation of $f_J(2220)$ at the
level of several standard deviations in the
$\pi^+\pi^-$, $K^+K^-$, $K_SK_S$, $p\bar{p}$ (Bai, 1996a) and
$\pi^0\pi^0$ (Bai, 1998) channels.  Mass distributions in the region
of the $f_J(2220)$ are reproduced from (Bai, 1996a) in
Fig.~\ref{figV:xiBai}.  The product branching ratios for these channels are
marginally consistent with those determined by Mark~III.

Stringent limits have been placed on the two-photon coupling of the
$f_J(2220)$ by the CLEO collaboration in the reactions
$\gamma\gamma\to K_SK_S$ (Godang, 1997) and
$\gamma\gamma\to \pi^+\pi^-$ (Alam, 1998).  Of course, one expects
the two-photon width of glueballs to be small.  These results determine
the ``stickiness'' (Chanowitz, 1984) of the $f_J(2220)$ to be 100 times
as large as the $f_2(1270)$.  Despite the clear presence of the
$f_2^\prime(1525)$, Godang {\it et al.}
(Godang, 1997) see very few events for $K_SK_S$ masses above 2~GeV/$c^2$.

Narrow structures have been reported at 2220~MeV/$c^2$ in peripheral
hadron production.  GAMS reported (Alde, 1986) a small but
significant signal decaying to $\eta\eta^\prime$ in $\pi^-p\to\eta\eta^\prime n$
interactions at 38~GeV/$c$ and at 100~GeV/$c$.  The angular distribution
argues strongly that $J\geq2$. The LASS group (Aston, et al, 1988d) report
a narrow $J^{PC}=4^{++}$ state decaying to $K\bar{K}$ in both the reactions
$K^-p\to K^+K^-\Lambda$ and $K^-p\to K_SK_S\Lambda$, at 11~GeV/$c$ beam
momentum.  Both the mass and width of the GAMS and LASS states are consistent
with the $f_J(2220)$ as seen in $J/\psi$ radiative decay.  A moments analysis
of the LASS results make it clear that spins greater than $J=2$ are required
to describe the data.

The production in high mass $K\bar{K}$ states in peripheral hadronic reactions
prompted a recent study of $s\bar{s}$ quark model states with $J\geq2$ and
$C=P=+$ (Blundell and Godfrey, 1996). 
These are the $L=3$, (i.e. $^3F_2$ or $^3F_4$)
states, and the quark model does indeed predict 
rather large widths.  Although these states
may explain at least some of the structure observed in the 2.2~GeV/$c^2$
region, it would be difficult to identify them
with one having a width as small as the $f_J(2220)$.

Since the $f_J(2220)$ lies above the $p\bar{p}$ threshold, it is possible
to search for it in $p\bar{p}$ annihilation in flight.  This is particularly
interesting in light of the positive result observed by BES
(Bai, 1996a) for $J/\psi\to\gamma f_J(2220)$ followed by
$f_J(2220)\to p\bar{p}$.  Several annihilation in flight searches have in
fact been carried out, including
$p\bar{p}\to \pi^+\pi^-$ (Hasan and Bugg, 1996),
$p\bar{p}\to K^+K^-$ (Sculli, 1987; Bardin, 1987),
and from the JETSET collaboration
$p\bar{p}\to K_SK_S$ (Evangelista, et al, 1997),
$p\bar{p}\to \phi\phi$ (Evangelista, et al, 1998), and
$p\bar{p}\to p\bar{p}\pi^+\pi^-$ (Buzzo, et al, 1997).
No evidence for a narrow state at 2220~MeV/$c^2$ is seen in any of these
experiments.  Since the branching ratio $B[J/\psi\to\gamma f_J(1710)]$ is
not known, and in principle is unconstrained, it is not possible to make a
model-independent consistency check of these data.  However, if we combine
the JETSET result (Evangelista, 1997)
$$ B[f_J(2220)\to p\bar{p}] \times B[f_J(2220)\to K_SK_S] \leq 7.5\times10^{-5}
   ~~~~~~~ (95\% C.L.)$$
with the values from BES (Bai, 1996a)
\begin{eqnarray*}
B[J/\psi\to\gamma f_J(2220)]\times B[f_J(2220)\to K_SK_S]
 &=&(2.7\pm1.1)\times10^{-5}\\
B[J/\psi\to\gamma f_J(2220)]\times B[f_J(2220)\to p\bar{p}]
 &=&(1.5\pm0.6)\times10^{-5}
\end{eqnarray*}
then we can infer the lower bound
$$B[J/\psi\to\gamma f_J(2220)]\geq(2.3\pm0.6)\times10^{-3}$$
This is not only an inordinately large branch for a radiative decay, it
also implies that all the branches reported by BES,
\mbox{$B[J/\psi\to\gamma f_J(2220); f_J(2220)\to X]\approx1.5\times10^{-4}$},
represent only about 10\%
of the total decay modes of the $f_J(2220)$.
We conclude that the evidence for a narrow state $f_J(2220)$ is rather
suspect and the branching ratio to $p\bar{p}$ must be checked.

\subsection{$J^{PC}=0^{-+}$ and $1^{++}$ states in the $E$ region}
\label{secV:Eiota}

\begin{figure}
\begin{minipage}{3.0in}
\centerline{\epsfig{file=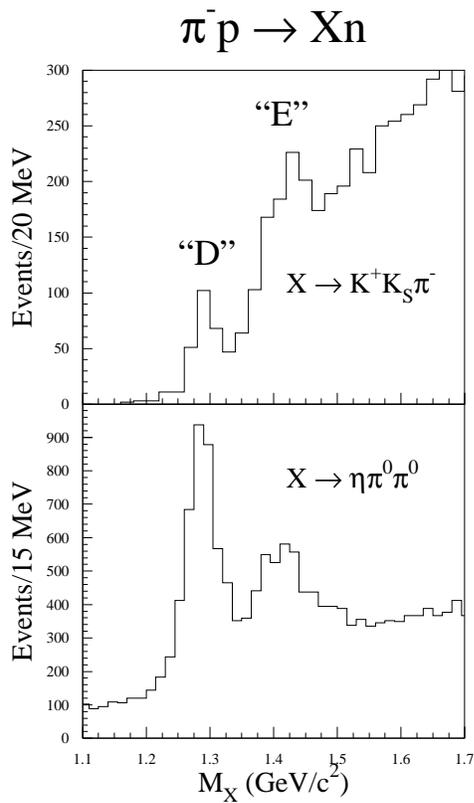,width=3.0in}}
\end{minipage}
\hfill
\begin{minipage}{3.0in}
\caption{Invariant mass distributions for states produced in peripheral
production with high energy $\pi^-$ beams.  In (a) is shown the $K^+K_S\pi^-$
distribution at 18~GeV/$c$ (Cummings, 1995) and (b) shows the $\eta\pi^0\pi^0$
result at 100~GeV/$c$ (Alde, et al., 1997).
In each case, peaks in the the $D$ and $E$ regions are clear.}
\label{figV:histDE}
\end{minipage}
\end{figure}

In principle, the $K\bar{K}\pi$ final state is a good way to study mesons
which couple to $s\bar{s}$ but which are forbidden to decay to $K\bar{K}$.
This would include mesons with $J^P={\rm odd}^+$ or $J^P={\rm even}^-$.
Indeed, experiments which produce $K\bar{K}\pi$ show significant structure
in the mass spectra.  As an example, Fig.~\ref{figV:histDE}(a) histograms the
$K\bar{K}\pi$ mass for the reaction $\pi^-p\to K^+K_S\pi^-n$ at 18~GeV/$c$
(Cummings, 1995).
Relatively narrow structures are clear at
$\sim1300$~MeV/$c^2$ and $\sim1400$~MeV/$c^2$. These structures are
traditionally referred to as the ``$D$'' and ``$E$'' regions, respectively.
The names $D$ and $E$ were historically used to mean $J^{PC}=1^{++}$ states
within these peaks.  In fact, we know that these regions are more complex.
The $D$ region is actually well understood, and contains the
$f_1(1285)$ and $\eta(1295)$ mesons.  However, the structure within the
$E$ region is considerably more complicated, and remains controversial.
The name given to $J^{PC}=0^{-+}$ strength within this peak was originally
called $\iota(1450)$ and for that reason many still refer to this controversy
as the $E/\iota$ puzzle.
The discovery of the $\iota(1450)$ led to 
enormous excitement that the first glueball may have
been found (Scharre, 1980; Ishikawa, 1981; Edwards, 1982a, Aihara, 1986a)
but this is no longer the generally accepted viewpoint. There are a number of
experimental inconsistencies which make this interpretation difficult.
Furthermore, lattice gauge theory would have great difficulty accomodating
such a low mass glueball with these quantum numbers.

A large part of the difficulty is that these structures lie very close to
the $K^\star\bar{K}$ threshold\footnote{We use the notation $K^\star\bar{K}$
to denote the combination of $K^\star(892)\bar{K}+{\rm complex~conjugate}$.},
nominally just below 1400~MeV/$c^2$.  In fact,
it is clear that the background beneath the $D$ and $E$ peaks in
Fig.~\ref{figV:histDE}(a) rises very sharply right at this point.  In principle,
this would favor $J^{PC}=1^{+\pm}$ states since they would decay to an
$S$-wave $K^\star\bar{K}$ pair.  However, the limited phase space leads to
very small relative momenta between the $K$ and $\bar{K}$.  
As a consequence, coupling through
the $a_0(980)$, which has a great affinity for $K\bar{K}$, is
very likely, leading to $S$- and $P$-wave $a_0(980)\pi$ decays, that is
$J^{PC}=0^{-+}$ and $J^{PC}=1^{++}$ states.  It is therefore not surprising
that the $K\bar{K}\pi$ system can get very complicated.

A byproduct of the nearness to $K^\star\bar{K}$ threshold, and a possible tool
for unraveling this structure, is suggested by the decay $a_0(980)\to\eta\pi$.
That is, one might expect similar structure in $\eta\pi\pi$ final states.
Figure~\ref{figV:histDE}(b) plots the $\eta\pi\pi$ invariant mass distribution
for the reaction $\pi^-p\to\eta\pi^0\pi^0n$ at 100~GeV/$c$ (Alde, 1997).
The same $D$ and $E$ structures are apparent here as well.  We note that
the $E$ peak in the mass spectrum does not appear very clearly in peripheral
hadroproduction of states decaying to $\eta\pi^+\pi^-$ (Fukui, 1991;
Manak, 1997), because of additional quasi-twobody states such as $\eta\rho$.

Besides peripheral hadroproduction, $K\bar{K}\pi$ and $\eta\pi\pi$ final states
in the $E$ region have been studied in $J/\psi$ radiative decay, central
production, and in two-photon collisions.  Although there is still significant
controversy, one can identify a number of experimental consistencies:
\begin{itemize}
\itemsep=0in
\topsep=0in
\item The $J^{PC}=0^{-+}$ strength is resolved into two components.  One,
decaying to $\eta\pi\pi$ and to $K\bar{K}\pi$ through $a_0(980)\pi$, has
mass between 1400 and 1420~MeV/$c^2$. The other decays to $K^\star\bar{K}$
and has mass between 1470 and 1480~MeV/$c^2$.
(Note that the Particle Data Group (Caso, 1998) tabulates all the
$J^{PC}=0^{-+}$ decays under the single heading $\eta(1440)$.)
\item There is at least one $J^{PC}=1^{++}$ component to the $E$, called
$f_1(1420)$.  This is seen most clearly in single-tagged two-photon collisions,
which identifies the quantum numbers unambiguously.  It is also clearly seen
in central production and in $p\bar{p}$ annihilation at rest in gaseous
hydrogen.
\item A second $J^{PC}=1^{++}$ state, the $f_1(1510)$, has been identified.
It decays to $K^\star\bar{K}$ and is produced in peripheral hadroproduction with
both $K^-$ and $\pi^-$ beams. (See Sec.~\ref{secIV:ss}.)
\item There is no evidence for the $\eta(1440)$ in untagged two photon
collisions.  It is considerably more ``sticky'' than the $\eta$ or
$\eta^\prime$.  This large stickiness is attributed to a very large
$J/\psi$ radiative decay width, and this may be the strongest evidence for
significant gluonic degrees of freedom in these states.
\end{itemize}
There is also fair evidence for a $J^{PC}=1^{+-}$ state, the $h_1(1380)$,
decaying to $K^\star\bar{K}$ and produced in $K^-p$ interactions and in
$p\bar{p}$ annihilations at rest. (See Sec.~\ref{secIV:ss}.)

Even putting aside various experimental inconsistencies (which we detail
below), there is already serious difficulty accomodating these results.
In the quark model, one might expect two states in the $E$ region, the
$s\bar{s}$ partners to the $f_1(1285)$ and $\eta(1295)$.  The evidence, is,
however, that there are two states of each $J^{PC}$.  Accomodating the
overpopulation with states having gluonic degrees of freedom, however, is
problematic since this is at odds with nearly all models. A variety of
models have been suggested (for example, see Longacre, 1990) but it
is difficult to find clear testable predictions of these models outside of the
$E$ region.  It may be that
the proximity of this region to the $K^\star\bar{K}$ threshold is at the
heart of the difficulties, both in the experiments and in their interpretation.

We separately discuss in greater detail 
the individual evidence and possible interpretation 
for the $J^{PC}=1^{++}$ and $0^{-+}$ states. 

\subsubsection{$J^{PC}=1^{++}$}

The $f_1(1420)$ and the $f_1(1510)$ are well separated in mass and well
resolved in the different experiments although there are no experiments
(or production reactions) in which
{\em both} states are observed.
The proximity to $K^\star\bar{K}$ threshold makes it plausible that a
$K^\star\bar{K}$ $L=0$ molecular bound state is mixing with the $s\bar{s}$
state, but so far there is no good explanation of why the different
production mechanisms so strongly favor one component over the other.

In experiments performed since around 1980, the lower mass is
favored by all measurements {\em other than} peripheral hadron production
(Caso, 1998) including central production, two-photon collisions,
$p\bar{p}$ annihilation, and $J/\psi$ radiative decay.  The higher mass comes
only from production in
$\pi^-p\to Xn$ (Birman, 1988; Cummings, 1995)
and $K^-p\to Xn$ (Aston, 1988c; King, 1991) reactions.
(A measurement of $\gamma\gamma^\star\to\pi^+\pi^-\pi^0\pi^0$
(Bauer, 1993) suggests a mass
near 1510~MeV/$c^2$, but the final state is not fully reconstructed.)

A natural way to study the mainly $s\bar{s}$
partner of the $f_1(1285)$ would be hadronic
peripheral production in $K^-p\to K\bar{K}\pi\Lambda$.  That is, one would
consider hypercharge exchange leading to $K^\star\bar{K}$ final states, so
the intermediate state couples to $s\bar{s}$ on both the input and output
channels.  This experiment was in fact carried out thoroughly by the
LASS collaboration (Aston, 1988c) using $K^-p\to K_SK^\pm\pi^\mp\Lambda$
reaction at 11~GeV/$c$. 
A clean sample of 3900~events was obtained with
the $K_S$ and $\Lambda$  both clearly identified with decay vertices
separated from the primary interaction vertex.  The $K^\star\bar{K}$ mass
spectrum clearly shows the $D$ region heavily suppressed relative to the $E$,
suggesting that the $E$ is dominated by $s\bar{s}$.  A partial wave
analysis of the $E$, shown in Fig.~\ref{fig:ss-axials},
shows a predominance of $J^P=1^+$, and also shows that this
cannot be the result of a single resonance.  By symmetrizing the final state on
the basis of charge conjugation, the $J^P=1^+$ strength is clearly resolved into
a $J^{PC}=1^{++}$ state at $1530\pm10$~MeV/$c^2$, and
a $J^{PC}=1^{+-}$ state at $1380\pm20$~MeV/$c^2$.  The experimenters explicitly
point out that these are good candidates for the mainly $s\bar{s}$ members of
their respective nonets.
Furthermore, these results are confirmed by a BNL/MPS experiment (King, 1991)
which studied the reaction $K^-p\to K_SK^+\pi^-(\Lambda,\Sigma^0)$ at
8~GeV/$c$.

Single-tagged two-photon production (Sec.~\ref{secIII:gg}) of $K\bar{K}\pi$
final states
(Aihara, 1986b;
 Gidal, 1987b;
 Hill, 1989;
 Behrend, 1989b)
all show clear evidence of the $J^{PC}=1^{++}$ state, although the mass is
generally more consistent with 1420 than 1510~MeV/$c^2$, with a
$\gamma\gamma^\star$ width (times $K\bar{K}\pi$ branching ratio) of
$1.7\pm0.4$~keV (Caso, 1998).
On the other hand, single-tagged production of $\eta\pi\pi$
(Gidal, 1987a; Aihara, 1988) shows a clear signal for
$f_1(1285)$ and a $\gamma\gamma^\star$ width (times $\eta\pi\pi$ branching
ratio) of $1.4\pm0.4$~keV (Caso, 1998).  This suggests a
consistency with the $f_1(1285)$ and $f_1(1420)$ being $n\bar{n}$ and
$s\bar{s}$ partners, but the comparison is likely misleading.  It is quite
difficult to make good comparisons of the $\gamma\gamma^\star$ width alone,
because of the proximity to threshold and to the inherent difficulties
in predicting the width at all.  In fact, two-photon couplings to
molecular states are not likely to be profoundly different (Barnes, 1985b).

Radiative $J/\psi$ decay into $K\bar{K}\pi$ shows considerable strength in
the $E$ region, although it is dominated by $J^{PC}=0^{-+}$.
(See the following section.) However, a partial wave analysis
(Bai, 1990; Augustin, 1992) shows a signal for $J^{PC}=1^{++}$
consistent with a mass closer to 1420~MeV/$c^2$, and
$B(J/\psi\to f_1(1420)\gamma)\times
 B(f_1(1420)\to K\bar{K}\pi)=8.3\pm1.5\times10^{-4}$ (Caso, 1998).
A small $J^{PC}=1^{++}$ contribution decaying to $K\bar{K}\pi$ has been seen
in $p\bar{p}$ annihilation in gaseous hydrogen by the Obelix collaboration
(Bertin, 1997a) with the reaction
$p\bar{p}\to K^\pm K^0\pi^\mp\pi^+\pi^-$ (where the $K^0$ is not seen), with
mass $1425\pm8$~MeV/$c^2$.
There is also recent evidence from the Crystal Barrel (Abele, 1997b)
for the $J^{PC}=1^{+-}$ state observed by LASS (Aston, 1988c).
Central production of $K\bar{K}\pi$
(Armstrong, 1989b, 1992; Barberis, 1997c)
and of $\eta\pi\pi$ (Armstrong, 1991b)
has been performed many times, and a clear peak appears at 1420~MeV/$c^2$,
consistent with $J^{PC}=1^{++}$.  The behavior of this signal with transverse
momentum suggests that it, like the $f_1(1285)$ is a conventional $q\bar{q}$
meson.

These results have led some to question the existence of two separate
states (Close and Kirk, 1997b).  It is clearly of high importance to
observe these two states simultaneously.  There is a chance that the
next generation of $\gamma\gamma^\star$ measurements will be able to see
both states and resolve them separately.

\subsubsection{$J^{PC}=0^{-+}$}

Once again, we might normally expect one $J^{PC}=0^{-+}$ state in the $E$
region, namely the $s\bar{s}$ partner of the $\eta(1295)$.  These two states
would be taken as linear combinations of the radial excitations of the ground
states $\eta$ and $\eta^\prime$.  There again seems to be clear evidence of two
$J^{PC}=0^{-+}$ states in the $E$, but these are seen simultaneously in the
same experiments, unlike the $f_1(1420)$ and $f_1(1510)$.
There are other oddities about the $J^{PC}=0^{-+}$ as well.

Table~\ref{tabV:Eps} shows results from several experiments which see
pseudoscalar resonances in the $E$ region, decaying to $K\bar{K}\pi$.
Nearly all measurements of this final state, including peripheral production
with pion beams, $p\bar{p}$ annihilation, and $J/\psi$ radiative decay,
determine two states. The masses differ by about 50~MeV/$c^2$, with the
heavier decaying to $K^\star\bar{K}$, except for (Augustin, 1992).
Furthermore, the mass of the lower state agrees very well with measurements of
states decaying to $\eta\pi\pi$, again in a number of different types of
experiments.  Most recently, measurements of the reactions
$\pi^-p\to\eta\pi^0\pi^0$ at 100~GeV/$c$ (Alde, 1997) and
$\pi^-p\to\eta\pi^+\pi^-$ at 18~GeV/$c$ (Manak, 1997) show that most of
the $J^{PC}=0^{-+}$ $\eta\pi\pi$ signal is concentrated in
$\eta(\pi\pi)_S$ instead of $a_0(980)\pi$.
\begin{table}
\caption{Masses, in MeV/$c^2$, of $J^{PC}=0^{-+}$ $K\bar{K}\pi$ resonances
in the $E$ region.}
\label{tabV:Eps}
\begin{center}
\begin{tabular}{llcc}
\hline
                  &                &\multicolumn{2}{c}{Quasi-Two Body Mode}\\
Reaction          & Reference      & $a_0\pi$     &   $K^\star\bar{K}$\\
\hline
$p\bar{p}\to K^\pm K^0\pi^\mp\pi^+\pi^-$ & Bertin, et al., 1997a 
                                   & $1407\pm5$   &   $1464\pm10$\\
$J/\psi\to\gamma K_SK^\pm\pi^\mp$      & Bai, et al., 1990 
                                   & $1416\pm10$  &   $1490\pm20$\\
$J/\psi\to\gamma K\bar{K}\pi$          & Augustin, et al., 1992 
                                   & $1459\pm5$   &   $1421\pm14$\\
$\pi^-p\to K_SK_S\pi^0$ at 21 GeV/$c$  & Rath, et al., 1989 
                                   & $1413\pm5$   &   $1475\pm4$\\
$\pi^-p\to K^+K_S\pi^-$ at 18 GeV/$c$  & Cummings, 1995 
                                   & $1412\pm2$   &   $1475\pm6$\\
\hline
Particle Data Group & Caso, et al., 1998 & $1418.7\pm1.2$ & $1473\pm4$\\
\hline
\end{tabular}
\end{center}
\end{table}

Clearly, this leads us to guess that one of these states is the $s\bar{s}$
partner of the $\eta(1295)$, and the other is some manifestation of non
$q\bar{q}$ degrees of freedom.  That would imply, however, that one of the
states (presumably the one that decays to $K^\star\bar{K}$) should appear
in untagged $\gamma\gamma$ collisions, while the other might show some
anomalous dependence on transverse momentum in central $pp$ and $\pi p$
collisions.  In fact, there is {\em no} evidence that {\em either} state
is produced in either of these reactions.

The most stringent limits on two-photon production of an $s\bar{s}$
pseudoscalar meson in the $E$
region were obtained by (Behrend, 1989b) in the reaction
$\gamma\gamma\to K_SK^\pm\pi^\mp$. They find
$\Gamma_{\gamma\gamma}[\eta(1440)]\times
 B[\eta(1440)\to K\bar{K}\pi]<1.2$~keV at 95\% C.L.
They further determine that this implies that the $\eta(1440)$ is at least
20 times as ``sticky'' as the $\eta^\prime$.  Unless there is a fortuitous
cancellation due to quark mixing angles, this would argue that the
$\eta(1440)$ (or both pseudoscalars in this region) have large glue content,
and it would leave the $s\bar{s}$ partner of the $\eta(1295)$ unidentified.
There may be more to this than meets the eye, however. The $\eta(1295)$
has also been unidentified in two-photon collisions (Caso, 1998).

There is some evidence of $J^{PC}=0^{-+}$ production in $K_SK^\pm\pi^\mp$
final states in $pp$ and $\pi^+p$ central collisions (Armstrong, 1992)
but the signal in the $E$ region is dominated by $1^{++}$.

It is obviously very difficult to draw a clear, consistent picture from
the $J^{PC}=0^{-+}$ results in the $E$ region.  It may simply be that its
proximity to $K^\star\bar{K}$ threshold brings in more complicated
mechanisms that can be treated with the formalisms presently at our
disposal.

\subsection{Other Puzzles}

Thus far we've discussed the states that have received the most attention 
in recent years.  In addition to these, numerous other extraneous 
states have been reported which have received much less attention of 
late, primarily because there has been no new information  on them. In 
this subsection, for compeleteness, we briefly discuss some 
additional examples.

\subsubsection{Extra $J^{PC}=2^{++}$ States}
\label{secV:gT}

The ground state $2^{++}$ nonet, consisting of the 
$a_2(1325)$, $f_2(1270)$, $f_2'(1525)$,  and $K^*(1430)$,
has been complete for some time.
However, several additional isoscalar $2^{++}$
states have been observed which are inconsistent
with quark model predictions.  
These include the three $g_T$ states in $\phi\phi$ at 2011, 2297 and 2339 MeV;
the $f_2(1565)$ seen in $\bar{p}N$ annihilation;
and the $f_2(1430)$ and $f_2(1480)$ observed in $\pi\pi$ and $K\bar{K}$ spectra
between the $f_2(1270)$ and the $f_2'(1525)$.

\paragraph{OZI suppression and states in $\pi^-p\to\phi\phi n$}

In the high mass region three states, somtimes known as $g_T$, have been
observed in the OZI-suppressed reaction $\pi^-p\to\phi\phi n$ at 22~GeV/$c$
by a BNL group
(Etkin, 1978a, 1978b, 1982a, 1985, 1988) using the Multi Particle Spectrometer
(MPS) facility at the AGS.
The three are distinguishable by their decay couplings to different
$\phi\phi$ partial waves. The two higher mass states have also been seen as
peaks in the two $2^{++}$ components of inclusive production from $\pi^-$ Be
interaction at 85~GeV/$c$ in the WA67 experiment at CERN (Booth, 1986).

The $\phi\phi$ mass spectrum observed by the BNL group
shows a broad enhancement from threshold to 2.4 GeV, while
the experimental acceptance remains flat up to 2.6 GeV.
A partial-wave analysis of the bump reveals that it consists of
three distinct $2^{++}$ states,  $2^{++}$ $f_2(2010)/g_{_T}$,
$f_2(2300)/g'_{_T}$ and $f_2(2340)/g''_{_T}$ (Longacre, 1986; Etkin, 1985).  
Using the notation $L_S$ where $L$ is the orbital angular momentum
and $S$ is the total intrinsic spin for $\phi\phi$, the states are
$g_{_T}$ with $M=2011\pm70$~MeV/$c^2$ and $\Gamma = 202\pm65$~MeV/$c^2$
(about 98\% $S_2$);
$g'_{_T}$ with $M=2297\pm28$~MeV/$c^2$ and $\Gamma=149\pm41$~MeV/$c^2$
(about 25\% $D_2$ and 69\% $D_0$);
and $g''_{_T}$ with $M=2339\pm55$~MeV/$c^2$ and $\Gamma=319\pm$~MeV/$c^2$
(about 37\% $S_2$ and 59\% $D_0$).  
From a study of the production characteristics, it was concluded that the
$g_T$ states are produced by one-pion-exchange processes.
If so, the $g_T$ states should also couple to $\pi\pi$ channels but are
difficult to observe due to large backround in these channels.
As they are produced in the OZI-forbidden channel (Landberg, 1996), they are
thought to be candidates for tensor glueballs, although this interpretation
is controversial (Lindenbaum and Lipkin, 1984).
 
The WA67 group at the CERN $\Omega$-Spectrometer (Booth, 1986) studied
inclusive $\phi\phi$ production from $\pi^-Be$ interactions at 85 GeV/$c$.
They see general enhancement at the $\phi\phi$ threshold followed by
a second peak at 2.4 GeV. Assuming that they see the second
and third $g_T$ states, they have fitted their mass spectrum
with two Breit-Wigner forms, one with 50-50\% $S$- and $D$-waves
and the other 100\% $D$-wave over a smooth background.  The resulting
masses and widths are $2231\pm10$~MeV/$c^2$ and $133\pm50$~MeV/$c^2$ for the
second $f_2(2300)/g_T'$ and $2392\pm10$~MeV/$c^2$ and $198\pm50$~MeV/$c^2$ for
the third $f_2(2340)/g_T''$, respectively.  They have also carried out a
joint moment analysis and find that the $\phi\phi$ system up to 2.5~GeV/$c^2$
is mainly $2^{++}$ (Booth  1986, Armstrong 1989b) although the statistics are
limited.

One might expect the $g_T$'s to couple to the $\rho\rho$ and $\omega\omega$
channels as well, if they are indeed glueballs.  
The GAMS group (Alde, 1988b) observes two $2^{++}$ resonances in
$\pi^-p\to\omega\omega n$ at 38 GeV/$c$.  The state at $1956\pm20$~MeV/$c^2$
with width $220\pm60$~MeV/$c^2$ is not inconsistent with the lightest $g_T$.
 
The DM2 collaboration (Bisello, 1986) carried out an analysis
of the $\phi\phi$ system produced in $J/\psi$ radiative decays.
Although $2^{++}$ is found to be the main wave, no threshold
enhancement in the $\phi\phi$ system is observed,
in contrast to the hadronic production.
However, they find a narrow peak at around 2.2 GeV with a preferred
spin-parity of $0^-$.  The Mark III collaboration also studied their 
$\phi\phi$ spectrum in $J/\psi$ radiative decays, with possible structures
in the 2.1$-$2.4 GeV mass region (Mallik, 1986; Toki, 1987; Blaylock, 1987).
However, no spin-parity has been given for the $\phi\phi$ structures and it is
not clear if they are to be associated with their $0^{-+}$ structures near
threshold in $\rho\rho$ and $\omega\omega$,
also seen in $J/\psi$ radiative decays, or with the BNL $g_T$ states.

As these states are above $p\bar{p}$ threshold, they can in principle be
observed in annihilation-in-flight reactions.  The JETSET collaboration
(Evangelista, 1998) studied the reaction $\bar{p}p\to\phi\phi$ with antiproton
beams between 1.1 and 2.0~GeV/$c$ momentum.  This experiment is sensitive to
intermediate states formed in the annihilation channel with masses between
2.1 and 2.4~GeV/$c^2$.  There is no evidence for the $f_2(2300)/g_T'$ or
$f_2(2340)/g_T''$.

Finally, the $\phi\phi$ system has been studied in $pp$ central production
by the WA102 Collaboration (Barberis, 1998).  Some weak structure is observed
in the $\phi\phi$ mass distribution, but the statistics are poor.  The
angular distribution favors $J^{PC}=2^{++}$.

\paragraph{The $f_2(1565)$}

This state has a long history going back to the 1960's
(Bettini 1966; Conforto 1967).
In 1990 ASTERIX at LEAR presented evidence for the production
of a resonance $f_2(1565)$ in P-wave $p\bar{p}\to\pi^0\pi^+\pi^-$ annihilation in 
hydrogen gas (May, 1989, 1990a).  A state with $M=1565\pm10$~MeV/$c^2$ and
$\Gamma=170\pm20$~MeV/$c^2$ was observed decaying to $\pi^+\pi^-$, recoiling
against the $\pi^0$.  No enhancement was visible in the $\pi^\pm\pi^0$ invariant
mass indicating that it is an $I=0$ resonance.  A Dalitz plot analysis
showed clear evidence for $J^{PC}=2^{++}$.  This resonance cannot be
identified with the $f_2'(1525)$ meson which decays mostly to $K\bar{K}$.
Otherwise it would be produced strongly in the final state  $K\bar{K}\pi$ 
where it has not been observed (Conforto, 1967).
In a separate analysis (May 1990b), selecting initial $p\bar{p}$ $S$-states,
no indication of a resonance at 1.5 GeV was observed.

The Crystal Barrel experiment at LEAR subsequently studied all neutral events
from $\bar{p} p$ annihilation (Aker, 1991) which gives information
on the $3\pi^0$, $\eta\eta\pi^0$, and $\eta\eta'\pi^0$ channels.  
The $f_2(1270)$ and $f_2(1565)$ resonances are clearly visible in the
$\pi^0\pi^0$ invariant mass projections.
As more data was acquired and the analysis matured, however, it became clear
that the prominent feature in $\pi^0\pi^0$ near 1500~MeV/$c^2$ was in fact
the $f_0(1500)$.  Still, the analysis requires the presence of some $2^{++}$
strength decaying to $\pi^0\pi^0$ in the same region (Amsler, 1998).
Recently, the OBELIX collaboration provided new evidence for the $f_2(1565)$
in $\bar{n}p\to\pi^+\pi^-\pi^+$ (Bertin, 1998).

\paragraph{The $f_2(1430)$ and $f_2(1480)$}

Evidence for these two states was found in the data on the 
double-Pomeron-exchange reaction in an experiment R807 at CERN ISR
(Akesson, 1986; Cecil, 1984).
The reaction concerns the exclusive $\pi^+\pi^-$ production
in $pp \to p_f(\pi^+\pi^-)p_s$ at $\sqrt{s} = 63$~GeV.  The recoil protons have
been detected with $-t$ less than 0.03~(GeV/c)$^2$, thus ensuring
nearly pure Pomeron exchanges at both vertices.  The resulting
$\pi\pi$ spectrum exhibits a set of remarkable bump-dip
structures near 1.0, 1.5 and 2.4 GeV, respectively (Akesson 1986).
Another striking feature is that the $\rho$(770) and the $f_2(1270)$ are not
seen in the data.  One may expect that the I=1 $\rho$(770) should not have
been seen in a pomeron-pomeron interaction; however, the apparent
absence of the $f_2(1270)$ is noteworthy.  
The second drop-off
near 1.4 GeV is partly due to the $f_0(1370)$ and a $D$-wave
structure which is attributed to the $f_2(1480)$.
The data in fact favours a $2^{++}$ structure above the $f_2(1270)$
with mass and width of (1480 $\pm$ 50) and (150 $\pm$ 40)~MeV respectively.
A full understanding
of the nature of the $f_2(1480)$ will probably also require  
an explanation of absence of the $f_2(1270)$ in the R807 data.
An explanation may follow from Close and Kirk's (Close 1997b) glueball filter.

\subsubsection{Structure in $\gamma\gamma\to VV$}

Structures in $\gamma\gamma\to VV'$ have generated considerable interest.
A recent review of results from the ARGUS experiment (Albrecht, 1996) includes
a thorough discussion of this reaction.
 
This subject originated from the original observation by the TASSO collaboration
(Brandelik, 1980) of structures in the $\gamma\gamma\to\rho^0\rho^0$ cross
section. This was subsequently confirmed by other experiments 
(Burke, 1981; Althoff, 1982; Behrend, 1984; Aihara, 1988), 
where one would normally expect the rate to be suppressed in the threshold
region due to the reduced phase space.
Results are now available for numerous final states, including
$\rho^0\rho^0$, $\rho^+\rho^-$, $\omega\omega$,
$\omega\rho^0$, $K^{0*}\bar{K}^{0*}$, $K^{+*}\bar{K}^{-*}$
$\rho^0\phi$, $\omega\phi$, and $\phi\phi$.
The cross sections for the different final states vary in 
their relative size and 
the energy at which the cross sections peak also varies from one final
state to another.  

The large difference in cross section between the $\rho^0\rho^0$ and
$\rho^+\rho^-$ channels, $\sigma(\gamma\gamma\to \rho^0\rho^0)
\simeq 4 \sigma(\gamma\gamma\to \rho^+\rho^-)$
(Albrecht, 1996; Behrend, 1989a) rules out a simple $s$-channel
resonance explanation where one expects the decay of a conventional resonance
into $\rho^+\rho^-$ to occur with a rate two times as often as that into 
$\rho^0\rho^0$.  The two $\rho^0$ mesons can only be in a state with
$I=0$, $I=2$ or a mixture of the two. The large ratio of the $\rho^0\rho^0$
to $\rho^+\rho^-$ cross section cannot be accounted for by a pure $I=0$ or a
pure $I=2$ state with the same spin-parity quantum numbers but from 
interference between the two.  The interference is observed to be 
constructive in $\gamma\gamma\to\rho^0\rho^0$ and destructive in
$\gamma\gamma\to \rho^+\rho^-$.  A demonstration that the $\rho\rho$
cross section is predominantly resonant, together with this
isospin argument, would imply the existence of exotic $I=2$ states,
possibly $q\bar{q}q\bar{q}$ resonances (Achasov, 1982; Li, 1983).

A number of models have been invoked to explain the 
structure in the original $\gamma\gamma\to\rho^0\rho^0$ 
and have made predictions for other channels:
$q\bar{q}q\bar{q}$ exotica were first suggested by
(Jaffe, 1977a, 1977b, 1978) and were explored as an explanation of the
structure in $\gamma\gamma\to VV'$ by (Achasov, 1982)
and by Li and Liu (Li, 1983).
The vector dominance model with factorization in the $t$-channel
attempts to identify specific $t$-channel exchange and 
extract them from photoproduction data (Alexander 1982).    
Perturbative QCD with Coulombic rescattering corrections were examined 
by Brodsky (1987).  
In the one meson exchange model (Achasov, 1988; T\"ornqvist 1991)
the structure in
$\gamma\gamma\to\rho\rho$ is explained by bound states formed by one
pion exchange potentials analogous to those used in nuclear physics.
These models do reasonably well for the process for which
they were constructed
but for the most part they fail to explain subsequent $VV'$ data.

%To really understand the $\gamma \gamma$ cross sections it would
%be worthwhile to backtrack by first considering the more conventional 
%contributions such as the $s$-channel process $\gamma\gamma\to M\to VV'$ where
%$M$ is a conventional $q\bar{q}$ state, either real or virtual,
%the $\pi$ and $K$ exchange contributions, and threshold enhancements,
%perhaps due to the fact that the $\rho$'s are broad and couple directly
%to $\gamma$'s as in the vector dominance model.  
%Once the conventional physics has been included we can see if there is
%a need for more complicated and more exotic explanations.
%Even then, we should be careful how we treat $q^2\bar{q}^2$ contributions.
%As shown by Weinstein and Isgur (Weinstein, 1982; 1983)
%it is a dynamical question whether specific $J^P$ $q^2\bar{q}^2$ states
%in fact exist so it is important that a similar dynamical calculation 
%be done for vector mesons.  Even if
%stable states do not exist it is quite possible that four quark configurations
%generate potentials between vector mesons that give rise to the
%observed cross sections. See also (Bajc, 1996).
%The CLEO collaboration's large data sample  currently being analyzed  
%will undoubtably shed light on some of these questions.

\subsubsection{The $C(1480)$}

The Lepton-F collaboration at Serpukhov (Bityukov, 1987) examined the reaction
$\pi^-p\to\phi\pi^0n$ with a 32~GeV/$c$ beam.  A very strong peak in the cross
section was observed near 1500~MeV/$c^2$ in $\phi\pi$ mass.  The production
cross section is substantial. Interpreted as a resonance, called
$C(1480)$, the peak has mass $1480\pm40$~MeV/$c^2$, width $130\pm60$~MeV/$c^2$,
and $\sigma(\pi^-p\to Cn)\times B(C\to\phi\pi^0)=40\pm15$~nb. The existence of
such a narrow, isovector state is clearly peculiar given this decay mode, and
a number of interpretations have been offered
(Kubarovski, 1988; Kopeliovich, 1995).

This state has not been observed in other reactions, including
$pp$ central production (Armstrong, 1992) and $p\bar{p}$ annihilation at
rest (Reifenr\"{o}ther, 1991).  Recent data taken by the E852 collaboration
including $K$ identification will check this reaction with pion beams,
however, and photoproduction experiments are planned (Dzierba, 1994).

\subsection{Missing States}
\label{secV:missing}

Clearly one can only discuss ``overpopulation'' if all the expected
$q\bar{q}$ states have in fact been identified.  This is not the case
for some multiplets.  A particularly glaring example are the missing
isoscalar and isovector $J^{PC}=2^{--}$ (i.e. $\omega_2$ and $\rho_2$)
states.  These would be the $^3D_2$ partners of the relatively well
established $^3D_3$ ($\rho_3(1690)$ and $\omega_3(1670)$) and
$^3D_1$ ($\rho(1700)$ and $\omega(1600)$) $q\bar{q}$ combinations.
The strange $J^P=2^-$ members appear to be the two $K_2$ states near
1800~MeV/$c^2$ (Aston, 1993), i.e. $^3D_2$ and $^1D_2$, although
they need to be confirmed (PDG, Caso, 1998).

It will be important to fill in both the orbitally and radially
excited multiplets, although as we go higher in mass the states become
broader as well as more numerous. As an example of how a search
for the missing states would proceed we examine the missing states of the
$L=2$ meson multiplet.  Quark model predictions for these states are
listed in Tab.~\ref{tabV:missing}.
\begin{table}
\caption{Quark Model predictions for the properties of the 
missing $L=2$ mesons. The masses and widths are given in MeV.}
\label{tabV:missing}
$$\matrix{ 
\noalign{\hrule}
\noalign{\vskip 0.1cm}
\noalign{\hrule}
\noalign{\vskip 0.1cm}
\hbox{Meson State} & \hbox{Property}  & \hbox{Prediction}  \cr
\noalign{\vskip 0.1cm}
\noalign{\hrule}
\noalign{\vskip 0.1cm}
\eta_2 (1^1D_2 )& \hbox{Mass} 			& 1680  		  \cr
		& \hbox{width} 			& \sim 400 		  \cr
		& BR(\eta_2 \to a_2\pi)		& \sim 70\%		\cr
		& BR(\eta_2\to \rho\rho)	& \sim 10\%		\cr
		& BR(\eta_2\to K^*\bar{K}+c.c.)	& \sim 10\%		\cr
\noalign{\vskip 0.1cm}
\noalign{\hrule}
\noalign{\vskip 0.1cm}
\eta_2'\;(1^1D_2 )& \hbox{Mass}			& 1890			\cr
		& \hbox{width}			& \sim 150		\cr
		& BR(\eta'_2 \to K^*\bar{K}+c.c.) & \sim 100\%		\cr
\noalign{\vskip 0.1cm}
\noalign{\hrule}
%\noalign{\vskip 0.1cm}
%K_2 \; (1^1D_2 )& \hbox{Mass} 		& 1780 			\cr
%		& \hbox{width}		& \sim 300		\cr
%		& BR( K_2 \to K^* f(1280))& \sim 30\%		\cr
%		& BR( K_2 \to \rho K)	& \sim 20\%		\cr
%\noalign{\vskip 0.1cm}
%\noalign{\hrule}
\noalign{\vskip 0.1cm}
\omega_1 \; (1^3D_1 ) & \hbox{Mass}	& 1660			\cr
		& \hbox{width}		& \sim 600		\cr
		& BR( \omega_1 \to B\pi ) & \sim 70 \%		\cr
		& BR( \omega_1 \to \rho \pi ) & \sim 15\%	\cr
\noalign{\vskip 0.1cm}
\noalign{\hrule}
%\noalign{\vskip 0.1cm}
%K_2 \; (1^3D_2 )& \hbox{Mass} 		& 1810 			\cr
%		& \hbox{width}		& \sim 300		\cr
%		& BR( K_2 \to K^* (1420)\pi) & \sim 50 \%	\cr
%		& BR( K_2 \to K^* \pi)	& \sim 30 \%		\cr
%\noalign{\vskip 0.1cm}
%\noalign{\hrule}
\noalign{\vskip 0.1cm}
\rho_2 \; (1^3D_2) & \hbox{Mass}	& 1700			\cr
		& \hbox{width}		& \sim 500		\cr
		& BR( \rho_2 \to [a_2\pi]_S) & 	\sim 55\%	\cr
		& BR( \rho_2 \to \omega \pi )&  \sim 12\%	\cr
		& BR( \rho_2 \to \rho \rho ) &  \sim 12\%	\cr
\noalign{\vskip 0.1cm}
\noalign{\hrule}
\noalign{\vskip 0.1cm}
\omega_2\; (1^3D_2) & \hbox{Mass}	& 1700			\cr
		& \hbox{width}		& \sim 250		\cr
		& BR(\omega_2 \to \rho \pi) & \sim 60 \%	\cr
		& BR(\omega_2 \to K^*\bar{K} ) & \sim 20 \%     \cr
\noalign{\vskip 0.1cm}
\noalign{\hrule}
\noalign{\vskip 0.1cm}
\phi_2\; (1^3D_2) & \hbox{Mass}		& 1910			\cr
		& \hbox{width}		& \sim 250		\cr
		& BR (\phi_2 \to K^*\bar{K}+c.c.) & \sim 55\% 	\cr
		& BR (\phi_2 \to \phi \eta ) &\sim 25\%		\cr
\noalign{\vskip 0.1cm}
\noalign{\hrule}
\noalign{\vskip 0.1cm}
\noalign{\hrule} }$$
\end{table}

Consider the $\eta_2 (1^1D_2)$.  There is some evidence that this state
has in fact been observed (Sec.~\ref{secIV:is}).
We expect it to be almost degenerate in
mass with its non-strange isovector partner, the $\pi_2(1670)$.
From Tab.~\ref{tabV:missing} we see that it is expected to
be rather broad and it decays predominantly through the $a_2(1320)$ isobar
which in turn decays to $\rho\pi$.  The $4\pi$ final state is complicated
to reconstruct.  Since $a_2\to\eta\pi$, other final states should be checked.
The $\rho_2 (1^3D_2)$ will also  decay dominantly to a $4\pi$ final state.
The $\omega_2$ decays to the simpler  $\rho\pi$ final state with a moderate
width but since it has a similar mass as the $\pi_2 (1680)$  which also decays
to $\rho\pi$ it is possible that it is masked by the $\pi_2$.

Similar situations exist for the other $J^P=2^-$ mesons.  Clearly, it will
be important to search thoroughly through data that is already in hand, in
order to try and identify the missing states.  However, it is at least
equally important to understand phenomenologically why these states are
produced less copiously than their $J^P=1^-$ and $J^P=3^-$ counterparts,
if that is indeed the case.

One can perform a similar analysis of other multiplets.  
However, as we go higher in mass there are more channels available for decay
so that the meson widths become wider and wider (Barnes 1997).
In general, given
how complicated the meson spectrum is, it appears that a good starting 
place would be the strange mesons.  The reason for this is that in
the strange meson sector we don't have the additional problem of 
deciding whether new states are glueballs or conventional mesons and
we don't have the additional complication of mixing between isoscalar
states due to gluon annihilation.  Following this a detailed survey of $\phi$
states would be useful since they form a bridge between the heavy
quarkonia ($c\bar{c}$ and $b\bar{b}$) and the light quark mesons and would\
help us understand the nature of the confinement potential.

These of course are guidelines for the next generation of experiments,
which we now discuss.

% ======================================================================
%\include{rev-v6-6}
% ======================================================================
\newpage
\section{FUTURE DIRECTIONS}

To understand the remaining puzzles will require
new data with significantly higher statistics.
It is also important that the data come from different processes and
channels to produce hadronic states with different quantum numbers and
production mechanisms.

Among the highest priority goals of hadron spectroscopy is to establish
the existence, and to study the properties, of gluonic degrees of freedom
in the hadron spectrum.  New evidence discussed in earlier chapters just
begins to scratch the surface of this field.  There is surely much new
physics to be gleaned from a new generation of experiments and theoretical
calculations and modeling.
Another important step is to find some of the missing $q\bar{q}$ states,
including both the orbitally and radially excited multiplets.

It is unlikely that all this could be done simply by bump hunting.
Rather, we will need experiments with unprecedented statistics and
uniform acceptance so that high quality
partial wave analyses can filter by $J^{PC}$.  A useful
guide to the expected properties of excited quarkonia has been produced by
Barnes, Close, Page, and Swanson (Barnes 1997).
As an example, in  Sec.~\ref{secV:missing},
we discussed the Quark Model predictions for the properties of the missing
$L=2$ mesons.

We point out that
a number of results have not yet appeared in the journals.  A good
source for these are the proceedings of the Hadron\ '97 conference
(Chung and Willutzki, 1998).  In particular, we refer the reader to results
on light quark spectroscopy from Fermilab in central $pp$ collisions from E690
(Berisso, et al., 1998; Reyes, et al., 1998) and in high energy photoproduction
from E687 (LeBrun, 1998).

A number of complementary new facilities and experiments are on the horizon
(Seth, 1998).  A partial list of new and planned facilities, is given in
Table~\ref{tabVI:newfacs}.
These are described in some detail in the following sections.
\begin{table}[hbt]
\caption{Some Future Facilities for Studying Quark Gluon Spectroscopy}
\label{tabVI:newfacs}
\begin{center}
\begin{tabular}{ccccc}
\hline
            &            &            & Principle           & Approximate\\
Facility    & Beams      & Energy     & Reactions           & Date\\
\hline
DA$\Phi$NE  & $e^+e^-$   & $\sim1$~GeV& $\phi$ decays       & 1998\\
CLEO-III    & $e^+e^-$   & 10~GeV     & $B$ decays; $\gamma\gamma\to X$ & 1998\\
BaBAR       & $e^+e^-$   & 10~GeV     & $B$ decays; $\gamma\gamma\to X$ & 1999\\
KEK         & $e^+e^-$   & 10~GeV     & $B$ decays; $\gamma\gamma\to X$ & 1999\\
COMPASS     & $p$        & 400~GeV           & Central Production  & 1999\\
RHIC        & $p$, Nuclei& $\sim200$~GeV/$A$ & Central Production  & 1999\\
JHF         & $p$        & 50~GeV     & $\{\pi,K,p\}\to X$  & $\sim$2004\\
CEBAF       &$e^-,\gamma$& 12~GeV     & $\gamma p\to Xp$    & $\sim$2004\\
BEBC        & $e^+e^-$   & 3-4~GeV    & $J/\psi\to\gamma X$ & $\sim$2004\\
\hline
\end{tabular}
\end{center}
\end{table}

\subsection{DA$\Phi$NE at Frascati}

The DA$\Phi$NE $\phi$ factory (Zallo 1992), a high luminosity $e^+e^-$
collider operating at $\sqrt{s}\sim 1$~GeV, is nearing completion at
INFN-Frascati.  The main goal is to study
CP-violation, but it will also examine the nature of the $a_0(980)$ and
$f_0(980)$ scalar mesons through the radiative transitions
$\phi\to\{a_0,f_0\}\gamma$.  See (Achasov, 1997a).

$\phi$ factories also offer the possibility of studying low mass $\pi\pi$ 
production via two photon production.  The combination of relatively high 
luminosity with detectors optimized for detecting low momenta and 
energies should allow very detailed measurements of both charged and 
neutral modes to be made from threshold up to nearly 1 GeV.  This process 
can shed additional light on low mass scalar resonances.

\subsection{$B$-factories at CESR, SLAC, and KEK}

Although the primary motivation for a $B$ factory (Goldberg and Stone, 1989;
Bauer 1990, 1992) is to study CP violation in the B meson system,
light meson spectroscopy will also be addressed several ways.
Final states produced in the strong {\em and} weak decays of
$b$ and $c$ quarks will elucidate the structure of the light meson daughters.
Further, these high luminosity facilities will study
two-photon processes with very high statistics.
Lastly, exclusive radiative processes $\Upsilon\to\gamma X$ providing
information complementary to $J/\psi$ radiative decay.

Three such facilities in the final stages of construction are the CLEO-III
upgrade at CESR, the new BaBAR detector at the SLAC asymmetric collider, and
a new facility at KEK.
CLEO-III is an upgraded detector to go along with a luminosity boost in the
CESR storage ring, a classic and well-understood symmetric $e^+e^-$ collider
facility.  The SLAC $B$-factory is an asymmetric system (designed to boost
$B$-mesons in the lab frame) which has some implications for detection of
lower mass sytems, such as those produced in two-photon annihilation.
The KEK facility (Kurokawa, 1997) is similar to the SLAC experiment.

We also note that the LEP collaborations have made contributions to 
the subject of light meson spectroscopy and we should expect this to 
continue for at least the next several years.

\subsection{COMPASS at CERN}

CERN is about to
commission a new experiment on fixed targets to be operated in the years
prior to turn-on of LHC.  The {\bf CO}mmon {\bf M}uon and {\bf P}roton
{\bf A}pparatus for {\bf S}tructure and {\bf S}pectroscopy (COMPASS) is a
large magnetic system designed to detect multiparticle final states using
either muon or proton beams (von~Harrach, 1998).  The muon beam program is
mainly directed at
measurements of spin structure functions, but the proton beams will be used
for a number of hadronic production experiments, in particular
central production.

Operating such a multi-purpose apparatus has specific challenges, mainly
trying to optimize sensitivies to the
different situations. However, using a common apparatus with a large directed
collaboration has lead to a state-of-the-art high rate detector with
excellent momentum and particle identification capabilities.  First runs
with hadron beams are expected in 1999.
Central production measurements including
a RICH detector for particle identification are presently planned for 2002.

\subsection{RHIC at BNL}

The $Z^4$ dependence of two-photon production from charged particles 
implies an enormous cross section in heavy ion interactions.  However the 
luminosiites at heavy ion colliders are substantially less than the 
$e^+e^-$ factories so that in the end these two factors tend to balance 
out leaving the rates for two-photon physics generally lower than present
$e^+e^-$ facilities.  Still, reasonable event totals may be expected when
integrating over the long running periods, with a suitably triggered
detector.
The real strength of heavy-ion colliders may be the ability 
to compare the gluon-rich pomeron-pomeron interactions with  gluon-poor
photon-photon interactions in the same experiment.  

The Relativistic Heavy Ion Collider (RHIC) at BNL is nearing completion, and
at least one detector (STAR) will be able to trigger on low
mass, low multiplicity events from two-photon interactions and central
production (Nystrand and Klein, 1988).

\subsection{The Japanese Hadron Facility}

Peripheral hadronic production experiments still have plenty to contribute,
particularly using state of the art detection systems.  This is particularly
true if they can look at production mechanisms and final states that have been
largely ignored in the past.  This has been aptly demonstrated by E852 at
BNL, for example.  It is also important to recognize the lessons taught by
the LASS experiment as SLAC which, among other things, demonstrated the 
effectiveness of a programmatic approach and the value of analyzing
many different channels in the same experiment under conditions of uniform and
well understood acceptance.  Future experiments should follow these models.

The most promising scenario for such developments is the Japanese Hadron
Facility (JHF) which has been proposed for KEK.  This would be a high
intensity (10~$\mu$A) 50~GeV proton synchrotron, including polarized beams.
For comparison, the AGS at Brookhaven, one of the primary workhorses of
spectroscopy through peripheral hadroproduction, produces a few $\mu$A
proton beam, but with energy around 25~GeV.  The higher energy is worth
large factors in the secondary $\pi^\pm$ and $K^\pm$ beams at 20~GeV or so.

\subsection{CEBAF at Jefferson Laboratory}
\label{sec:JLab}

Peripheral photoproduction is conspicuously absent among the studied
production mechanisms in hadron spectroscopy.  This is despite the fact
that one expects profound new results from such experiments, as discussed
in Sec.~\ref{secIII:photons}.  The reason, however, is obvious.  Until
recently, virtually no suitable facilities have existed for detailed work
in this area.

The Continuous Electron Beam Accelerator Facility (CEBAF) at the Thomas
Jefferson National Accelerator Facility (Jefferson Lab) is a very high 
intensity electron beam facility with 4---6 GeV electron beams.  The beam
energy was centered around the laboratory's original primary goal, i.e the
study nuclear structure.  However, because of landmark improvements in the
the RF accleration cavities, beams with energies up to $\sim$8~GeV 
will be possible with minimal cost.  In fact, a plan is in place to push the
electron beam energy initially to 12~GeV, and hopefully later to 24~GeV.
High energy photons can be produced by either thin-radiator bremstrahlung 
or through a variety of other means.

An aggressive experimental plan is underway which will keep pace with the
accelerator improvements.  This involves the construction of a new
experimental hall at the site, which would include a dedicated experimental
facility.  Initial designs of this facility take directly from the LASS and
E852 experiences, with necessary modifications for photon beams.
State-of-the-art high rate data acquisition electronics and computing will be
an integral part of this new experiment.

\subsection{A $\tau$-Charm factor at BEBC}

$\tau$-charm factories, $e^+e^-$ colliders operating in the energy region 
of 3---4 GeV, have been proposed although none have yet been approved.
$\tau$-charm factories will produce large numbers of  $J/\psi$ and $D$ mesons.
The BEPC ring in China, presently home to the BES experiment, has worked well
and is considered a prototype for a higher luminosity facility at that
laboratory.  Present designs call for between a factor of 10 and 100
improvement in the luminosity over the present storage ring.  The project is
not yet approved, however.

Hadron spectroscopy can be pursued in several ways at the
$\tau$/charm Factory. These include $J/\psi$ and $\psi'$ decays,
decays of $\tau \to \nu_\tau + hadrons$ where the hadrons are
produced by a virtual W boson, and
semileptonic and leptonic decays of $D^\pm$, $D^0$.
Probably the most important question in light meson spectroscopy is the
existence of gluonic excitations, and as demonstrated earlier in this
review, $J/\psi\to\gamma X$ is a critical reaction for the
complete understanding of these states.

With the high statistics available it may even be possible to perform a 
partial wave analysis of the $\chi_{cJ}$ decay products produced in 
$\psi' \to \gamma \chi_{cJ}$ radiative decays.  For example $\chi_{c1}\to 
\pi H$ is sensitive to the hybrid exotic sector $H \; (J^{PC}=1^{-+})$ 
while $\chi_{c0} \to f_0(980) X$ would be a source of $0^{++}$ mesons.

In addition to high statistics searches for gluonic excitations a $\tau$-charm
factory will also study the properties of the $a_1(1270)$, $a_1'$, $\rho$, 
$\rho'$, $K^*$, $K_1(1270)$, and $K_1(1400)$
mesons via hadronic decays of the $\tau$ lepton.  These
observations will help disentangle the radially excitations of the vector
mesons.

% ======================================================================
%\include{rev-v6-7}
% ======================================================================
\newpage
\section{FINAL COMMENTS}

In this review we hope to have conveyed the sense that
meson spectroscopy is a lively, exciting subject with even the most 
basic questions still unresolved. We find it remarkable, that after over 20 
years of QCD we still do not know what are the physical states of the 
theory. 
Understanding QCD, and non-Abelian gauge theories in general, is one 
of the most important problems facing high energy and nuclear physics.  We 
outlined the important issues in light meson spectroscopy, the puzzles, 
and the open questions.  We have attempted so show where the field is 
heading with the next generation of experiments and how they can advance 
our knowledge of mesons spectroscopy.  In many cases, it is not only our 
lack of experimental data but a lack of answers to theoretical questions 
that has hindered progress in the field.  We have attempted to point 
these out.

Our survey of established meson states shows the basic validity of the 
constituent quark model.  However, the survey also 
highlights some possible discrepancies 
with the predictions of the quark model which may point to the need 
to go beyond the $q\bar{q}$ states and include gluonic excitations and
multiquark states in the light-quark sector.  In this sense we may be on
the verge of opening up a new frontier of hadron spectroscopy in the 
1.5 to 2.5 GeV mass region of the meson sector where the vast majority
of the complications seem to occur;  many $q\bar{q}$ states remain
to be discovered.  It is important to find them and then to pin down
the details of the $q\bar{q}$ spectrum to pave the ground for the search
for exotic states.  

Most of the significant deviations from the quark model occur in the
light-quark isoscalar sector.
Much of the current excitement is  with the scalar mesons.
Even if we relegate the $f_0(980)$ (along with the isovector $a_0(980)$)
to multiquark status, we still have to contend with the $f_0(1500)$ and
the $f_J(1710)$ (if indeed $J=0$).  It is quite possible that these two
states are mixtures of the ground state glueball and the $s\bar{s}$
scalar meson, but this will have to wait for better experiments and
clearer phenomenology to establish a consistent pattern.

The long outstanding $E/\iota$ problem remains with us.  The evidence
clearly points to two separate states each for $J^{PC}=1^{++}$ and $0^{-+}$,
where exactly one of each are expected based on $q\bar{q}$ degrees of
freedom.  Ascribing hybrid, glueball, or multiquark status to the extra
states is problematic, though, because no clear picture emerges from the
current set of experimental data.  New high statistics measurements of
$\gamma\gamma$ and $\gamma\gamma^\star$ production may be very helpful.
Certainly a better phenomenological understanding of reactions near
the $K^\star\bar{K}$ threshold is crucial for progress here.

Despite its long history, the $f_J(2220)$ is still enigmatic.  It may or
may not be a peculiar, narrow meson representing the $2^{++}$ glueball, and
it may or not have an underlying broad structure which is a manifestation
of $2^{++}$ or $4^{++}$ strangeonium.
We also continue to deal with a possible overpopulation of $2^{++}$
states, in particular the various $f_2$ states observed in $\phi\phi$ decay.

Finally, there are exotic mesons which necessarily imply a meson
state beyond $q\bar{q}$ models.  Evidence is finally beginning to sort
itself out in the $\eta\pi$ channel, and does now seem to point towards
an exotic $J^{PC}=1^{-+}$ state near 1380~MeV/$c^2$.  It is difficult
to accomodate this, however, in present models of excited gluonic degrees
of freedom.  Some evidence is now emerging for higher mass exotic states,
but more data will be necessary, particularly in reactions which are
expected to {\em enhance} the production rate.  A good candidate laboratory
is peripheral photoproduction, and new facilities are on the horizon.

Hadron Spectroscopy is undergoing a renaissance, 
taking place at many facilities worldwide.  We anxiously look forward to
new results.

% ======================================================================
%\include{rev-v6-k}
% ======================================================================
%\newpage
\section*{Acknowledgments}

This research was supported in part by the Natural Sciences and Engineering 
Research Council of Canada, and by the National Science Foundation in the US.
The authors thank
Gary Adams,
Ted Barnes,
Bob Carnegie,
Suh-Urk Chung,
John Cummings,
Bill Dunwoodie,
Alex Dzierba,
Richard Hemingway,
Nathan Isgur,
Joseph Manak,
Curtis Meyer,
Colin Morningstar,
Phillip Page,
Vladimir Savinov,
Eric Swanson,
Don Weingarten,
John Weinstein,
and Dennis Weygand for helpful conversations and communications.

% ======================================================================
%\include{rev-v6-r}
% ======================================================================
\newpage
%\begin{references}

\end{document}